\documentclass[fleqn,usenatbib]{mnras}

\usepackage[T1]{fontenc}

\DeclareRobustCommand{\VAN}[3]{#2}
\let\VANthebibliography\thebibliography
\def\thebibliography{\DeclareRobustCommand{\VAN}[3]{##3}\VANthebibliography}

\usepackage{graphicx}	
\usepackage{amsmath}	
\usepackage{amssymb}	
\usepackage{eso-pic}
\usepackage{pdflscape}
\usepackage{longtable}
\usepackage{threeparttablex}

\usepackage{newtxtext,newtxmath}

\AddToShipoutPictureBG*{%
  \AtPageUpperLeft{%
    \hspace{0.75\paperwidth}%
    \raisebox{-1.5\baselineskip}{%
      \makebox[0pt][l]{\textnormal{DES-2021-0655}}
}}}%

\AddToShipoutPictureBG*{%
  \AtPageUpperLeft{%
    \hspace{0.75\paperwidth}%
    \raisebox{-2.5\baselineskip}{%
      \makebox[0pt][l]{\textnormal{FERMILAB-PUB-22-526-PPD}}
}}}%

\defcitealias{li11}{L11}

\title[Core-collapse Supernovae in the Dark Energy Survey]{Core-collapse Supernovae in the Dark Energy Survey: Luminosity Functions and Host Galaxy Demographics}

\author[DES Collaboration]{
\parbox{\textwidth}{
\Large
M.~Grayling,$^{1}$
C.~P.~Guti\'errez,$^{2,3,1}$
M.~Sullivan,$^{1}$
P.~Wiseman,$^{1}$
M.~Vincenzi,$^{4,1}$
L.~Galbany,$^{5,6}$
A.~M\"oller,$^{7}$
D.~Brout,$^{8}$
T.~M.~Davis,$^{9}$
C.~Frohmaier,$^{4,1}$
O.~Graur,$^{4}$
L.~Kelsey,$^{4,1}$
C.~Lidman,$^{10,11}$
B.~Popovic,$^{12}$
M.~Smith,$^{1}$
M.~Toy,$^{1}$
B.~E.~Tucker,$^{11}$
Z.~Zontos,$^{1}$
T.~M.~C.~Abbott,$^{13}$
M.~Aguena,$^{14}$
S.~Allam,$^{15}$
F.~Andrade-Oliveira,$^{16}$
J.~Annis,$^{15}$
J.~Asorey,$^{17}$
D.~Bacon,$^{4}$
E.~Bertin,$^{18,19}$
S.~Bocquet,$^{20}$
D.~Brooks,$^{21}$
A.~Carnero~Rosell,$^{22,14,23}$
D.~Carollo,$^{24}$
M.~Carrasco~Kind,$^{25,26}$
J.~Carretero,$^{27}$
M.~Costanzi,$^{28,24,29}$
L.~N.~da Costa,$^{14}$
M.~E.~S.~Pereira,$^{30}$
J.~De~Vicente,$^{17}$
S.~Desai,$^{31}$
H.~T.~Diehl,$^{15}$
P.~Doel,$^{21}$
S.~Everett,$^{32}$
I.~Ferrero,$^{33}$
D.~Friedel,$^{25}$
J.~Frieman,$^{15,34}$
J.~Garc\'ia-Bellido,$^{35}$
M.~Gatti,$^{36}$
D.~Gruen,$^{20}$
J.~Gschwend,$^{14,37}$
G.~Gutierrez,$^{15}$
S.~R.~Hinton,$^{9}$
D.~L.~Hollowood,$^{38}$
K.~Honscheid,$^{39,40}$
D.~J.~James,$^{8}$
K.~Kuehn,$^{41,42}$
N.~Kuropatkin,$^{15}$
G.~F.~Lewis,$^{43}$
U.~Malik,$^{11}$
M.~March,$^{36}$
F.~Menanteau,$^{25,26}$
R.~Miquel,$^{44,27}$
R.~Morgan,$^{45}$
R.~L.~C.~Ogando,$^{37}$
A.~Palmese,$^{46}$
F.~Paz-Chinch\'{o}n,$^{25,47}$
A.~Pieres,$^{14,37}$
A.~A.~Plazas~Malag\'on,$^{48}$
M.~Rodriguez-Monroy,$^{49}$
A.~K.~Romer,$^{50}$
A.~Roodman,$^{51,52}$
E.~Sanchez,$^{17}$
V.~Scarpine,$^{15}$
I.~Sevilla-Noarbe,$^{17}$
E.~Suchyta,$^{53}$
G.~Tarle,$^{16}$
C.~To,$^{39}$
D.~L.~Tucker,$^{15}$
and T.~N.~Varga$^{54,55,56}$
\begin{center} (DES Collaboration) \end{center}
\textit{Affiliations are listed at end of paper}
}
}

\date{Accepted XXX. Received YYY; in original form ZZZ}

\pubyear{2022}

\begin{document}
\label{firstpage}
\pagerange{\pageref{firstpage}--\pageref{lastpage}}
\maketitle

\begin{abstract}
We present the luminosity functions and host galaxy properties of the Dark Energy Survey (DES) core-collapse supernova (CCSN) sample, consisting of 69 Type II and 50 Type Ibc spectroscopically and photometrically-confirmed supernovae over a redshift range $0.045<z<0.25$. We fit the observed DES $griz$ CCSN light-curves and K-correct to produce rest-frame $R$-band light curves. We compare the sample with lower-redshift CCSN samples from Zwicky Transient Facility (ZTF) and Lick Observatory Supernova Search (LOSS). Comparing luminosity functions, the DES and ZTF samples of SNe II are brighter than that of LOSS with significances of 3.0$\sigma$ and 2.5$\sigma$ respectively. 
While this difference could be caused by redshift evolution in the luminosity function, simpler explanations such as differing levels of host extinction remain a possibility. We find that the host galaxies of SNe II in DES are on average bluer than in ZTF, despite having consistent stellar mass distributions. We consider a number of possibilities to explain this -- including galaxy evolution with redshift, selection biases in either the DES or ZTF samples, and systematic differences due to the different photometric bands available -- but find that none can easily reconcile the differences in host colour between the two samples and thus its cause remains uncertain.
\end{abstract}




\section{Introduction}

Core-collapse supernovae (CCSNe) are among the most complex and diverse astrophysical events, demonstrating a wide range of spectroscopic and photometric properties \citep{Filippenko97, GalYam17, Modjaz19}. Type II SNe show hydrogen features in their spectra due to the outer hydrogen envelope of the progenitor star, while stripped envelope SNe (i.e., SNe Ib, Ic, IIb) present different features depending on the degree to which the outer hydrogen and helium envelopes have been stripped away by processes such as stellar winds \citep[e.g.,][]{Woosley93} and binary interaction \citep[e.g.,][]{Nomoto95}. While it is generally understood that CCSNe result from the cessation of fusion in the cores of massive stars after the formation of iron leading to the core collapsing, there is a great deal that remains uncertain about the exact mechanisms of this process. Studying the properties of populations of CCSNe can help constrain our knowledge of the processes involved in the explosion \citep[][]{li11, richardson14}.

The most straightforward population diagnostic is the luminosity function, the distribution of peak luminosities observed across the SN population. An accurate knowledge of the luminosity function is important in simulating CCSN explosions to ensure they exhibit the range of properties of the observed population. Luminosity functions are also important when simulating sky surveys to ensure simulated SNe are created with realistic properties. Simulations of this nature are used for a number of applications, for example calculating SN rates \citep[e.g.,][]{bazin09, Graur17a, Frohmaier21}, optimising observing strategies and preparing for upcoming surveys \citep[e.g.,][]{jones17, Villar18} and modelling the contamination of CCSNe in cosmological samples of SNe Ia \citep[e.g.,][]{Maria2019, Maria2021}. 

The host galaxies of SNe provide further demographic information about their properties and sample the stellar populations from which the progenitor star is drawn. CCSNe are generally found across a wide variety of star-forming host galaxy environments \citep[e.g.][]{Anderson2010, Graur17b}, while the most luminous transients \citep[e.g., superluminous supernovae, broad line SNe Ic, see e.g.][]{Angus2016, Perley16, Modjaz2020} and rapidly-evolving transients \citep[RETs, ][]{Wiseman21RET} are typically found in low mass, low metallicity and/or strongly star-forming environments. SNe Ib/c have also been shown to more closely trace underlying star formation in their host galaxies than SNe II \citep[e.g.,][]{Anderson2009, Galbany18}.

Several studies have examined CCSNe luminosity functions, primarily in the local universe. For example, \citet[][hereafter L11]{li11} produced luminosity functions for all SNe in the Lick Observatory Supernova Search \citep[LOSS;][]{LOSS2000}, including 105 CCSNe, and \citet{richardson14} presented luminosity functions based on data from the Asiago Supernova Catalogue \citep{Barbon1989} supplemented by further SNe from other studies. Here, we measure luminosity functions based on SNe detected by the Dark Energy Survey (DES) Supernova Program \citep[DES-SN,][]{DES-SN}. DES-SN is a deep, untargeted, five-season rolling SN search survey over 27\,deg$^2$ of sky. This leads to a higher redshift sample than presented in previous work \citep{Smith20}, providing an opportunity to study any redshift evolution in the CCSN luminosity function and the effect of any evolution in the SN host galaxy populations \citep[e.g.][]{Elbaz11}: such an evolution may in turn lead to an evolution in the CCSN population. The Zwicky Transient Facility's (ZTF) Bright Transient Survey \citep{Perley20} provides an additional sample of CCSNe which lie in a redshift range between that of DES and LOSS. We also include this sample in our analysis to allow for further investigation of any redshift evolution.

In this paper, we present luminosity functions and host galaxy properties of CCSNe in DES, and compare these to samples from LOSS and ZTF. In Section~\ref{samples}, we detail these three samples and describe the selection of objects suitable for inclusion in a luminosity function. Section~\ref{LF_method} describes the method used to construct luminosity functions for both DES and ZTF, and presents the luminosity functions themselves. In Section~\ref{hosts} we discuss the host galaxy demographics of the samples and the correlations between different host properties and peak SN luminosity. We discuss our results in Section~\ref{discussion} and conclude in Section \ref{conclusion}. Throughout, we assume a flat $\Lambda$CDM cosmology with $\Omega_M = 0.3$ and H$_0 = 70$ km s$^{-1}$ Mpc$^{-1}$, and correct external samples to this cosmological model as required. All photometry has been corrected for the effects of Milky Way extinction using dust maps presented in \citet{Schlegel98} and re-calibrated in \citet{ext_map}, assuming $R_V = 3.1$. All quoted magnitudes are in the AB system \citep{Oke83}.

\section{Core-collapse Supernova Samples}
\label{samples}

We begin by presenting the different CCSN samples we have used for this analysis, detailing the selection criteria applied to the DES and ZTF samples and how we use the data presented in \citetalias{li11}. We do not compare with the luminosity functions presented in \citet{richardson14}. This is because the sample presented is from a wide variety of different instruments and surveys which makes it very difficult to correct for Malmquist bias based on the limiting magnitude of each survey as we do in this work (see Section \ref{Vmax}). \citet{Drout11} presents a collection of 25 SNe Ibc and includes peak absolute magnitudes, however, all of these objects are brighter than -17.5, which suggests that only the most intrinsically luminous objects are included. As a result, this sample is not included in our comparisons. A sample of CCSNe is available from SDSS-II \citep[e.g.,][]{Taylor14}, however, peak luminosities and Malmquist bias corrections are not available for these objects to allow for a comparison. Finally, \citet{Arcavi12} and \citet{Kiewe12} present SNe II and IIn, respectively, from the Caltech Core-Collapse Project \citep[CCCP]{Gal-Yam07}. However, these samples combined contain relatively few objects with only 12 that have estimated peak absolute magnitudes; a further 9 have absolute magnitudes but of the plateau phase of a SN II rather than peak and 5 have only lower limits for the peak. As a result, we do not compare to this sample in this analysis.

Throughout this analysis, we treat CCSNe as two general classes rather than subdividing further. This is to ensure sufficient numbers of SNe in each class, and to acknowledge the uncertainties in the photometric SN classification that we use in the DES-SN sample. We refer to Type II SNe to include all hydrogen-rich SNe and Type Ibc SNe to include all hydrogen-poor/stripped-envelope SNe. Although SNe IIb, such as the very luminous SN IIb DES14X2fna in the DES-SN sample \citep{Grayling21}, do show hydrogen features at early times, they also have a partially stripped outer hydrogen envelope and are included with SNe Ibc for this analysis. Table~\ref{sample_sizes} contains summary information for each sample.

\begin{table*}
\caption{Sample sizes for our luminosity functions after applying selection criteria.}
\centering
\begin{tabular}{ ccccccc }
	\hline
	Survey & \multicolumn{2}{c}{Total Sample Size} & \multicolumn{2}{c}{Sample After Quality Cuts} & \multicolumn{2}{c}{Sample After Redshift and Magnitude Cuts} \\
	 & SNe II & SNe Ibc & SNe II & SNe Ibc & SNe II & SNe Ibc \\
	\hline
	DES (spectroscopically confirmed CCSNe) & 52 & 18 & 33 & 13 & 31 & 11 \\
	DES (photometric CCSNe with host spec-z) & -- & -- & 56 & 42 & 38 & 39 \\
	LOSS & 69 & 36 & -- & -- & 37 & 21 \\
	ZTF & 349 & 162 & 214 & 105 & 174 & 89 \\
	\hline
\end{tabular}
\label{sample_sizes}
\end{table*}

\subsection{The DES core collapse supernova sample}

The DES-SN CCSN sample contains three categories of objects: those with a spectroscopic confirmation, those with a spectroscopic redshift of the host galaxy, and those with photometric redshift information for the host galaxy. We discuss each of these in turn.

\subsubsection{Spectroscopically confirmed CCSNe}
\label{DES_sample_snspec}

The DES-SN CCSN sample has 70 spectroscopically-confirmed CCSNe between redshifts $0.045<z<0.33$. These were obtained over a variety of telescopes and instruments during the course of the DES survey \citep[][]{smith2020}.

We apply the following selection criteria to ensure that the light curve can be analysed to measure the peak SN brightnesses required for the luminosity function:
\begin{itemize}
    \item Each SN must have photometric coverage before and after maximum to ensure an accurate interpolation of the peak luminosity.
    \item Each object must have a well-constrained explosion epoch inferred either from the date of last non-detection of the SN, or from spectral template matching using the Supernova Identification code \citep[\textsc{snid};][]{SNID} following the prescriptions of \citet{Gutierrez17a}. An explosion epoch is required to select an appropriate model spectral energy distribution (SED) at each epoch for the K-correction of observed photometry to the rest-frame, as the SED models we use are defined with respect to explosion. For objects with pre-explosion non-detections, we assume an explosion epoch halfway between the last non-detection and the first detection.
    \item Each object must have at least 9 detections which are deemed real by a supervised machined learning classifier \citep[see][for details of the classifier]{Goldstein15}. A limit of 9 is selected to maximise both the sample size and photometric coverage, as overall this cut eliminates only 4 objects with the next object only having 6 detections deemed real by the classifier.
\end{itemize}

This selection leaves 46 spectroscopically-confirmed CCSNe in DES. Of these, 33 are spectroscopically hydrogen-rich (Type II) and 13 hydrogen-poor/stripped-envelope (Type Ibc). 

\subsubsection{Photometric CCSNe with host spec-z}
\label{DES_sample_specz}

DES also detected a much larger number of transients which have no spectroscopic confirmation. We first investigate transients which have a spectroscopic measurement of the SN host galaxy redshift (spec-z); for example, from the OzDES survey \citep{Yuan2015, Childress2017, Lidman2020} or external redshift catalogues - \citet[][]{Maria2021} presents details of the different host galaxy redshift sources using galaxy associations from \citet{stacks}. To this sample, we apply a number of cuts to select objects that:

\begin{itemize}
    \item were detected in at least nine epochs based on the classifier detailed in \citet{Goldstein15} to ensure good photometric coverage, matching the cut applied to spectroscopically confirmed sample in Section \ref{DES_sample_snspec}.
    \item have an assigned host galaxy with a spectroscopic redshift less than 0.3 (as an initial redshift cut).
    \item were single-season transients, removing some obvious AGN.
\end{itemize}
These cuts gave a total of 1609 transients. We remove all spectroscopically-confirmed objects that are not CCSNe (e.g., SNe Ia, AGN) and remove the remaining AGN using the classifier discussed in Section 2.2.3 of \citet{Wiseman21RET}. We next remove SNe Ia from the sample employing the photometric SN classifier \textsc{SuperNNova} \citep[][]{supernnova}, using the trained model discussed in \citet{Vincenzi22}. We follow \citet{Wiseman21}, \citet{Moller22} and \citet{Vincenzi22} in removing all objects with a probability of being a SN Ia ($P_\textrm{Ia}$) of greater than 0.5\footnote{In most cases, $P_\textrm{Ia}$ is close to 0 or 1 meaning that our results are not sensitive to this threshold}. For the remaining objects, we apply the same quality cuts in terms of requiring coverage pre- and post-peak and a well-constrained explosion date as in Section \ref{DES_sample_snspec} and visually inspect the photometry of the remaining objects to ensure they are consistent with a CCSN; 51 objects are excluded as they are active in several observing seasons and do not resemble SNe, while 1 is excluded as it lacks any $r$-band data. We then classify remaining objects as SNe II or SNe Ibc using the light curve template fitter pSNid \citep{Sako11} -- 8 SNe are excluded as they are either classified as SNe Ia by pSNid or are consistent with both SNe II and Ibc. This leaves a sample of 98 photometrically-confirmed CCSNe from DES, of which 56 are SNe II and 42 SNe Ibc; combined with spectroscopically-confirmed objects, the total DES CCSN sample with spectroscopic redshifts has 89 SNe II and 55 SNe Ibc. The requirement for constraint of peak luminosity means that the quality cuts we apply are relatively strict, making this sample smaller than might be used for other purposes such as rate calculations.

To assess the suitability of pSNid for these purposes, we evaluate its performance for the full sample of 70 spectroscopically-confirmed SNe in DES; we obtain an estimated pSNid class for 69 of these. For SNe II, there are 8 misclassifications out of 51 - 6 are classified as SNe Ia and 2 as SNe Ibc by pSNid. For SNe Ibc, there are 2 misclassifications out of 18 objects, with one misclassified as a SN Ia and one as a SN II. This gives accuracy for these classes of 84 and 89 per cent respectively and an overall accuracy of 86 per cent. We opt to use SuperNNova to remove SNe Ia and then pSNid to separate SNe II from SNe Ibc, rather than simply using pSNid for all SNe, as SuperNNova has been shown to have very high performance upwards of 98 when separating SNe Ia from non-SNe Ia \citep[][]{supernnova, Vincenzi22}. Out of all DES transient candidates, of all objects classified as SNe Ia by SuperNNova around 5 per cent are classified as CCSNe rather than SNe Ia by pSNid. The good level of agreement between these classifiers is reassuring, but we favour SuperNNova due to its high performance. It is not possible to use SuperNNova for both tasks as a suitable SuperNNova model for multi-class classification trained on DES-like light curves is not currently available.

It should be noted that we do not apply a confidence threshold based on chi-squared ($\chi^2$) - for example, if the SN II template in pSNid is a much better fit to a given SN than either the SN Ia or Ibc templates, it was classified as a SN II regardless of the SN II template $\chi^2$ value when fitting the light curve. Given all the checks and cuts we apply to remove SNe Ia and other types of transients prior to using pSNid, we can be confident that all remaining objects are CCSNe and are therefore justified in not using a $\chi^2$ cut. Nevertheless, as a check, we analyse the properties of the sample excluding the 20 objects with the worst $\chi^2$ values and find that this does not impact the trends we observe.

\subsubsection{Photometric CCSNe with host photo-z}

Beyond the sample of CCSNe with spectroscopic host redshifts, there are a number of CCSNe in DES with no spectroscopic information, for example because the host was too faint to take a reliable spectrum. While we cannot directly include these objects in our luminosity function, it is important to understand any selection effects that arise from excluding these objects from the sample. 

Photometric redshifts (photo-zs) have been produced for three of the ten DES-SN fields from coadded photometry as outlined in \citet{DES3year}, based on the photometric redshift fitting code EAzY \citep{Brammer2008}. To produce a sample of DES CCSNe without spectroscopic host redshifts, we take all real transient candidates in the full DES-SN sample, select those located in hosts with photometric redshifts, and remove known AGN and other transient types such as variable stars using existing catalogues. We use \textsc{SuperNNova} to remove all likely SNe Ia using a model trained without spectroscopic redshift information - this cut leaves 45 objects (again, using $P_\textrm{Ia}>0.5$). Finally, we visually inspect each light curve to remove other types of transient that are clearly not SN-like in nature (e.g., AGN) and apply the same quality cuts as in Sections \ref{DES_sample_snspec} and \ref{DES_sample_specz}, leaving 25 CCSNe.

As we apply redshift cuts in this analysis and there are large uncertainties in photometric redshifts, the exact size of this sample is not fixed but typically varies between 3 and 8 -- this is discussed in detail in Appendix~\ref{appendix:host}. In brief, this analysis suggests that we obtain spectroscopic host redshifts for $\sim 75 - 90$ per cent of CCSNe observed by DES. This sample of objects with photometric redshifts is used only for selection efficiency checks.

\subsection{The ZTF Bright Transient Survey}

\citet{Perley20} presents a public catalogue of transients from the ZTF Bright Transient Survey with spectroscopic classifications. Excluding SNe Ia and super-luminous SNe, this sample consists of 511 CCSNe. For all of these objects, we gather publicly-available $g$ and $r$-band photometry from the Lasair\footnote{Available at \url{https://lasair.roe.ac.uk/}} transient broker \citep{Lasair}. We apply the same cuts as for the DES sample, only including objects with photometric coverage both pre and post-peak in both bands and with a well-constrained explosion date. This leaves a sample of 319 CCSNe from ZTF: 214 SNe II and 105 SNe Ibc, applying our broad classifications as described earlier.

\subsection{LOSS}

The Lick Observatory Supernova Search \citep[LOSS;][] {LOSS2000} was a galaxy-targeted SN survey that monitored approximately 5000 nearby galaxies for transients using the Katman Automatic Imaging Telescope (KAIT). Luminosity functions of different SN sub-types from LOSS were presented in \citetalias{li11}, and we re-categorise these into our broader classifications. Note that L11 includes SNe IIb with SNe II rather than SNe Ibc, hence the LOSS luminosity functions presented here will differ slightly from L11. With this classification scheme the LOSS sample contains 105 CCSNe: 69 SNe II and 36 SNe Ibc. It should be noted that \citet{Shivvers17} revisits the classification of the LOSS sample, with some object classes modified from \citetalias{li11}. However, because we are using broad labels of SNe II and Ibc, in all cases the new class falls into the same category as the original. In \citet{Shivvers17}, there are a small number of SNe which show hydrogen lines with only a single spectrum which are presented as having an uncertain class of either SN II or SN IIb. We class these objects as SNe II as robust classification of a SN IIb requires multiple spectra showing the transition from hydrogen to helium, although changing this has little impact on our results.

Note that for this analysis we correct the LOSS absolute magnitudes presented in \citetalias{li11} from H$_0 = 73$ km s$^{-1}$ Mpc$^{-1}$ to H$_0 = 70$ km s$^{-1}$ Mpc$^{-1}$ and also convert from Vega to AB magnitudes using conversions from \citet{Blanton07}. This is done to ensure consistency with DES and ZTF.

\subsection{Host galaxy properties}
\label{sec:host-galaxy-measurements}

We also assign every SN across the three samples to a host galaxy, and estimate the physical properties of those hosts. We use DES host galaxy associations and $griz$ photometry from \citet{stacks}. We perform galaxy SED fits based on the SED models produced by the spectral evolution code \textsc{P\'{E}GASE.2} \citep[][]{PEGASE1, PEGASE3} following the procedure as outlined in \citet{Smith20} and \citet{Kelsey20}, assuming a Chabrier initial mass function \citep[][]{Chabrier03}. This provides us with host galaxy stellar masses, star formation rates (SFRs) and rest-frame colours. These fits require an input redshift: for the sample with photometric redshifts, the larger redshift uncertainties must be accounted for. We calculate distributions of host properties for this sample which factor in redshift uncertainty using a method outlined in Appendix~\ref{appendix:host}.

For LOSS, several choices for stellar masses are available. Host galaxy stellar masses are presented in L11 using $K$ and $B$-band mass-to-luminosity ratios, and \citet{Graur17a} presents stellar mass and SFR values for LOSS hosts from Sloan Digital Sky Survey \citep[SDSS, ][]{SDSS} spectroscopy. In addition, the majority of LOSS hosts have stellar mass and SFRs calculated in the literature using a variety of different methods, including using near-infrared and far-ultraviolet flux \citep{Leroy19, Karachentseva20}. However, for consistency across our samples, we obtain $ugriz$ photometry for the hosts from SDSS and follow the same SED-fitting procedure described above. Each SN in LOSS is already matched to a host in Table 4 of \citet{LOSS1}, which we match to a corresponding SDSS galaxy.

Out of 58 LOSS CCSNe in our sample, 30 fall in the SDSS footprint and we are able to match to an SDSS host for 26 of these. To assess the quality of our method, we compare the stellar mass and SFR values calculated from SED fits to previously published literature values (see Appendix~\ref{appendix:SED}). We find that our inferred stellar masses from SED fitting are consistent with other methods although there is unsurprising scatter in the SFR values derived from SED fits that are known to be difficult to measure using only $ugriz$ data \citep{Childress13}. As rest-frame $U-R$ colour correlates with morphology and traces star formation \citep[][]{Lintott08, Trayford16}, we instead use rest-frame $U-R$ colour as a proxy for star-formation. We opt to use $U-R$ rather than SFR because it is more directly linked to the observed photometry and is not dependent on the star formation history (SFH) model used in the SED fits. By contrast, the SFR is estimated based on the average SFR over the 250\,Myr prior to the best-fit timestep in the SFH (see section 2.2.2 of Smith et al. 2020). It is thus dependent on the choice of that SFH (and sensitive to other assumptions that we make) and is not \textit{directly} linked to any observable. The $U-R$ colour is directly linked to the observed colours, modulo a $k$-correction (for which the best-fit SED is used). We also find that $U-R$ correlates well with our inferred specific star formation rate (sSFR) values.


For ZTF, 
we again search for host $ugriz$ photometry in SDSS, using a broad search radius of 50\arcsec\footnote{This was set to a large value to ensure that large, local galaxies were matched correctly} radius around the SN position, matching to the closest galaxy and then visually confirming the matches. Out of the 263 CCSNe in our ZTF sample, 212 objects lie within the SDSS footprint and we are able to match 203 of these to an SDSS galaxy.

\section{Core collapse SN luminosity functions}
\label{LF_method}

We construct luminosity functions for the DES and ZTF samples using the following procedure:

\begin{enumerate}
    \item We interpolate the observed photometry to obtain simultaneous observations in all photometric bands ($griz$ in the case of DES, $gr$ for ZTF) using Gaussian Processes (GP; \citealt{GP_Rasmussen}). We use the \textsc{python} package \textsc{george} \citep{hodlr}, following the process outlined in \citet{slsne}. Each photometric band is interpolated separately.
    \item We K-correct this interpolated observed photometry to the rest-frame, using SED models for SNe II from \citet{dessart_models}\footnote{Available at \url{https://www-n.oca.eu/supernova/home.html}} and for SNe Ibc from \citet{Levan2005}\footnote{Available at \url{https://c3.lbl.gov/nugent/nugent_templates.html}}. At each epoch with observations, we interpolate the time series SED to obtain a model SN SED. We then warp the model SED to colour-match it to our GP-interpolated photometry in all bands and use this spectrum for the K-correction.
    \item This K-corrected rest-frame photometry is then again interpolated using GPs in order to estimate the peak luminosity of each object as well as its corresponding uncertainty.

\end{enumerate}
    
We make an additional selection on peak absolute magnitude and redshift for each survey to produce the luminosity functions. The absolute magnitude limit of our combined sample is set by DES as it is the highest redshift survey and thus shallowest in terms of absolute magnitude: we exclude objects with a peak absolute magnitude fainter than $-16$\,mag in $R$-band. We also exclude objects brighter than $-19.5$\,mag in $R$-band to ensure a like-for-like comparison between the samples as these are only present in ZTF.

We make a redshift selection in the DES sample of $z<0.25$ to obtain a volume-limited sample above our absolute magnitude limit, and similarly use a redshift selection for the ZTF sample of $z<0.06$. This means that our ZTF sample does not overlap in redshift with the DES sample. This leaves 69 SNe II and 58 SNe Ibc in DES, 37 SNe II and 21 SNe Ibc in LOSS, and 177 SNe II and 89 SNe Ibc in ZTF, as detailed in Table \ref{sample_sizes}. 

\begin{figure}
\centering
\includegraphics[width = \columnwidth]{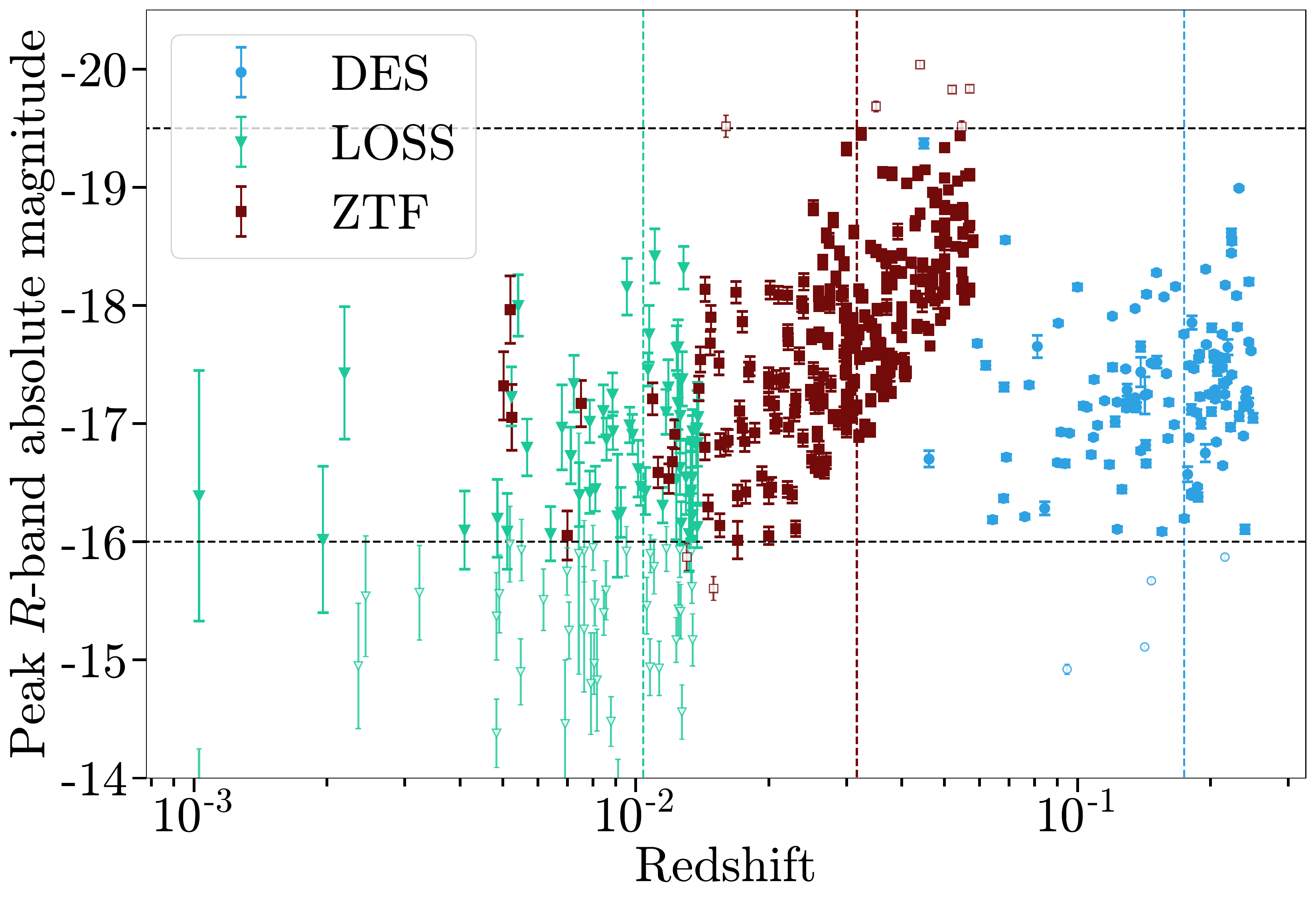}
\caption{Peak rest-frame CCSN $R$-band absolute magnitude, corrected for Milky Way extinction, plotted against redshift for the DES, LOSS and ZTF CCSN samples. Horizontal dashed lines indicate the median redshift of each SN sample, while vertical dashed lines represent the bounds of the absolute magnitude selection cut we apply. Closed symbols denote SNe included in the final samples and open circles are SNe excluded by the cut.}
\label{Mvsz_dist}
\end{figure}

Fig.~\ref{Mvsz_dist} shows our final samples, including those objects removed by this selection. As shown in Fig. \ref{Mvsz_dist}, Malmquist bias (i.e. bias towards more luminous objects at higher redshifts) is seen in our samples. The ZTF sample in particular shows a strong trend towards more luminous supernovae at higher redshift - the redshift cut at 0.25 and greater depth of DES means that it is less affected by this, while the local nature of LOSS means that this sample has good completeness over the absolute magnitude range we are studying.

\subsection{Correcting for Malmquist Bias}
\label{Vmax}

We correct for Malmquist bias using a simple $V_\text{max}$ correction \citep{schmidt1968}. This weights fainter objects, which would not be detected (or followed up) over the full survey volume, higher in the luminosity function calculation. For a volume-limited sample with an upper redshift limit $z_{\text{survey}}$, each SN has an upper redshift limit $z_{\text{max}}$, beyond which the object would fall below the detection limit of the survey. We calculate the weight $w$ each object makes to the luminosity function according to
\begin{equation}
    w= 
    \begin{cases}
        \Big(\frac{d_c(z_{\text{survey}})}{d_c(z_{\text{max}})}\Big)^3 & \text{if } z_{\text{max}} < z_{\text{survey}} \\
         1 & \text{otherwise}
    \end{cases}
\label{eq:weighting}
\end{equation}
where $d_c$ is the comoving distance. Thus, an intrinsically luminous SN that could have been detected over the full survey volume is given a weight of 1, while a fainter SN is assigned an increased weight.

This approach assumes that each survey has a magnitude limit above which it is complete. For the DES sample we use limits of $m=23.5$ and $m=24.5$ for the shallow and deep fields respectively \citep{Kessler15}. In reality, completeness in SN surveys is more complex than a simple cut-off, and thus this assumption introduces some uncertainty in the analysis; however, we find that altering these limits within $\pm0.5$\,mag has no significant effect on the luminosity distributions.

For ZTF we consider the 97, 93 and 75 per cent spectroscopic completeness limits of $18$, $18.5$ and $19$\,mag respectively \citep{Perley20}. In brief, we find that we obtain consistent luminosity functions with a limit of either 18.5 or 19\,mag and that a limit of 18\,mag causes the sample to miss fainter supernovae (see Appendix \ref{appendix:ztf_lim}). As a result, we use 19\,mag as the magnitude limit in this analysis to maximise the sample size.

Finally, we experimented with using the $V_\mathrm{max}$ correction for the LOSS sample. However, as expected we found that the sample is complete in the absolute magnitude range we are studying.

We now form the luminosity functions for the three SN samples. 
We incorporate the weighting (equation~\ref{eq:weighting}) into our cumulative distributions using
\begin{equation}
    C(M_n) = \frac{\sum_{i=1}^nw_i}{\sum_{i=1}^Nw_i},
\end{equation}
where $C(M_n)$ is the cumulative density up to absolute magnitude $M$, $n$ is the index position of in the sorted distribution of $M$ values, $i$ is the index of each supernova and $N$ is the total number of objects. In this section we use the two-sample Kolmogorov–Smirnov (KS) test to compare different luminosity functions -- this weighted cumulative distribution is incorporated into all KS tests in this section.

\subsection{CCSN luminosity functions}
\label{ccsn_lf}

\begin{figure}
\centering
\includegraphics[width = \columnwidth]{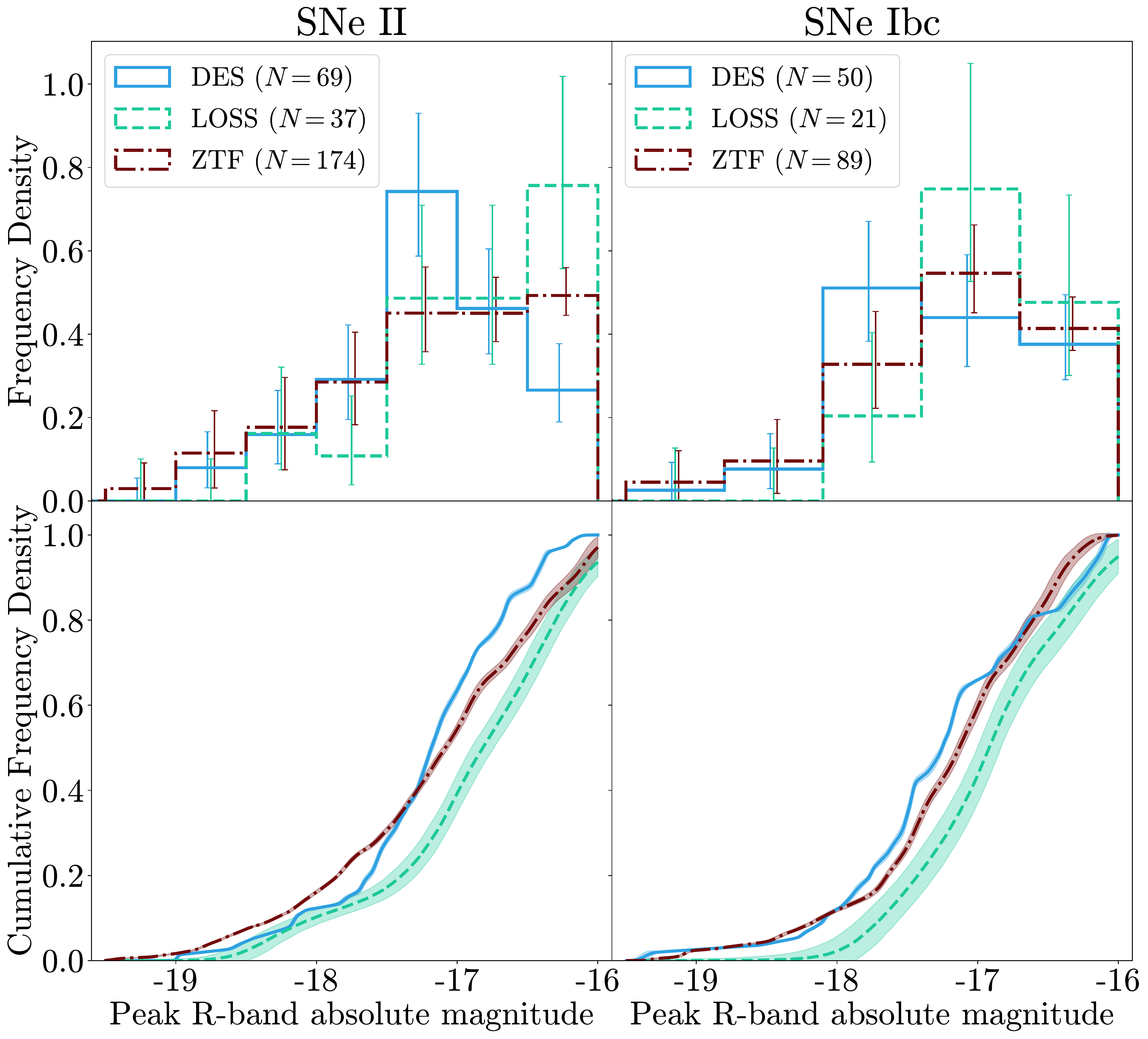}
\caption{SNe II and Ibc  $R$-band luminosity functions for the DES, LOSS and ZTF samples. Each event has been weighted by equation~\ref{eq:weighting} ($V_\text{max}$ correction). Histogram uncertainties are from the Poisson distribution confidence limits of \citet{Gehrels86}, while CDF uncertainties are derived from a Monte Carlo approach based on the measurement uncertainty of each value.}
\label{lfs}
\end{figure}

The left two panels of Fig.~\ref{lfs} shows the SN II luminosity functions. Histogram uncertainties in the upper panel represent the expected uncertainties from Poisson statistics and are derived from confidence limits presented in \citet{Gehrels86}, while the cumulative density function (CDF) uncertainties in the lower panel represent the statistical uncertainties in the individual measurement and are estimated from a Monte Carlo (MC) approach described as follows: 

\begin{itemize}
    \item The measured values of peak absolute magnitude and their uncertainties are used as the mean and standard deviations of a Gaussian distribution.
    \item 1000 randomised CDFs are generated using the Gaussian distribution of each data point.
    \item The mean and standard deviation of the CDFs at each value are calculated -- these are the values and uncertainties plotted.
\end{itemize}

Uncertainties here will depend on both the uncertainty on the luminosity of each SN and also on the sample size, as for a smaller sample changing an individual measurement will have a larger effect on the CDF. As can be seen, LOSS overall has a larger uncertainty in the CDF than DES in this figure. This is because DES photometry has lower uncertainties than LOSS photometry, meaning that the Gaussian distributions of each point are narrower.

Table~\ref{lf_ks} shows the results of two-sample KS tests between the different samples, with the $p$-values converted to a significance in $\sigma$. The DES sample overall appears brighter than both LOSS and ZTF, with significances of 3.0$\sigma$ and 1.8$\sigma$ respectively. The ZTF sample is also brighter than LOSS at a significance level of 2.5$\sigma$.

\begin{table}
\caption{Two-sample KS tests between the SN II and SN Ibc luminosity functions.}
\centering
\begin{tabular}{ cccc }
	\hline
	Survey 1 & Survey 2 & \multicolumn{2}{c}{KS test significance} \\
	& & SNe II & SNe Ibc \\
	\hline
	DES & LOSS & 3.0$\sigma$ & 1.9$\sigma$ \\
	DES & ZTF & 1.8$\sigma$ & 1.1$\sigma$ \\
	LOSS & ZTF & 2.5$\sigma$ & 1.8$\sigma$ \\
	\hline
\end{tabular}
\label{lf_ks}
\end{table}

The right two panels of Fig.~\ref{lfs} shows the luminosity function of SNe Ibc in DES, ZTF and LOSS. DES appears slightly more luminous than both LOSS and ZTF although at low significance levels of 1.9$\sigma$ and 1.1$\sigma$. 

\subsection{Parameterised luminosity functions}
\label{sec:lf_params}

In order to allow these luminosity functions to be used in simulations going forward, we fit a number of different distributions to the histograms presented in Section \ref{ccsn_lf}. We do this for our newly derived DES and ZTF luminosity functions; LOSS luminosity functions are already presented in \citetalias{li11}. These fits are only possible where the distributions peak above $-16$ and begin to decline again as otherwise we cannot constrain the location of the peak of the luminosity function. For SNe II in ZTF, the distribution does not obviously begin to decline above $-16$ which makes this difficult. As a result, for this fit we include two extra SNe which have a peak $R$-band magnitude below $-16$ which allows the peak of the distribution to be constrained. This fit is included for completeness, but we emphasise that the results for SNe II in ZTF should be considered with the strong caveat that these two extra objects have significant weight in determining the location of the peak of the distribution.

We consider both Gaussian and Lorentzian fits to the luminosity functions. We also consider skewed Gaussian distributions but find we are not able to constrain the skewness parameter $\gamma$ with the available data. The parameter values for these fits are shown in Table \ref{lf_params}. The exact parameter values for the distributions are sensitive to the binning of the histogram. The values shown in this table are based on the bin edges presented in Fig. \ref{lfs}, from $-19.5$ to $-16$ in steps of 0.5 for SNe II and 0.7 for SNe Ibc. For these fits, we use the mean of the absolute magnitudes in each bin for the x-coordinates. In this table, we present the uncertainty in each parameter when fitting to distributions with these bins (fit error). To take into account the how varying the binning will affect the parameter values, we also present a binning error; this is defined as the standard deviation of the parameter values measured when considering all possible bin widths from 0.10 to 1.0 in steps of 0.01.

To assess the quality of each fit, we also present reduced $\chi^2$ values. These are calculated assuming a $\sqrt{N}$ uncertainty in the histogram, rather than the uncertainties based on \citet{Gehrels86} presented in Fig. \ref{lfs}, in order to provide symmetric uncertainties for the fitting process. For SNe II in DES, a Gaussian distribution provides a reasonable fit although is too broad around the peak and underestimates the number of brighter SNe. A Lorentzian distribution better fits the sharp peak and brighter tail of the luminosity function. For SNe Ibc in DES, both distributions provide similar fits although the Gaussian has a lower reduced $\chi^2$. In contrast, for SNe II in ZTF a Gaussian better represents the luminosity function and a Lorentzian overestimates the number of brighter SNe. As for SNe Ibc in ZTF, a Gaussian has a lower reduced $\chi^2$ and better represents the luminosity function around peak although again underestimates the number of SNe in the brighter tail.

\begin{table*}
\caption{Reduced $\chi^2$ values and parameter values for different model fits to our calculated luminosity functions, including mean $\mu$ and width $\sigma$. The fit error represents the uncertainty in the fit to the distributions using the binning detailed in Section \ref{sec:lf_params}. The binning error represents the uncertainty in the parameter values based on the binning of the data and is defined as the standard deviation of the parameter values measured when considering all possible bin widths from 0.10 to 1.0 in steps of 0.01. }
\centering
\begin{tabular}{ cccccccc }
	\hline
	Survey & SN Type & Model Type & Reduced $\chi^2$ & Parameter & Value & Fit Error & Binning Error \\
	\hline
	DES & II & Gaussian & 1.44 & $\mu$ & $-17.10$ & 0.13 & 0.07 \\
	& & & & $\sigma$ & 0.70 & 0.13 & 0.06 \\
	& & Lorentzian & 0.41 & $\mu$ & $-17.10$ & 0.05 & 0.05 \\
	& & & & $\sigma$ & 0.53 & 0.08 & 0.11 \\
	DES & Ibc & Gaussian & 1.59 & $\mu$ & $-17.05$ & 0.19 & 0.18 \\
	& & & & $\sigma$ & 0.72 & 0.16 & 0.28 \\
	& & Lorentzian & 2.88 & $\mu$ & $-16.96$ & 0.25 & 0.18 \\
	& & & & $\sigma$ & 0.72 & 0.29 & 0.38 \\
	ZTF$^*$ & II & Gaussian & 1.98 & $\mu$ & $-16.85$ & 0.09 & 0.15 \\
	& & & & $\sigma$ & 1.02 & 0.07 & 0.11 \\
	& & Lorentzian & 5.53 & $\mu$ & $-16.73$ & 0.13 & 0.12 \\
	& & & & $\sigma$ & 0.84 & 0.15 & 0.14 \\
	ZTF & Ibc & Gaussian & 2.11 & $\mu$ & $-16.98$ & 0.14 & 0.12 \\
	& & & & $\sigma$ & 0.78 & 0.11 & 0.10 \\
	& & Lorentzian & 3.22 & $\mu$ & $-16.95$ & 0.10 & 0.10 \\
	& & & & $\sigma$ & 0.60 & 0.11 & 0.11 \\
	\hline
\end{tabular}
\begin{tablenotes}
\item $^*$Please note, for SNe II in ZTF two extra objects with a peak $R$-band absolute magnitude below -16 were included in order to constrain the peak of the distribution.
\end{tablenotes}
\label{lf_params}
\end{table*}

\section{Host galaxy demographics}
\label{hosts}

The host galaxy properties of a SN provide insight into the environment in which the progenitor star exploded. In this section, we explore the demographics of the host galaxies of our SNe in detail. When considering any differences in the samples we perform both two-sample KS and Anderson-Darling (AD) tests to assess.

\subsection{Host stellar mass}
\label{host_mass}

\begin{figure}
\centering
\includegraphics[width = \columnwidth]{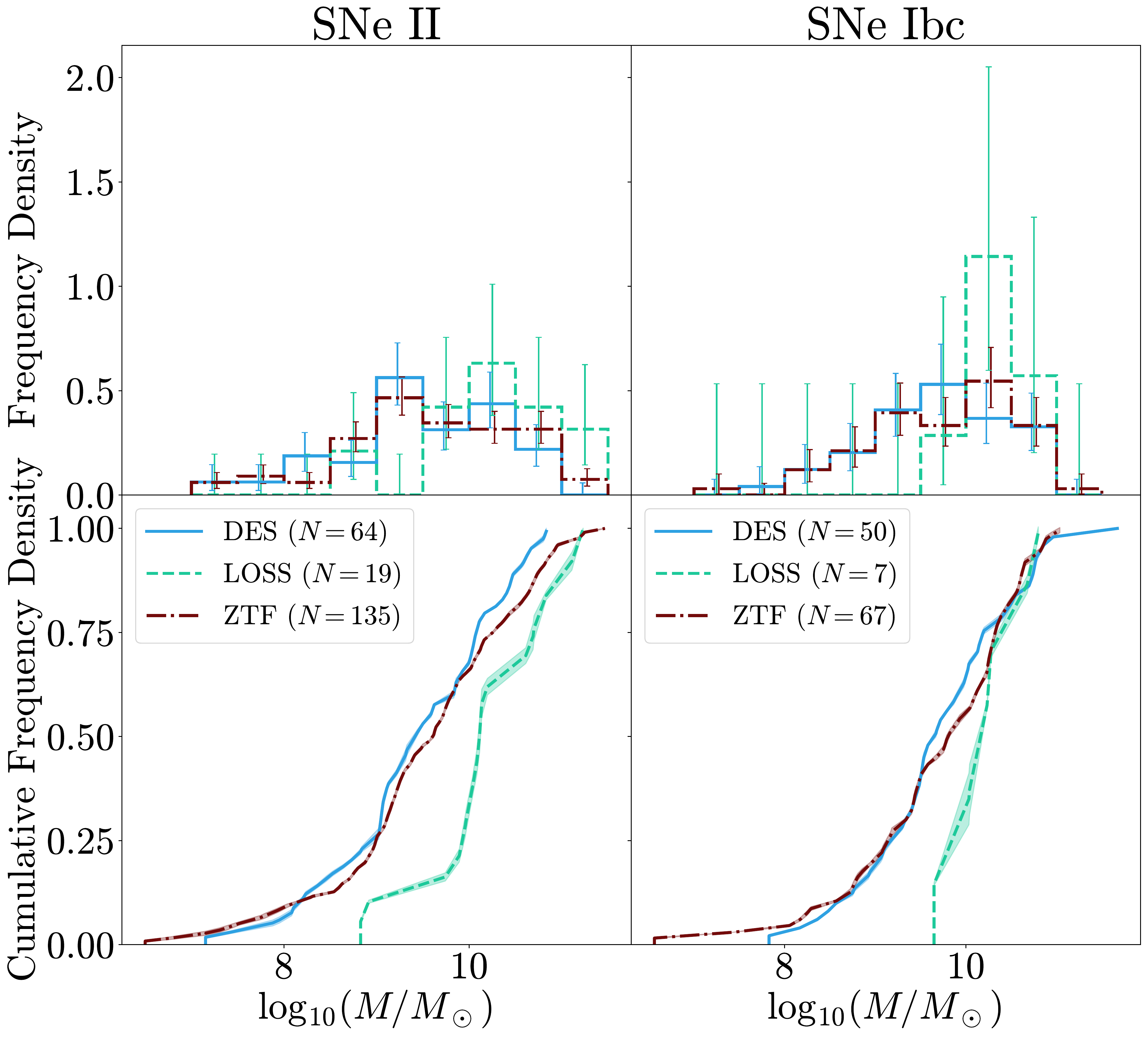}
\caption{Host galaxy stellar mass distributions and cumulative distributions for SNe II and SNe Ibc for the DES, LOSS and ZTF samples. Histogram uncertainties here (and throughout the paper) are estimated from the Poisson distribution, while CDF uncertainties are estimated from the Monte Carlo approach described in Section \ref{ccsn_lf}.}
\label{host_mass_dist}
\end{figure}

Fig.~\ref{host_mass_dist} shows the distribution of host galaxy stellar masses across our three samples for SNe II and SNe Ibc. DES and ZTF appear consistent with each other, but show discrepancies with LOSS. This difference is expected: LOSS is a galaxy targeted SN survey that monitored massive, luminous galaxies so low mass galaxies will be underrepresented in the LOSS sample. We perform a two-sample KS and AD tests for each combination of samples with results in Table~\ref{galaxy_ks} which reinforce our interpretations above.

\subsection{Host rest-frame colours}
\label{sec:host-galaxy-colours}

\begin{figure}
\centering
\includegraphics[width = \columnwidth]{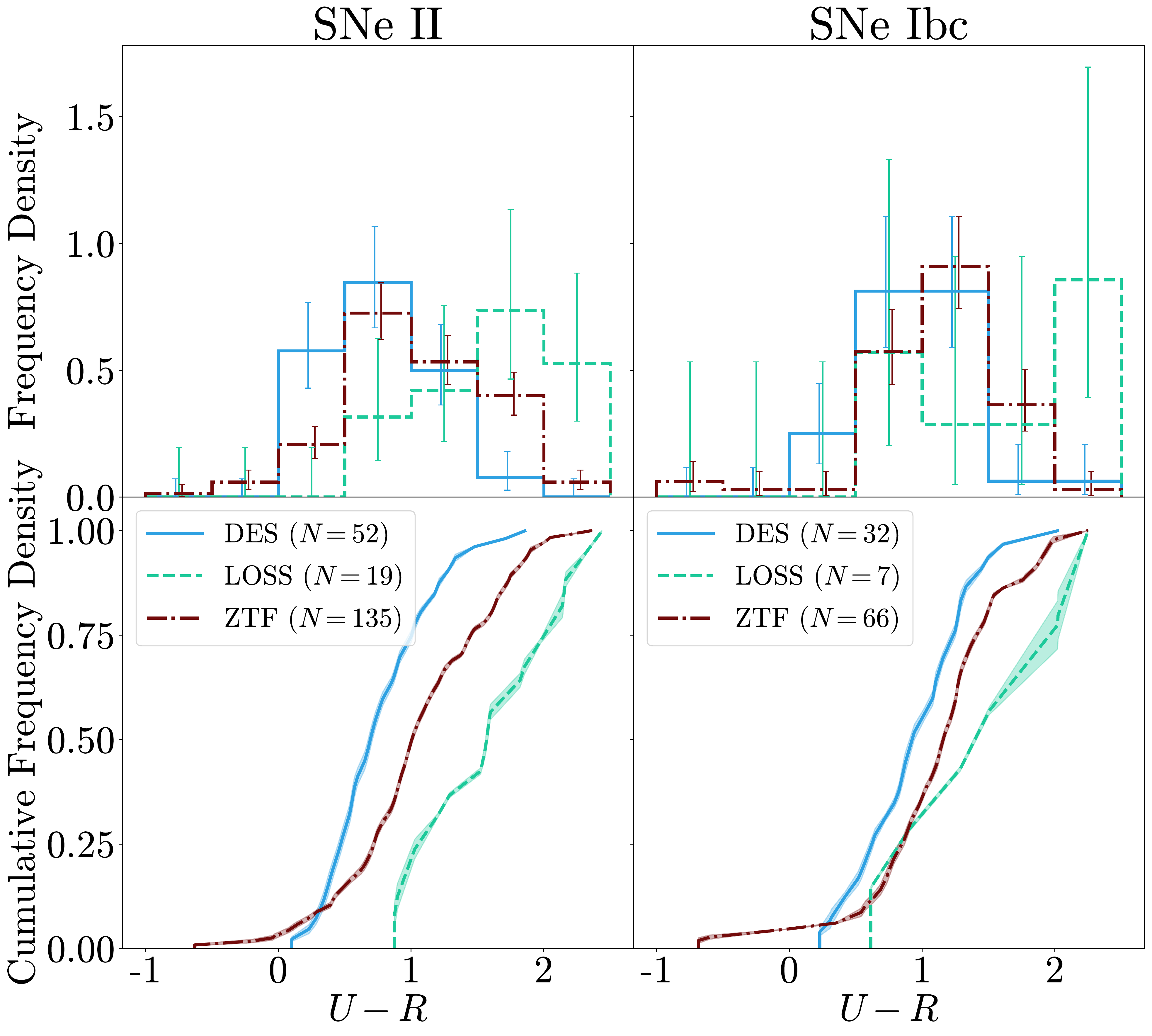}
\caption{As Fig.~\ref{host_mass_dist}, but for the host galaxy rest-frame $U-R$ colour in place of stellar mass.}
\label{all_host_UR_dist}
\end{figure}

Fig.~\ref{all_host_UR_dist} shows the distribution of rest-frame $U-R$ colours for the host galaxies in our three samples. For SNe II, we see differences between the three samples: the high-redshift DES sample has the bluest host galaxies, followed by the lower-redshift ZTF sample and then the local LOSS sample. Two-sample KS and AD tests show that the differences between the samples have significances in excess of 3$\sigma$ respectively. For the redshift range considered here, $griz$ does not cover rest-frame $U$-band meaning some extrapolation is involved in calculating $U-R$ for DES hosts. However, we see similar results when using rest-frame $B-V$ which is covered by $griz$. For SNe Ibc, the distributions visually suggest a similar finding however the offset between DES and ZTF is reduced and significance levels are below 2$\sigma$ for $U-R$ and below 2.3$\sigma$ for $B-V$.

\begin{figure}
\centering
\includegraphics[width = \columnwidth]{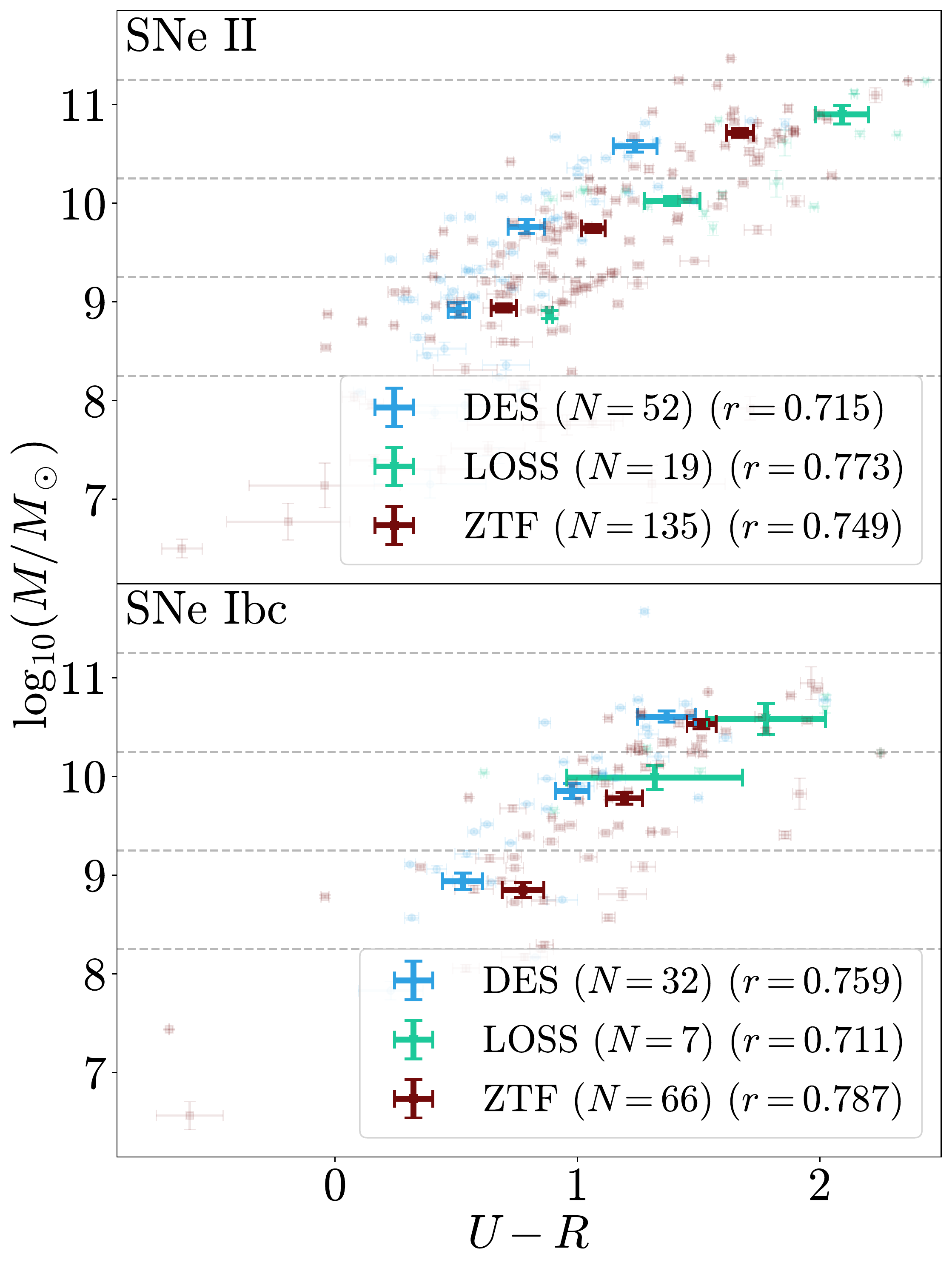}
\caption{Host galaxy stellar mass plotted against host galaxy rest-frame $U-R$ colour for each sample along with correlation coefficients for each. Thick data points represent the mean and standard error for stellar mass and colour for each sample in stellar mass bins of $8.25 < \log(M/M_\odot) < 9.25$, $9.25 < \log(M/M_\odot) < 10.25$ and $10.25 < \log(M/M_\odot) < 11.25$. The horizontal dashed lines mark these bin boundaries.}
\label{M_vs_U_R}
\end{figure}

Fig.~\ref{M_vs_U_R} shows host galaxy stellar mass plotted against host galaxy rest-frame $U-R$ colour for each of the three samples, with the thicker symbols showing the mean and standard error for each property across galaxies in bins of $8.25 < \log(M/M_\odot) < 9.25$, $9.25 < \log(M/M_\odot) < 10.25$ and $10.25 < \log(M/M_\odot) < 11.25$ for each sample. Across DES, ZTF and LOSS we see strong correlations between host stellar mass and host $U-R$ colour. This plot also shows that the difference we see in rest-frame $U-R$ colour between DES and ZTF is observed across the range of host galaxy masses in the DES sample, i.e., at fixed stellar mass the DES host galaxy sample is bluer, with this difference more pronounced for SNe II.

\begin{table*}
\caption{The results of two-sample KS and AD tests between the distributions of host galaxy stellar mass, rest-frame $U-R$ and $B-V$ colour. Also shown are metallicity values derived from stellar mass in Section \ref{metallicity} and $U-R$ (SFRcorr) and $U-R$ (Zcorr), the rest-frame $U-R$ colour corrected for SFR and metallicity evolution with redshift introduced in Section \ref{zevolution}.}
\centering
\begin{tabular}{ ccccccc }
	\hline
	Property & Survey 1 & Survey 2 & \multicolumn{2}{c}{KS test significance} & \multicolumn{2}{c}{AD test significance} \\
	& & & SNe II & SNe Ibc & SNe II & SNe Ibc \\
	\hline
	Stellar mass & DES & LOSS & 3.2$\sigma$ & 2.0$\sigma$ & 3.4$\sigma$ & 1.9$\sigma$ \\
	 & DES & ZTF & 0.4$\sigma$ & 0.5$\sigma$ & 0.8$\sigma$ & 0.3$\sigma$ \\
	 & LOSS & ZTF & 3.4$\sigma$ & 1.5$\sigma$ & 2.9$\sigma$ & 1.4$\sigma$ \\
	$U-R$ & DES & LOSS & 4.7$\sigma$ & 1.9$\sigma$ & 4.8$\sigma$ & 2.6$\sigma$ \\
	 & DES & ZTF & 3.4$\sigma$ & 1.6$\sigma$ & 3.8$\sigma$ & 1.8$\sigma$ \\
	 & LOSS & ZTF & 2.7$\sigma$ & 1.4$\sigma$ & 3.7$\sigma$ & 2.0$\sigma$ \\
	$B-V$ & DES & LOSS & 4.3$\sigma$ & 2.1$\sigma$ & 4.8$\sigma$ & 2.7$\sigma$ \\
	& DES & ZTF & 3.6$\sigma$ & 1.9$\sigma$ & 3.9$\sigma$ & 2.3$\sigma$ \\
	& LOSS & ZTF & 2.6$\sigma$ & 1.5$\sigma$ & 3.7$\sigma$ & 2.2$\sigma$ \\
	Metallicity & DES & LOSS & 4.0$\sigma$ & 2.4$\sigma$ & 4.0$\sigma$ & 2.2$\sigma$ \\
	& DES & ZTF & 1.4$\sigma$ & 0.9$\sigma$ & 1.8$\sigma$ & 0.8$\sigma$ \\
	& LOSS & ZTF & 3.5$\sigma$ & 1.5$\sigma$ & 3.0$\sigma$ & 1.5$\sigma$ \\
	$U-R$ (SFRcorr) & DES & LOSS & 4.3$\sigma$ & 1.4$\sigma$ & 4.7$\sigma$ & 2.0$\sigma$ \\
	& DES & ZTF & 3.0$\sigma$ & 0.9$\sigma$ & 3.4$\sigma$ & 1.3$\sigma$ \\
	& LOSS & ZTF & 2.8$\sigma$ & 1.4$\sigma$ & 3.6$\sigma$ & 2.0$\sigma$ \\
	$U-R$ (Zcorr) & DES & LOSS & 3.8$\sigma$ & 1.6$\sigma$ & 4.5$\sigma$ & 2.4$\sigma$ \\
	& DES & ZTF & 2.4$\sigma$ & 0.9$\sigma$ & 2.8$\sigma$ & 1.0$\sigma$ \\
	& LOSS & ZTF & 2.7$\sigma$ & 1.4$\sigma$ & 3.6$\sigma$ & 2.1$\sigma$ \\
	\hline
\end{tabular}
\label{galaxy_ks}
\end{table*}

\subsection{Relations between SN and Host Properties}
\label{sn_vs_host}

We next consider the relations between the properties of the SNe and the properties of the host galaxies for the three samples.

\subsubsection{SNe II/Ibc host properties comparison}

\begin{figure}
\centering
\includegraphics[width = \columnwidth]{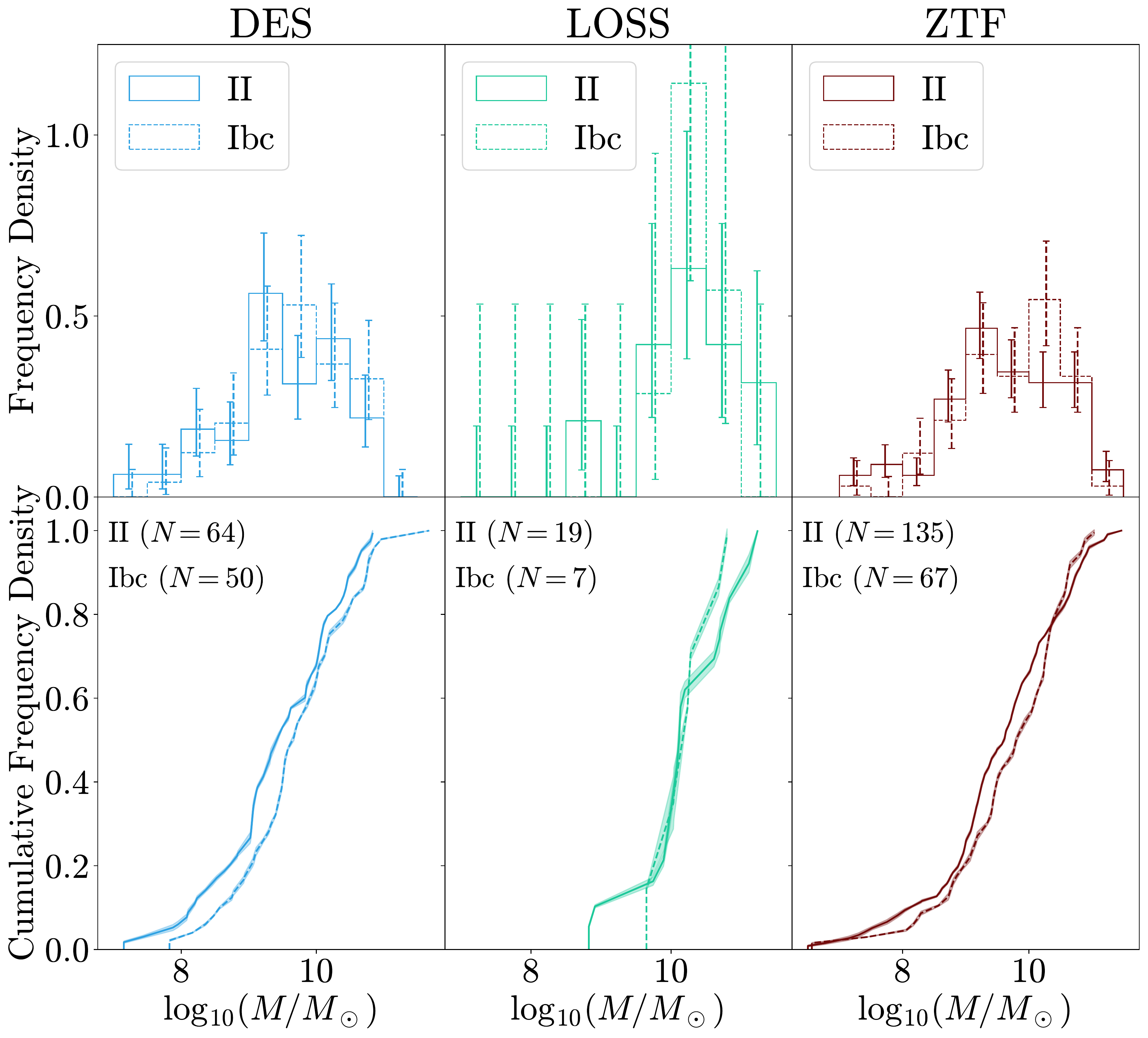}
\caption{Host galaxy stellar mass distributions and cumulative distributions for each of the DES, LOSS and ZTF samples showing the properties of the hosts of SNe II and SNe Ibc for each sample.}
\label{M_IIvsIbc}
\end{figure}

\begin{figure}
\centering
\includegraphics[width = \columnwidth]{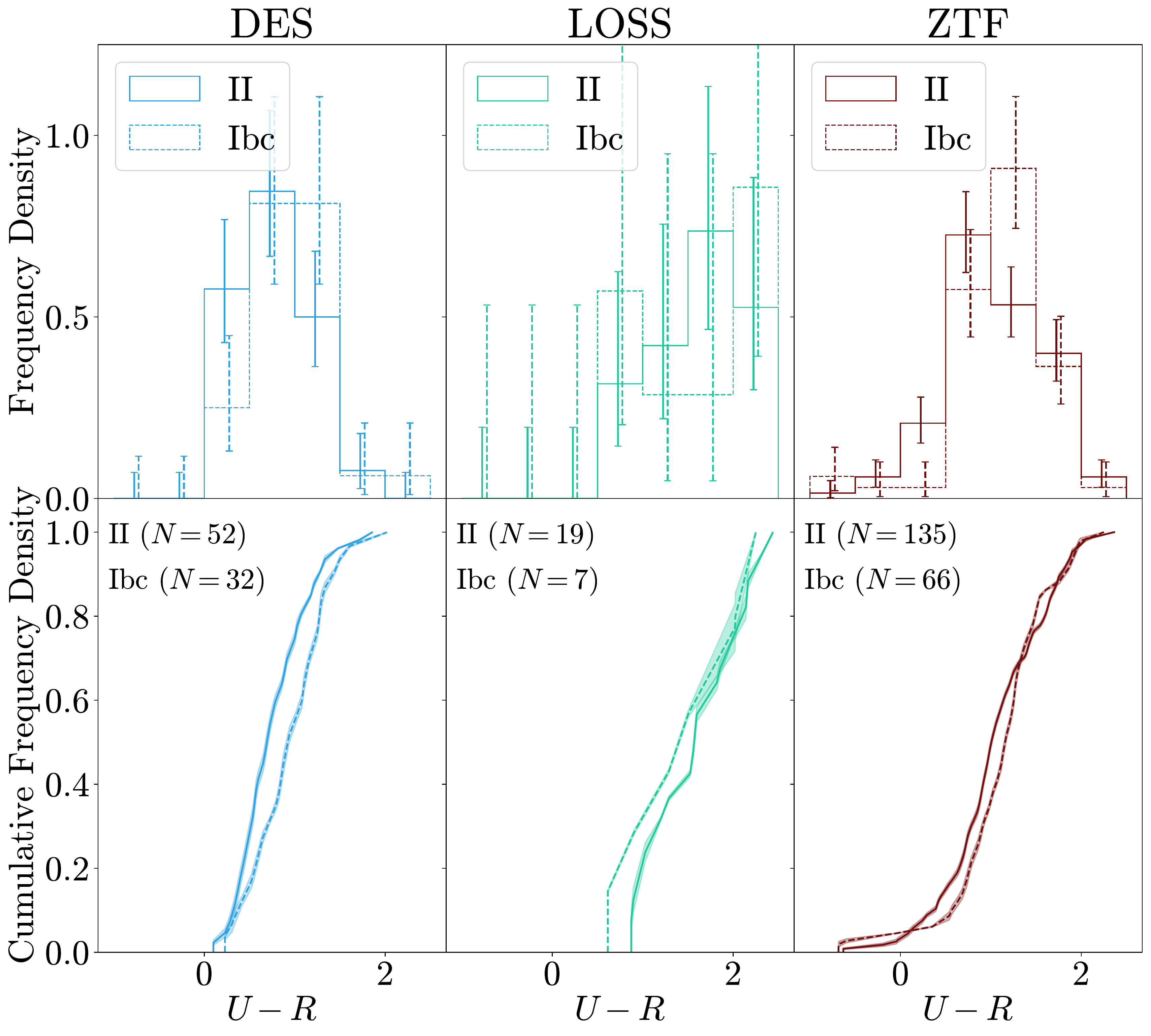}
\caption{As Fig.~\ref{M_IIvsIbc}, but for host galaxy rest-frame $U-R$ colour instead of stellar mass.}
\label{U_R_IIvsIbc}
\end{figure}

Fig.~\ref{M_IIvsIbc} and Fig. \ref{U_R_IIvsIbc} show the distributions of host stellar mass and $U-R$ colour comparing the host galaxies of SNe II and Ibc, and Table \ref{II_Ibc_host_ks} shows the results of two-sample KS tests between these distributions. For host stellar mass, we do not see any significant differences between the hosts of SNe II and SNe Ibc. We also see no significant difference in host $U-R$ colour for LOSS and ZTF, the latter consistent with the findings of \citet{Perley20}. For the DES sample, the hosts of SNe Ibc appear slightly redder with a significances of 2.0$\sigma$ from the KS and AD tests. Taking $U-R$ as a proxy for star formation rate, this could indicate that SNe Ibc are exploding in galaxies with less star formation than SNe II. SNe Ibc have been shown to trace galaxy star formation more closely than SNe II \citep[e.g.,][]{Anderson2009, Galbany18} which would makes this result surprising, albeit with the caveats that these differences refer to local properties rather than the global host properties we present here and that the significance level is not high.

\begin{table}
\caption{The results of two-sample KS tests between the SNe II and SNe Ibc host properties in each survey.}
\centering
\begin{tabular}{ cccc }
	\hline
	Property & Survey & KS test significance & AD test significance \\
	\hline
	Stellar mass & DES & 1.1$\sigma$ & 1.4$\sigma$ \\
	& LOSS & 0.1$\sigma$ & 0.2$\sigma$ \\
	& ZTF & 1.2$\sigma$ & 1.3$\sigma$ \\
	$U-R$ & DES & 2.0$\sigma$ & 2.0$\sigma$ \\
	& LOSS & 0.1$\sigma$ & 0.2$\sigma$ \\
	& ZTF & 1.8$\sigma$ & 1.5$\sigma$ \\
	\hline
\end{tabular}
\label{II_Ibc_host_ks}
\end{table}

\subsubsection{SN/host correlations}

We also consider relations between the properties of the host galaxy and the properties of the SN, looking at correlations between peak SN luminosity and host stellar mass and rest-frame colour. Table \ref{host_sn_correlations} shows the Pearson correlation coefficients ($r$) between these properties for each of the DES, LOSS and ZTF samples.

For both SN II and SN Ibc samples, we see no obvious or significant trends between the SN luminosities and the properties of the galaxies that host them -- the correlations seen for SNe Ibc in LOSS are not statistically significant and correspond to only 7 galaxies. 
\citet{Gutierrez18} also finds no relation between stellar mass and peak SN luminosity for SNe II, and \citet{Wiseman21RET} finds a lack of strong evidence for a relation between peak transient luminosity and host mass and sSFR for RETs.

\begin{table}
\caption{Correlation coefficients between peak SN $R$-band absolute magnitude and host galaxy stellar mass and rest-frame $U-R$ colour for SNe II and Ibc in DES, LOSS and ZTF.}
\centering
\begin{tabular}{ cccc }
	\hline
	Property & Survey & SN Type & Correlation with SN peak \\
	& & &  $R$-band absolute magnitude (r) \\
	\hline
	Stellar mass & DES & II & 0.09 \\
	& & Ibc & 0.03 \\
	& LOSS & II & -0.28 \\
	& & Ibc & 0.57 \\
	& ZTF & II & -0.12 \\
	& & Ibc & 0.14 \\
	$U-R$ & DES & II & -0.06 \\
	& & Ibc & 0.07 \\
	& LOSS & II & -0.12 \\
	& & Ibc & 0.55 \\
	& ZTF & II & 0.07 \\
	& & Ibc & 0.21 \\
	\hline
\end{tabular}
\label{host_sn_correlations}
\end{table}

\section{Discussion}
\label{discussion}

In this section we explore our results, considering and assessing a number of potential causes for some of the noteworthy trends that we observe.

\subsection{Impact of photometric misclassification}
\label{class_error}

We begin by discussing the potential impact of misclassification of the sample of photometric CCSNe with host spec-z in DES. SNe Ia have been removed using SuperNNova model presented in \citet{Vincenzi22}, which has a high degree of accuracy upward of 98 per cent. As discussed in Section \ref{DES_sample_specz}, the pSNid model used to split this sample into SNe II and Ibc has an accuracy of 86 per cent on the sample of DES CCSNe with spectroscopic classifications, with similar performance on each of the two classes. While this method works well, it does leave open the possibility that a small proportion of SNe in this sample are assigned to the wrong class.

\begin{figure}
\centering
\includegraphics[width = \columnwidth]{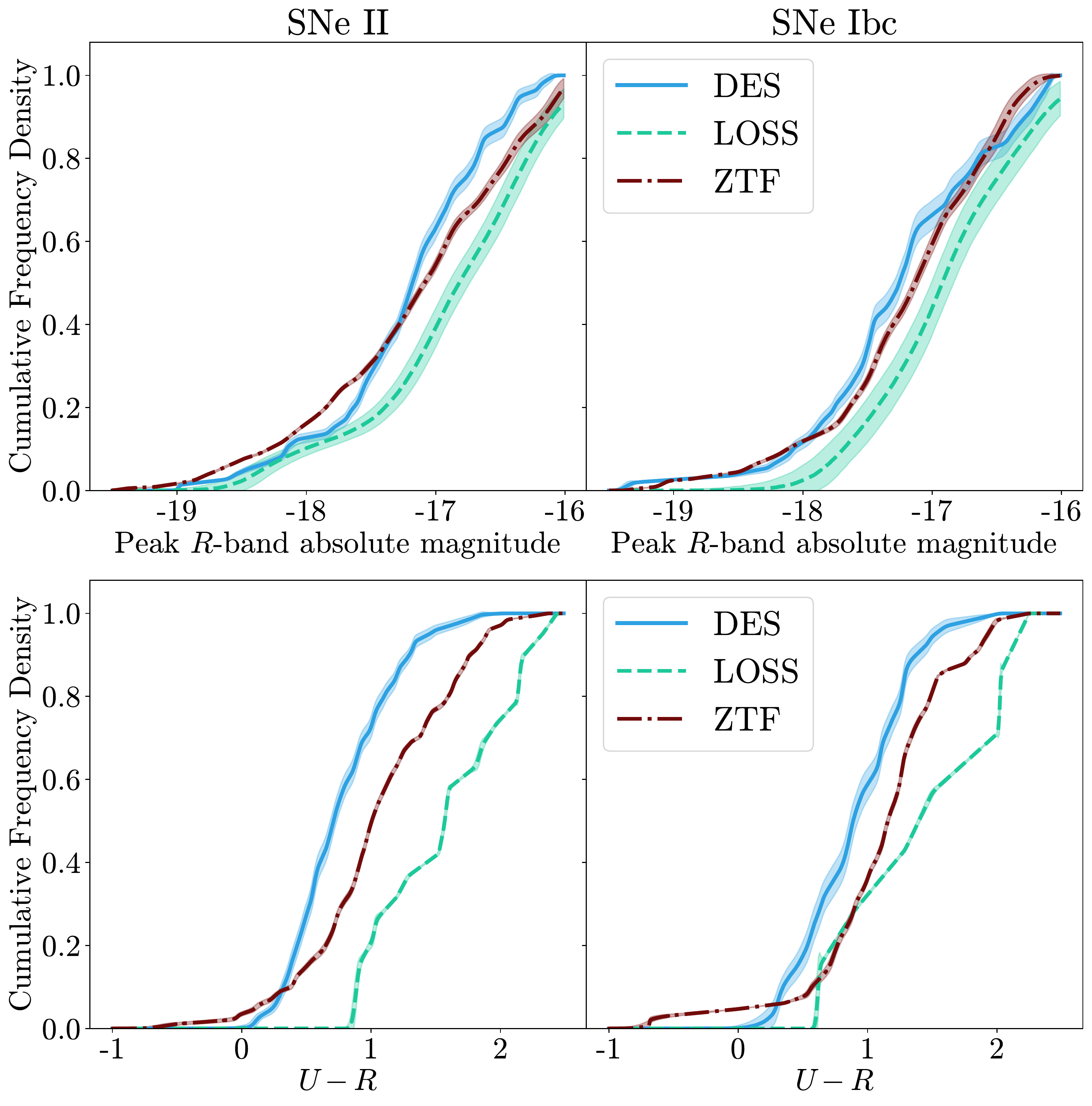}
\caption{CDFs for the luminosity functions and rest-frame host galaxy $U-R$ colours for each of the DES, LOSS and ZTF samples, incorporating a 14 per cent misclassification rate for photometrically classified SNe in DES as outlined in Section \ref{class_error}.}
\label{class_error_plots}
\end{figure}

To investigate what effect this may have on our analysis, we repeat the Monte Carlo process for CDF uncertainty outlined in Section \ref{ccsn_lf} but this time in each iteration we flip 14 per cent of the classes, corresponding to the expected error rate, for the photometrically classified DES CCSNe (SNe II are changed to SNe Ibc and vice versa) to see what effect this has on the final CDF. Fig. \ref{class_error_plots} shows the luminosity functions and host galaxy $U-R$ distributions for the three samples, with the randomised class flipping applied to DES photometrically classified SNe. Overall, these distributions appear very similar to those in Fig. \ref{lfs} and \ref{all_host_UR_dist} and the incorporation of the class flipping has little effect. We also try restricting the sample to only spectroscopically confirmed supernovae from DES and see the same trends for SNe II, although there are too few spectroscopically confirmed SNe Ibc to make this comparison. 

Considering the overall samples after quality cuts, but before redshift and magnitude cuts, the final ZTF sample consists of 174 SNe II and 89 SNe Ibc, a ratio of $\sim$2:1. In contrast, the photometrically confirmed sample with hosts spectroscopic redshifts from DES consists of 56 SNe II and 42 SNe Ibc at a ratio of $\sim$1.33:1. At first glance, this suggests that pSNid is classifying too many objects as SNe Ibc. However, it is important also to consider the spectroscopically confirmed sample from DES - as this sample is based on targeted follow-up, it would not be expected to follow the same ratio of classes as an untargeted sample such as ZTF. Combining both these DES samples, there are 89 SNe II and 55 SNe Ibc, a ratio of $\sim$1.75:1 which is much closer to ZTF. After redshift and magnitude cuts, this ratio shifts further from 2:1, but the relatively small sample sizes compared with ZTF mean this is not surprising. Overall, we consider the results presented in this analysis robust to the potential misclassification of SNe II and Ibc.

\subsection{Difference in the luminosity function}

As part of our analysis, we have carried out two-sample KS tests between the luminosity functions of DES, LOSS and ZTF. For SNe II, DES is brighter than LOSS at a significance level of 3.0$\sigma$ and appears brighter than ZTF although only at a significance of 1.8$\sigma$. For SNe Ibc, DES also appears brighter than both LOSS and ZTF although at a significances of only 1.9$\sigma$ and 1.1$\sigma$. Although the significance levels are not high, these differences raise the possibility of underlying differences in the luminosity functions of these samples. If there is a difference, one natural explanation would be redshift evolution in the underlying stellar populations and progenitor stars. However, we first consider other, simpler explanations.

\subsubsection{Incompleteness}

The most straight forward explanation for any difference between DES and LOSS is a lack of completeness in the DES sample due to lower sensitivity to fainter SNe. Fig.~\ref{Mvsz_dist} shows the peak absolute magnitudes of all objects in our samples plotted against redshift, prior to making any selection in absolute magnitude. Fig.~\ref{lfs} shows the main differences between DES and LOSS for SNe II are in the [$-16$, $-16.5$] luminosity bin and for SNe Ibc are in the [$-16$, $-16.7$]. DES is not complete in this range and the distribution is affected by the $V_\text{max}$ correction, whereas the LOSS sample is not affected by this correction. The $V_\text{max}$ correction for DES gives a maximum weighting of ~2.9 but for ZTF this is much higher due to the lack of completeness in the sample, up to a maximum of 14.5 although this is typically around 2-3. We can mitigate for this with lower redshift cuts for DES and ZTF to obtain more complete samples -- doing so reduces the significances from the KS test due to the smaller sample size but overall the trends that we see appear unchanged. This suggests that incompleteness is not the cause of potential differences between the samples.

\subsubsection{Host Properties}

An alternative possibility is that any difference in luminosity function could be explained by a difference in host properties between the samples. For example, the host galaxies of the LOSS sample are significantly more massive and redder than that of DES, likely because of the galaxy-targeted nature of LOSS (Section~\ref{hosts}). However, as there are no significant correlations between either host colour or mass and peak SN luminosity (Section \ref{sn_vs_host}) this is unlikely to cause any differences in the luminosity function. 

\subsubsection{Host Extinction}
\label{extinction}

The difference in luminosity function could also result from differing levels of host galaxy extinction between the two samples. This could be due to both global and local host properties; for example, on average we might expect a higher level of host extinction in more massive, redder, dustier host galaxies and SNe closer to the central dusty regions of the host.

The DES hosts are, on average, bluer than those of LOSS and ZTF, which could indicate higher levels of host extinction in LOSS and ZTF that might explain any differences we see. To explore this possibility, we compare the luminosity functions of LOSS and ZTF with only SNe in DES that are in redder host galaxies. When we make cuts at either $U-R > 0.5$, $U-R > 0.75$ or $U-R > 1.0$, while the KS test significances are reduced by the smaller sample size we find that the same overall trends are observed as for the full sample. This would indicate that differing levels of host extinction do not cause any differences we see, though without measurements of the host extinction we cannot rule this out as a possibility -- local environment properties are likely to play a significant part in the level of extinction. We also consider the possibility of differing SN radial distributions across the three samples leading to differing levels of extinction, but do not find any significant differences in the physical separation between SN and host.

\subsubsection{Metallicity}
\label{metallicity}

Differences in metallicity may also explain potential differences in the luminosity functions; for example, DES SNe may occur in lower metallicity environments than LOSS or ZTF. As previously mentioned, host metallicity affects the supernova population as the most luminous classes of supernovae preferentially occur in low-mass, low-metallicity environments. Metallicity varies with stellar mass, star formation rate, redshift (e.g. \citealt[][]{Zahid13, Yates12, Curti20}) and also radially within a galaxy \citep{Parikh21}. There are a number of reasons why the DES hosts could be expected to be lower metallicity than either the LOSS or ZTF hosts: the DES hosts are lower stellar mass than those of LOSS, they are bluer and hence more star forming than hosts in ZTF or LOSS (although the effect of increased SFR on metallicity will vary depending on galaxy mass) and they are at higher redshift. Metallicity differences are a possible cause of any differences in luminosity function.

While we do not have metallicity values calculated from host galaxy spectroscopy, we can get an indication of global host galaxy metallicity using the relation between stellar mass and metallicity given in equation 4 of \citet{Zahid13}. We calculate global galaxy metallicity using following approach:

\begin{itemize}
    \item We fit a straight line to the relation between the redshifts and mass-metallicity relation parameters quoted in Table 1 of \citet{Zahid13}. We use only the samples from SDSS, the Smithsonian Hectospec Lensing Survey \citep[SHELS;][]{Geller14} and the DEEP2 survey \citep{Newman13} quoted here as the higher redshift samples have very uncertain values for these parameters.
    \item For a galaxy at a given redshift, we use these linear fits to estimate the mass-metallicity relation parameters at that redshift and then use the relation at that redshift to convert our measured stellar mass from SED fits to a metallicity.
\end{itemize}

The results of two-sample KS and AD tests between the global host metallicities of each of our samples are shown in Table \ref{galaxy_ks} - as for stellar mass, DES and ZTF are consistent while both show differences to LOSS. Of course, in reality there will be a large degree of scatter around the mass-metallicity relation. However, this indicates differing host metallicity could explain differences between DES and ZTF but not between DES and LOSS.

We can also probe metallicity looking at the decline rates during the plateau phase after maximum light of SNe II. Theoretical models suggest that the metallicity of the progenitor star may affect the decline rate during the `plateau' phase of the SN light curve \citep[][]{Dessart13}, however,  observations do not show this dependence \citep[][]{Anderson16}. The absence of correlations could be related to the lack of SNe~II in low-luminosity hosts. Nevertheless, some relations can be established when SNe II in faint hosts are included. \citet{Gutierrez18} find that slow-decliner SNe~II (i.e. SNe with lower s2 values) occur preferentially in low-luminosity (and therefore low-metallicity) hosts. For SNe II in DES and ZTF, we calculate the decline rate of this phase of the light curve \citep[corresponding to s2 in][]{Anderson14} and find that the decline rates calculated are consistent across the two samples. This suggests that there is not a significant metallicity difference between the two samples, indicating that this is unlikely to explain any differences between DES and ZTF.

\subsubsection{Summary}

The notion of a luminosity function which evolves with redshift is an interesting one - the differences we see in the luminosity functions of SNe II in DES and LOSS and SNe Ibc in DES, LOSS and ZTF raise this as a possibility, with the caveat that the significances are not especially high. Any differences could be explained by a lack of completeness in the DES sample, however this will have been significantly mitigated for by the $V_{max}$ correction and making a lower redshift cut does not change the trends we see. Greater dust extinction from redder host galaxies is another possible explanation given that LOSS and ZTF hosts are bluer than DES, but selecting only DES SNe in redder hosts or ZTF and LOSS SNe in bluer hosts does not change the trends we see which suggests that this is not the case - despite this, without measurements of the host extinction we cannot rule this out as a possibility. Differing metallicity also does not seem to explain the differences as we see consistent global host galaxy metallicities between DES and ZTF using the mass-metallicity relation of \citet{Zahid13} and consistent decline rates after peak for SNe II.

\subsection{Host galaxy colour discrepancy}

Section~\ref{sec:host-galaxy-colours} uncovered a puzzling trend: SNe II in DES on average occur in bluer galaxies than those in ZTF and LOSS. A difference in host galaxy properties between DES/ZTF and LOSS can be explained, at least in part, by the differences in targeting between the surveys. However, the difference in host rest-frame colour between DES and ZTF is not so easily understood. In this section we explore possible explanations for this difference.

\subsubsection{DES spectroscopic selection bias}

The DES sample in Fig.~\ref{all_host_UR_dist} contains only CCSNe with a spectroscopic host redshift, obtained from a variety of sources \citep{Maria2021}. Typically, galaxy redshifts are measured through the presence of narrow emission lines in their spectra, which will generally be stronger in bluer, star-forming galaxies. This may lead to a bias towards bluer galaxies in the DES sample, although \citet{Maria2021} finds that the difference in spectroscopic selection efficiency in DES between red and blue galaxies is small. By contrast, ZTF has an automated SN spectroscopic follow-up programme which provides redshift information for 93 per cent of observed transients with $m < 18.5$\,mag and 100 per cent with $m < 17$\,mag. As a result, any possible bias affecting DES would not affect ZTF.

\begin{figure}
\centering
\includegraphics[width = \columnwidth]{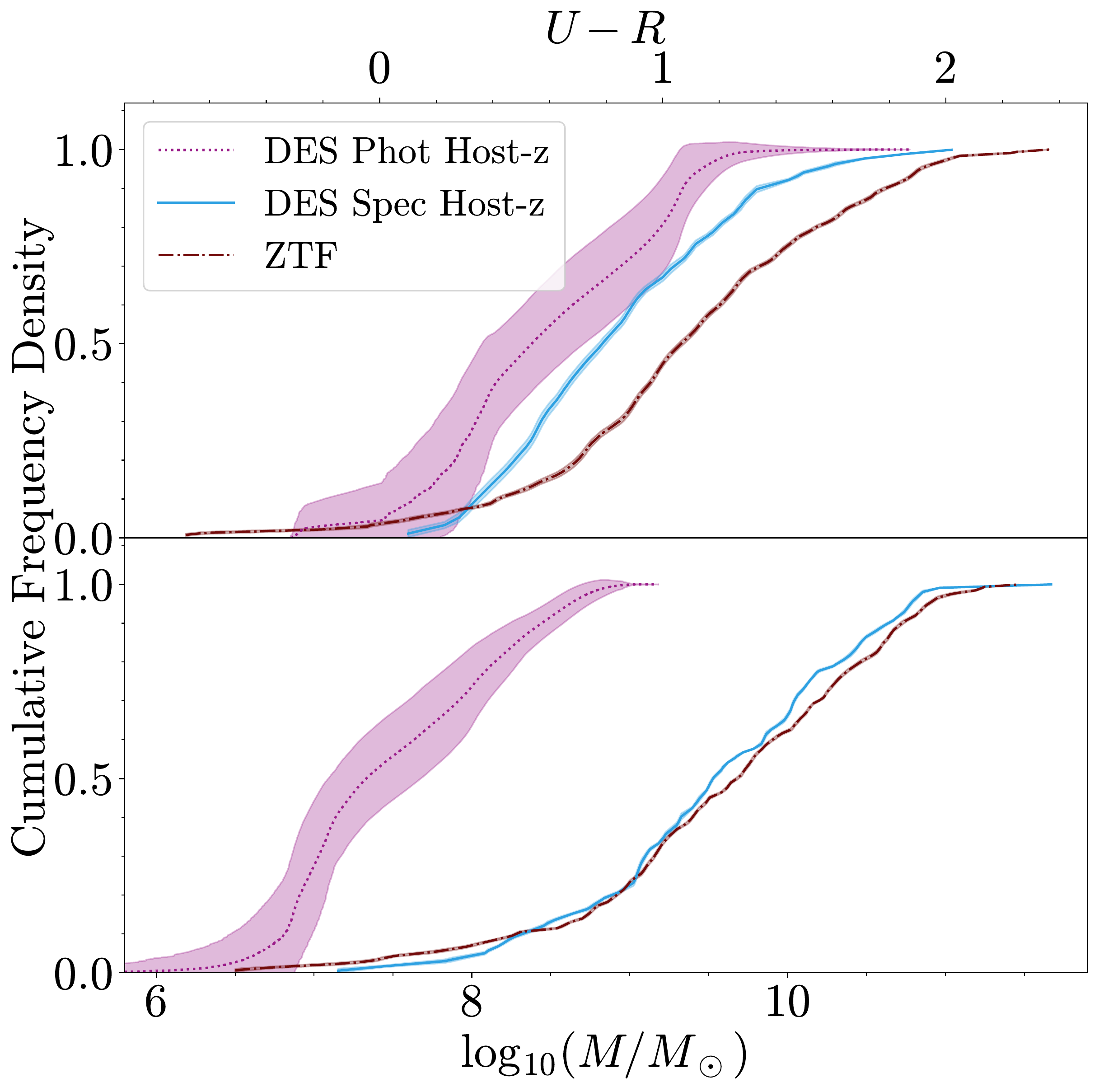}
\caption{Host galaxy rest-frame $U-R$ (top) and stellar mass (bottom) cumulative distributions for all CCSNe in the the DES sample of objects without spectroscopic host redshifts, compared with the DES sample with spectroscopic host redshifts as well as the ZTF sample.}
\label{all_host_phot_phot}
\end{figure}

We compare the DES samples with host spec-zs (both the spectroscopically confirmed and photometric with spec-z samples) and ZTF samples to the DES CCSN sample with only photo-zs. (Fig.~\ref{all_host_phot_phot}). However, rather than explaining the difference in host colour, this photometric sample appears bluer than the DES spec-z sample. Considering the host stellar mass distribution for this sample, this is not unexpected: the DES hosts without spectroscopic redshifts are low stellar mass galaxies which are typically bluer and more strongly star-forming than higher mass galaxies. In summary, the difference in host rest-frame colour cannot be easily explained by a simple spectroscopic selection bias in DES. 

\subsubsection{ZTF spectroscopic selection effects}
An alternative explanation is some selection bias in ZTF which favours SNe in redder hosts. The ZTF BTS sample has a very high level of spectroscopic completeness; however, spectroscopy is not captured exclusively by the ZTF spectroscopic instrument SEDMachine (SEDM) -- in cases where the SEDM spectra are unreliable other instruments may be used, and in cases where an object is first classified by another survey ZTF do not take an additional spectrum. To understand whether this may introduce selection effects, we examine the sub-sample of ZTF SNe only with a classification reported by SEDM. However, we find no significant difference in this population: a two-sample KS test between the $U-R$ host colour distributions of DES and only ZTF SNe classified by SEDM has a significance of 3.2$\sigma$ and 3.4$\sigma$ from KS and AD tests, almost unchanged from the full ZTF sample.

\subsubsection{Redshift evolution of the host galaxies}
\label{zevolution}

Another possible explanation is redshift evolution, with a period of $\sim$1--2\,Gyr between most of the ZTF and DES SNe exploding. ZTF hosts are therefore on average older and less strongly star-forming. The host galaxy SFRs can be corrected for redshift evolution following the method of section~4.2 of \citet{Taggart21}, based on the star-forming sequences of thousands of galaxies outlined in \citet{Salim2007} and \citet{Noeske2007}, correcting the SFR values to $z=0$. However, we do not measure SFR directly, and instead measure $U-R$ colour. We adapt the \citet{Taggart21} method to $U-R$ using the following steps:

\begin{itemize}
    \item We calculate the SFR correction for each galaxy; this correction will be the same for sSFR as well.
    \item We fit a linear relationship between $U-R$ colour and sSFR for all host galaxies in our sample.
    \item We use the gradient of this line to convert the sSFR correction into a $U-R$ correction.
\end{itemize}
We can then compare the distributions of these corrected colours with those of the ZTF hosts. 

\begin{figure}
\centering
\includegraphics[width = \columnwidth]{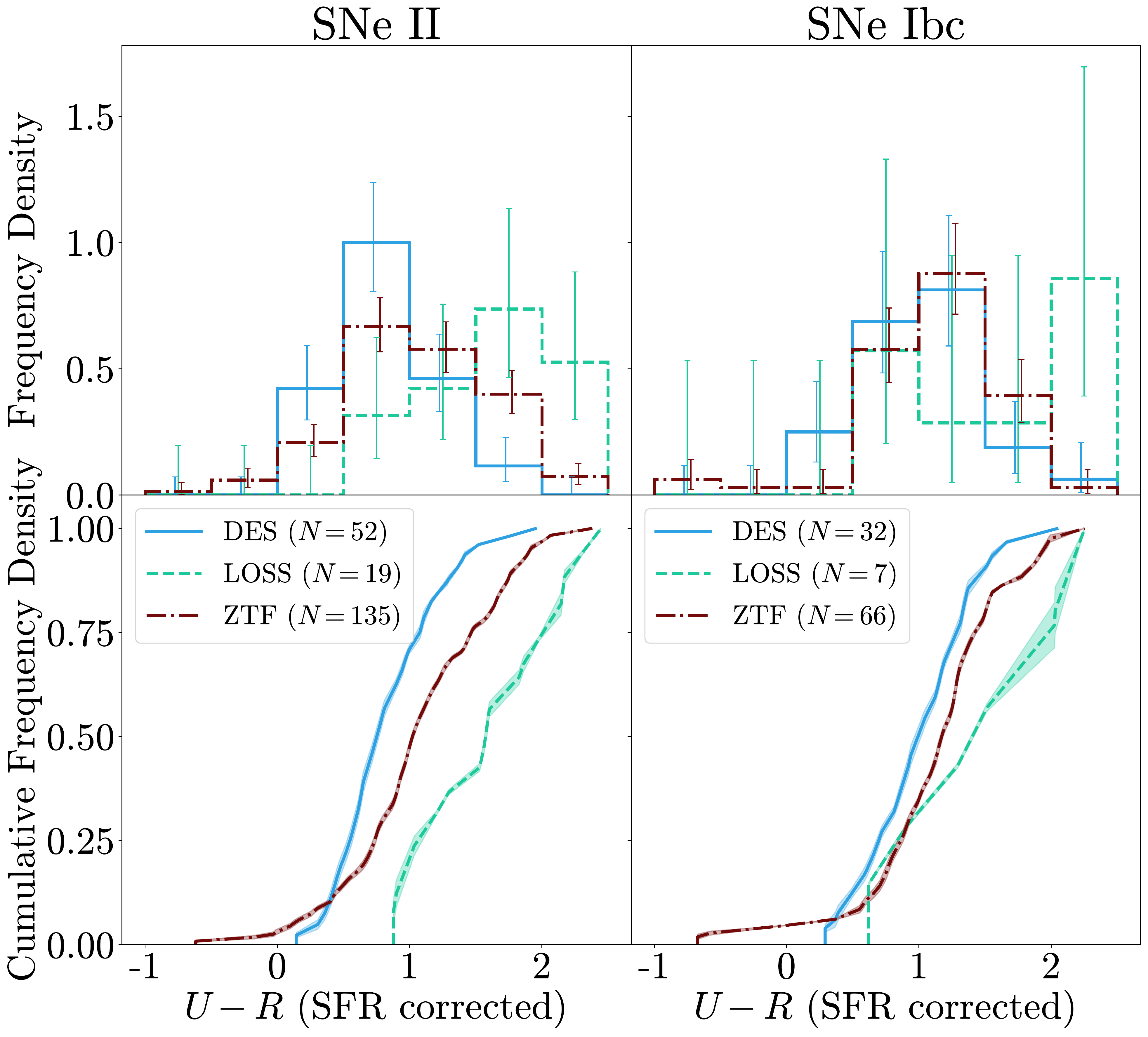}
\caption{Host galaxy rest-frame $U-R$ colour distributions and cumulative distributions, corrected for the effects of SFR evolution with redshift, for both SNe II and SNe Ibc for DES, LOSS and ZTF samples.}
\label{U_R_corr_dist}
\end{figure}

Fig.~\ref{U_R_corr_dist} shows the distributions of rest-frame $U-R$ colour corrected for the evolution of SFR, hereafter $U-R$ (SFRcorr). The correction factor between typical DES and ZTF redshifts is $\simeq0.03$--$0.04$\,mag and thus the effect is small: the significance of the difference between DES and ZTF from the two-sample KS test is reduced by only 0.4$\sigma$. Based on this, the difference in host colour seems unlikely to be caused by redshift evolution of the underlying galaxy populations. 

However, another redshift evolution we should consider is the evolution of the mass-metallicity relation with redshift. The metallicity of the host galaxies will have an effect on the emission lines produced, which will in turn affect galaxy colour. To investigate this, we use the following process to correct $U-R$ colour for the effects of metallicity evolution:

\begin{itemize}
    \item Fit a relation between our metallicity values inferred from \citet{Zahid13} discussed in Section \ref{metallicity} and our rest-frame host galaxy $U-R$ colours. Unlike the SFR correction, this relationship is not linear. Instead, we fit an exponential relation with a linear term of the form $y = mx + c + e^{A(x-x_0)}$ where $m$, $c$, $A$ and $x_0$ are the fitting parameters.
    \item Compare the metallicity difference for each galaxy of a given mass between its actual redshift and $z=0$.
    \item Use the fitted relation between metallicity and $U-R$ to estimate how much this change in metallicity would affect the rest-frame $U-R$ colour.
    \item Modify our calculated $U-R$ colours by this correction factor to calculate the rest-frame $U-R$ colour corrected for metallicity evolution, hereafter $U-R$ (Zcorr).
\end{itemize}

This correction involves the use of two relations which show a large degree of scatter, the mass-metallicity relation from \citet{Zahid13} and our relation between $U-R$ colour and metallicity. However, this does give an indication of the extent that evolving metallicity will have on $U-R$ colour.

\begin{figure}
\centering
\includegraphics[width = \columnwidth]{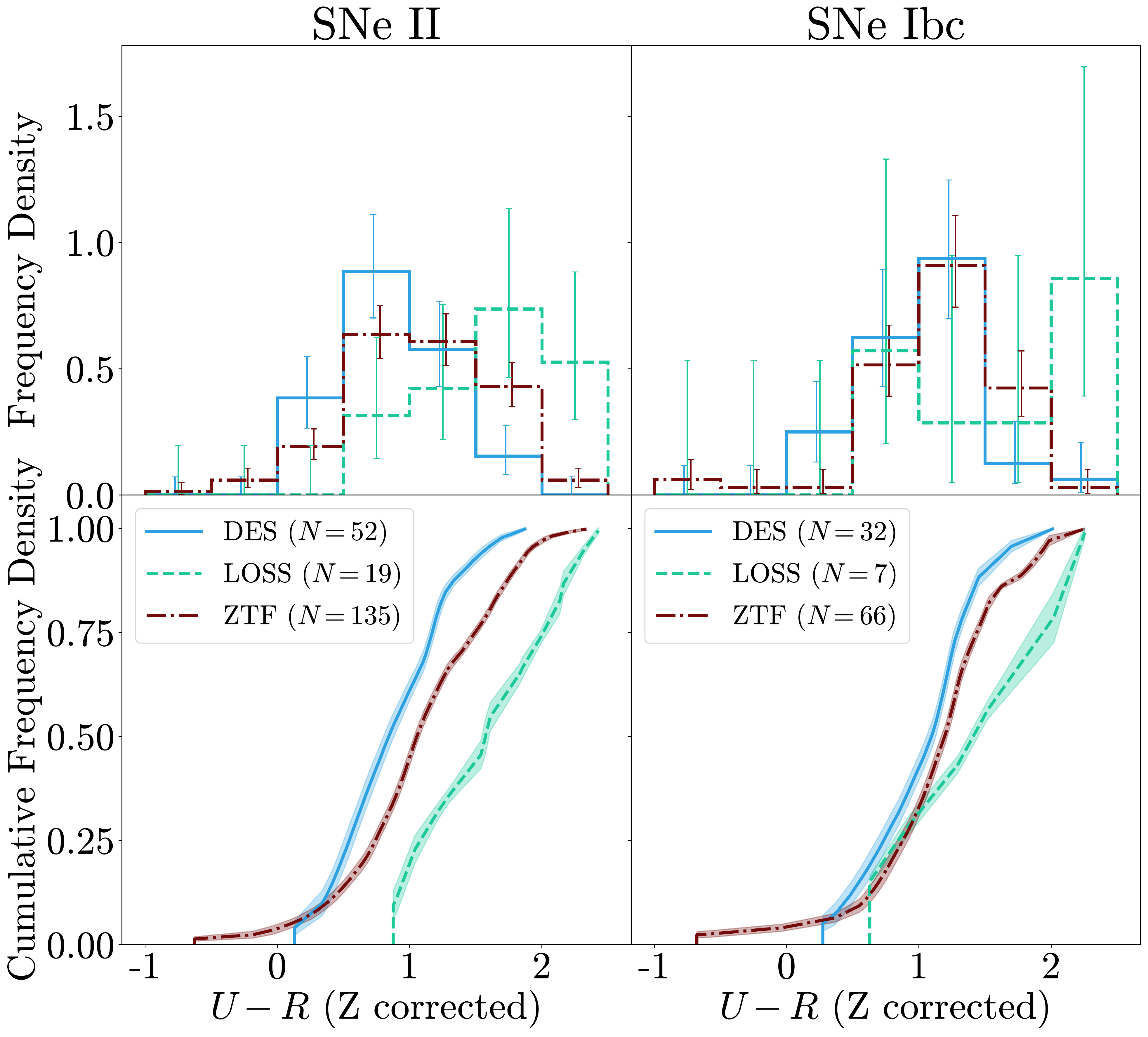}
\caption{Host galaxy rest-frame $U-R$ colour distributions and cumulative distributions, corrected for the effects of metallicity evolution with redshift, for both SNe II and SNe Ibc for DES, LOSS and ZTF samples.}
\label{U_R_Zcorr_dist}
\end{figure}

Fig. \ref{U_R_Zcorr_dist} shows the distributions of $U-R$ (Zcorr). This correction factor is larger than the previous correction for SFR evolution. For SNe II, this correction reduces the gap between DES and ZTF and the significance of this offset is reduced to 2.4$\sigma$ and 2.8$\sigma$ for both KS and AD tests, however the offset in $U-R$ between DES and ZTF across different galaxy masses as in Fig. \ref{M_vs_U_R} is still seen. Overall, metallicity evolution with redshift may explain some but not all of the offset in rest-frame $U-R$ colour between host galaxies in DES and ZTF.

\subsubsection{Systematic differences in photometry used}

An additional possibility is that the difference sources of host galaxy photometry between DES and ZTF is causing some systematic offset between the two samples. It may be that the inclusion of $u$-band data in the SED fits for the ZTF hosts is causing a systematic difference compared with the $griz$-only SED fits for the DES hosts (Section~\ref{sec:host-galaxy-measurements}). We remove the $u$-band data from the ZTF host photometry and repeat the SED fits using only $griz$, but find that the ZTF rest-frame colours from $griz$ fits are consistent with those from $ugriz$ fits and thus that the difference between DES and ZTF host colours remains.

Alternatively, there may be differences between DES and SDSS photometry which cause an offset when considering the same bands. To investigate this possibility, we match DES supernovae to the SDSS host catalogue using a 5\arcsec search radius, finding 47 objects with SDSS host galaxies. We then repeat the SED fits using SDSS photometry instead of DES. We find that rest-frame $U-R$ colours from SDSS photometry are consistent with those from DES with no systematic offset between the two. Overall, the difference in host colour does not seem to be caused by systematic differences in the data used to calculate host properties.

\subsubsection{Summary}

In summary, the difference in host galaxy rest-frame colour between the ZTF and DES samples is not obviously caused by selection biases in the two samples or systematic differences in the SED fitting for DES and ZTF, and metallicity evolution with redshift can only partially explain this offset. We further note that the difference in host colour is much more pronounced in the SN II host sample: if there were some overall systematic bias, we would expect to see the same effect in the SN Ibc sample as well. It remains unclear what may be driving the difference in host colour, and more data is required to study this in further detail.

\section{Conclusions}
\label{conclusion}

DES provides a large sample of high-redshift spectroscopically and photometrically-confirmed CCSNe. We derive rest-frame luminosity functions for the DES sample using SED models to K-correct to the rest-frame and GP-interpolations to estimate the peak luminosity. Using the deep $griz$ DES host photometry from \citet{stacks}, we calculate the host properties of the DES sample using SED fits. To examine any selection biases in the sample and investigate the possible effect of redshift evolution on the luminosity function and host properties, we also compare SN and host properties to a low redshift CCSN sample from LOSS and an intermediate redshift sample from ZTF. From this comparison, our main conclusions are as follows:

\begin{enumerate}
    \item We present luminosity functions of SNe II and SNe Ibc for DES, LOSS and ZTF, incorporating a $V_{max}$ correction to mitigate for the effects of Malmquist bias. Where we see a peak in the luminosity function, we fit Gaussian and Lorentzian distributions and present the parameter values to allow these to be used to simulate CCSN samples.
    \item We explore differences between the DES luminosity functions. The DES luminosity functions appears brighter than those of LOSS and ZTF, with differences of significance level 3.0$\sigma$ and 1.8$\sigma$ to each survey for SNe II and 1.9$\sigma$ and 1.1$\sigma$ for SNe Ibc. This could result from higher levels of host galaxy extinction in LOSS and ZTF, however selecting a subset of DES SNe which explode in redder host galaxies does not change the trends we see which suggests this is not the case. This raises the possibility of a luminosity function which evolves with redshift, although at the significance levels we calculate we cannot be sure that any differences are real. Were these effects real, we also cannot rule out causes such as differing host extinction without measurements of this.
    \item There are differences in the host galaxy properties of the LOSS CCSNe compared to the DES and ZTF CCSNe, but these are expected given that LOSS is a galaxy-targeted survey while DES and ZTF are untargeted.
    \item There are also differences in the host galaxy properties of DES CCSNe compared with those in ZTF. The host galaxy stellar masses of both samples are consistent across both SNe II and SNe Ibc. However for SNe II, DES host galaxies are significantly bluer than the ZTF hosts with a significance levels of 3.4$\sigma$ and 3.8$\sigma$ from two-sample KS and AD tests respectively.
    \item We explore correcting the host galaxy colours to account for redshift evolution, and study the possibility that this difference is caused by selection biases in the DES or ZTF samples or systematic differences in the data used, but find that none of the are able to adequately explain the differences.
    \item The host masses and rest-frame $U-R$ colours of SNe II compared to SNe Ibc are generally consistent in both the LOSS and ZTF samples. In the DES sample, hosts of SNe II appear bluer than those of SNe Ibc but only at a significance level of 2.0$\sigma$.
    \item Overall, we observe little environmental dependence on SN peak magnitude across the three samples.
\end{enumerate}

\section*{Acknowledgements}


This work was supported by the Science and Technology Facilities Council [grant number ST/P006760/1] through the DISCnet Centre for Doctoral Training. M.S. acknowledges support from EU/FP7-ERC grant 615929. P.W. acknowledges support from the Science and Technology Facilities Council (STFC) grant ST/R000506/1.


Funding for the DES Projects has been provided by the U.S. Department of Energy, the U.S. National Science Foundation, the Ministry of Science and Education of Spain, the Science and Technology Facilities Council of the United Kingdom, the Higher Education Funding Council for England, the National Center for Supercomputing Applications at the University of Illinois at Urbana-Champaign, the Kavli Institute of Cosmological Physics at the University of Chicago, the Center for Cosmology and Astro-Particle Physics at the Ohio State University, the Mitchell Institute for Fundamental Physics and Astronomy at Texas A\&M University, Financiadora de Estudos e Projetos, Funda{\c c}{\~a}o Carlos Chagas Filho de Amparo {\`a} Pesquisa do Estado do Rio de Janeiro, Conselho Nacional de Desenvolvimento Cient{\'i}fico e Tecnol{\'o}gico and
the Minist{\'e}rio da Ci{\^e}ncia, Tecnologia e Inova{\c c}{\~a}o, the Deutsche Forschungsgemeinschaft and the Collaborating Institutions in the Dark Energy Survey. 

The Collaborating Institutions are Argonne National Laboratory, the University of California at Santa Cruz, the University of Cambridge, Centro de Investigaciones Energ{\'e}ticas, Medioambientales y Tecnol{\'o}gicas-Madrid, the University of Chicago, University College London, the DES-Brazil Consortium, the University of Edinburgh, the Eidgen{\"o}ssische Technische Hochschule (ETH) Z{\"u}rich, Fermi National Accelerator Laboratory, the University of Illinois at Urbana-Champaign, the Institut de Ci{\`e}ncies de l'Espai (IEEC/CSIC), the Institut de F{\'i}sica d'Altes Energies, Lawrence Berkeley National Laboratory, the Ludwig-Maximilians Universit{\"a}t M{\"u}nchen and the associated Excellence Cluster Universe,
the University of Michigan, the National Optical Astronomy Observatory, the University of Nottingham, The Ohio State University, the University of Pennsylvania, the University of Portsmouth, SLAC National Accelerator Laboratory, Stanford University, the University of Sussex, Texas A\&M University, and the OzDES Membership Consortium.

Based in part on observations at Cerro Tololo Inter-American Observatory, National Optical Astronomy Observatory, which is operated by the Association of Universities for Research in Astronomy (AURA) under a cooperative agreement with the National Science Foundation.

The DES data management system is supported by the National Science Foundation under Grant Numbers AST-1138766 and AST-1536171. The DES participants from Spanish institutions are partially supported by MINECO under grants AYA2015-71825, ESP2015-66861, FPA2015-68048, SEV-2016-0588, SEV-2016-0597, and MDM-2015-0509, some of which include ERDF funds from the European Union. IFAE is partially funded by the CERCA program of the Generalitat de Catalunya.
Research leading to these results has received funding from the European Research
Council under the European Union's Seventh Framework Program (FP7/2007-2013) including ERC grant agreements 240672, 291329, and 306478. We acknowledge support from the Brazilian Instituto Nacional de Ci\^encia e Tecnologia (INCT) e-Universe (CNPq grant 465376/2014-2).

This manuscript has been authored by Fermi Research Alliance, LLC under Contract No. DE-AC02-07CH11359 with the U.S. Department of Energy, Office of Science, Office of High Energy Physics.

Based in part on data acquired at the Anglo-Australian Telescope, under program A/2013B/012. We acknowledge the traditional owners of the land on which the AAT stands, the Gamilaraay people, and pay our respects to elders past and present. Based on observations collected at the European Organisation for Astronomical Research in the Southern Hemisphere under ESO programmes 093.A-0749(A), 094.A0310(B), 095.A-0316(A), 096.A-0536(A), 095.D-0797(A), 198.A-0915(A).

This manuscript has been authored by Fermi Research Alliance, LLC under Contract No. DE-AC02-07CH11359 with the U.S. Department of Energy, Office of Science, Office of High Energy Physics. The United States Government retains and the publisher, by accepting the article for publication, acknowledges that the United States Government retains a non-exclusive, paid-up, irrevocable, world-wide license to publish or reproduce the published form of this manuscript, or allow others to do so, for United States Government purposes.

Funding for the Sloan Digital Sky Survey IV has been provided by the Alfred P. Sloan Foundation, the U.S. Department of Energy Office of Science, and the Participating Institutions. 

SDSS-IV acknowledges support and resources from the Center for High Performance Computing at the University of Utah. The SDSS website is www.sdss.org.

SDSS-IV is managed by the Astrophysical Research Consortium for the Participating Institutions of the SDSS Collaboration including the Brazilian Participation Group, the Carnegie Institution for Science, Carnegie Mellon University, Center for Astrophysics | Harvard \& Smithsonian, the Chilean Participation Group, the French Participation Group, Instituto de Astrof\'isica de Canarias, The Johns Hopkins University, Kavli Institute for the Physics and Mathematics of the Universe (IPMU) / University of Tokyo, the Korean Participation Group, Lawrence Berkeley National Laboratory, Leibniz Institut f\"ur Astrophysik Potsdam (AIP), Max-Planck-Institut f\"ur Astronomie (MPIA Heidelberg), Max-Planck-Institut f\"ur Astrophysik (MPA Garching), Max-Planck-Institut f\"ur Extraterrestrische Physik (MPE), National Astronomical Observatories of China, New Mexico State University, New York University, University of Notre Dame, Observat\'ario Nacional / MCTI, The Ohio State University, Pennsylvania State University, Shanghai Astronomical Observatory, United Kingdom Participation Group, Universidad Nacional Aut\'onoma de M\'exico, University of Arizona, University of Colorado Boulder, University of Oxford, University of Portsmouth, University of Utah, University of Virginia, University of Washington, University of Wisconsin, Vanderbilt University, and Yale University.

\section*{Data Availability Statement}

The data underlying this article, as well as the observed DES-SN photometry, are available in the article and in its online supplementary material. The ZTF photometry used can be accessed through the Lasair broker (https://lasair.roe.ac.uk/).

\section*{Affiliations}

$^{1}$ School of Physics and Astronomy, University of Southampton,  Southampton, SO17 1BJ, UK\\
$^{2}$ Finnish Centre for Astronomy with ESO (FINCA), FI-20014 University of Turku, Finland\\
$^{3}$ Tuorla Observatory, Department of Physics and Astronomy, FI-20014 University of Turku, Finland\\
$^{4}$ Institute of Cosmology and Gravitation, University of Portsmouth, Portsmouth, PO1 3FX, UK\\
$^{5}$ Institut d'Estudis Espacials de Catalunya (IEEC), 08034 Barcelona, Spain\\
$^{6}$ Institute of Space Sciences (ICE, CSIC),  Campus UAB, Carrer de Can Magrans, s/n,  08193 Barcelona, Spain\\
$^{7}$ Centre for Astrophysics \& Supercomputing, Swinburne University of Technology, Victoria 3122, Australia\\
$^{8}$ Center for Astrophysics $\vert$ Harvard \& Smithsonian, 60 Garden Street, Cambridge, MA 02138, USA\\
$^{9}$ School of Mathematics and Physics, University of Queensland,  Brisbane, QLD 4072, Australia\\
$^{10}$ Centre for Gravitational Astrophysics, College of Science, The Australian National University, ACT 2601, Australia\\
$^{11}$ The Research School of Astronomy and Astrophysics, Australian National University, ACT 2601, Australia\\
$^{12}$ Department of Physics, Duke University Durham, NC 27708, USA\\
$^{13}$ Cerro Tololo Inter-American Observatory, NSF's National Optical-Infrared Astronomy Research Laboratory, Casilla 603, La Serena, Chile\\
$^{14}$ Laborat\'orio Interinstitucional de e-Astronomia - LIneA, Rua Gal. Jos\'e Cristino 77, Rio de Janeiro, RJ - 20921-400, Brazil\\
$^{15}$ Fermi National Accelerator Laboratory, P. O. Box 500, Batavia, IL 60510, USA\\
$^{16}$ Department of Physics, University of Michigan, Ann Arbor, MI 48109, USA\\
$^{17}$ Centro de Investigaciones Energ\'eticas, Medioambientales y Tecnol\'ogicas (CIEMAT), Madrid, Spain\\
$^{18}$ CNRS, UMR 7095, Institut d'Astrophysique de Paris, F-75014, Paris, France\\
$^{19}$ Sorbonne Universit\'es, UPMC Univ Paris 06, UMR 7095, Institut d'Astrophysique de Paris, F-75014, Paris, France\\
$^{20}$ University Observatory, Faculty of Physics, Ludwig-Maximilians-Universit\"at, Scheinerstr. 1, 81679 Munich, Germany\\
$^{21}$ Department of Physics \& Astronomy, University College London, Gower Street, London, WC1E 6BT, UK\\
$^{22}$ Instituto de Astrofisica de Canarias, E-38205 La Laguna, Tenerife, Spain\\
$^{23}$ Universidad de La Laguna, Dpto. Astrofísica, E-38206 La Laguna, Tenerife, Spain\\
$^{24}$ INAF-Osservatorio Astronomico di Trieste, via G. B. Tiepolo 11, I-34143 Trieste, Italy\\
$^{25}$ Center for Astrophysical Surveys, National Center for Supercomputing Applications, 1205 West Clark St., Urbana, IL 61801, USA\\
$^{26}$ Department of Astronomy, University of Illinois at Urbana-Champaign, 1002 W. Green Street, Urbana, IL 61801, USA\\
$^{27}$ Institut de F\'{\i}sica d'Altes Energies (IFAE), The Barcelona Institute of Science and Technology, Campus UAB, 08193 Bellaterra (Barcelona) Spain\\
$^{28}$ Astronomy Unit, Department of Physics, University of Trieste, via Tiepolo 11, I-34131 Trieste, Italy\\
$^{29}$ Institute for Fundamental Physics of the Universe, Via Beirut 2, 34014 Trieste, Italy\\
$^{30}$ Hamburger Sternwarte, Universit\"{a}t Hamburg, Gojenbergsweg 112, 21029 Hamburg, Germany\\
$^{31}$ Department of Physics, IIT Hyderabad, Kandi, Telangana 502285, India\\
$^{32}$ Jet Propulsion Laboratory, California Institute of Technology, 4800 Oak Grove Dr., Pasadena, CA 91109, USA\\
$^{33}$ Institute of Theoretical Astrophysics, University of Oslo. P.O. Box 1029 Blindern, NO-0315 Oslo, Norway\\
$^{34}$ Kavli Institute for Cosmological Physics, University of Chicago, Chicago, IL 60637, USA\\
$^{35}$ Instituto de Fisica Teorica UAM/CSIC, Universidad Autonoma de Madrid, 28049 Madrid, Spain\\
$^{36}$ Department of Physics and Astronomy, University of Pennsylvania, Philadelphia, PA 19104, USA\\
$^{37}$ Observat\'orio Nacional, Rua Gal. Jos\'e Cristino 77, Rio de Janeiro, RJ - 20921-400, Brazil\\
$^{38}$ Santa Cruz Institute for Particle Physics, Santa Cruz, CA 95064, USA\\
$^{39}$ Center for Cosmology and Astro-Particle Physics, The Ohio State University, Columbus, OH 43210, USA\\
$^{40}$ Department of Physics, The Ohio State University, Columbus, OH 43210, USA\\
$^{41}$ Australian Astronomical Optics, Macquarie University, North Ryde, NSW 2113, Australia\\
$^{42}$ Lowell Observatory, 1400 Mars Hill Rd, Flagstaff, AZ 86001, USA\\
$^{43}$ Sydney Institute for Astronomy, School of Physics, A28, The University of Sydney, NSW 2006, Australia\\
$^{44}$ Instituci\'o Catalana de Recerca i Estudis Avan\c{c}ats, E-08010 Barcelona, Spain\\
$^{45}$ Physics Department, 2320 Chamberlin Hall, University of Wisconsin-Madison, 1150 University Avenue Madison, WI  53706-1390\\
$^{46}$ Department of Astronomy, University of California, Berkeley,  501 Campbell Hall, Berkeley, CA 94720, USA\\
$^{47}$ Institute of Astronomy, University of Cambridge, Madingley Road, Cambridge CB3 0HA, UK\\
$^{48}$ Department of Astrophysical Sciences, Princeton University, Peyton Hall, Princeton, NJ 08544, USA\\
$^{49}$ Centro de Investigaciones Energéticas, Medioambientales y Tecnológicas (CIEMAT), Av. Complutense, 22, E- 28040 Madrid, Spain\\
$^{50}$ Department of Physics and Astronomy, Pevensey Building, University of Sussex, Brighton, BN1 9QH, UK\\
$^{51}$ Kavli Institute for Particle Astrophysics \& Cosmology, P. O. Box 2450, Stanford University, Stanford, CA 94305, USA\\
$^{52}$ SLAC National Accelerator Laboratory, Menlo Park, CA 94025, USA\\
$^{53}$ Computer Science and Mathematics Division, Oak Ridge National Laboratory, Oak Ridge, TN 37831\\
$^{54}$ Excellence Cluster Origins, Boltzmannstr.\ 2, 85748 Garching, Germany\\
$^{55}$ Max Planck Institute for Extraterrestrial Physics, Giessenbachstrasse, 85748 Garching, Germany\\
$^{56}$ Universit\"ats-Sternwarte, Fakult\"at f\"ur Physik, Ludwig-Maximilians Universit\"at M\"unchen, Scheinerstr. 1, 81679 M\"unchen, Germany\\



\bibliographystyle{mnras}
\bibliography{refs} 




\appendix

\section{Estimating host properties using photometric redshifts}
\label{appendix:host}

For the DES-SN sample of CCSNe with only photometric host redshifts, we estimate host galaxy properties using the following Monte Carlo (MC) process:

\begin{itemize}
    \item For each object, we have 0.5th, 2.5th, 16th, 84th, 97.5th and 99.5th percentiles of the photometric redshift distribution. Studying the cumulative distributions of these values shows the distribution to be approximately Gaussian - as such, we model the redshift distribution of each host as a Gaussian, estimating the mean and standard deviation of this Gaussian by fitting a generalised error function to our cumulative distribution.
    \item We use the $griz$ photometry of each host from the DES deep coadded host images to estimate the properties of the host, following the same SED fitting process as for objects with spectroscopic redshift information. We do this for every redshift between the 0.5 and 99.5 percentiles in the redshift distribution of each host, in intervals of 0.001. This gives the properties the host galaxy would have were it located at each redshift in this distribution.
    \item For each host galaxy, we draw a random redshift from a Gaussian distribution using our estimates of mean and standard deviation. We then select the properties of each host galaxy at these redshifts and use these to produce a CDF for each host property for this randomised sample. (If a randomised redshift lies outside the redshift range of our spectroscopic sample, it is excluded from the sample.)
    \item We repeat this 10,000 times in an Monte Carlo process, examining the spread of the CDFs over all iterations to obtain a final CDF with an associated error. This allows us to include the large uncertainties in host redshift into a comparison of the host galaxy properties for different samples.
\end{itemize}

We find that only including SNe with a randomised redshift of less than 0.25 in each iteration gives a sample size that varies between 3 and 8, and typically 5 or 6. Photometric redshifts are currently only available for three out of the ten DES fields -- extrapolating, we would expect there to be between 10--27 CCSNe for which we have DES photometry but no spectroscopic host redshift, and which would otherwise be included in our luminosity functions. Our luminosity function sample contains 98 DES SNe with spectroscopic host redshifts, thus we have spectroscopic host redshifts for $\sim 75 - 90$ per cent of the CCSNe which should be included in our sample suggesting that these \lq missing\rq\ SNe should not have a significant effect on the luminosity function.

\section{SED-fitting analysis for LOSS host galaxies}
\label{appendix:SED}

\begin{figure*}
\centering
\includegraphics[width = \textwidth, angle=-90]{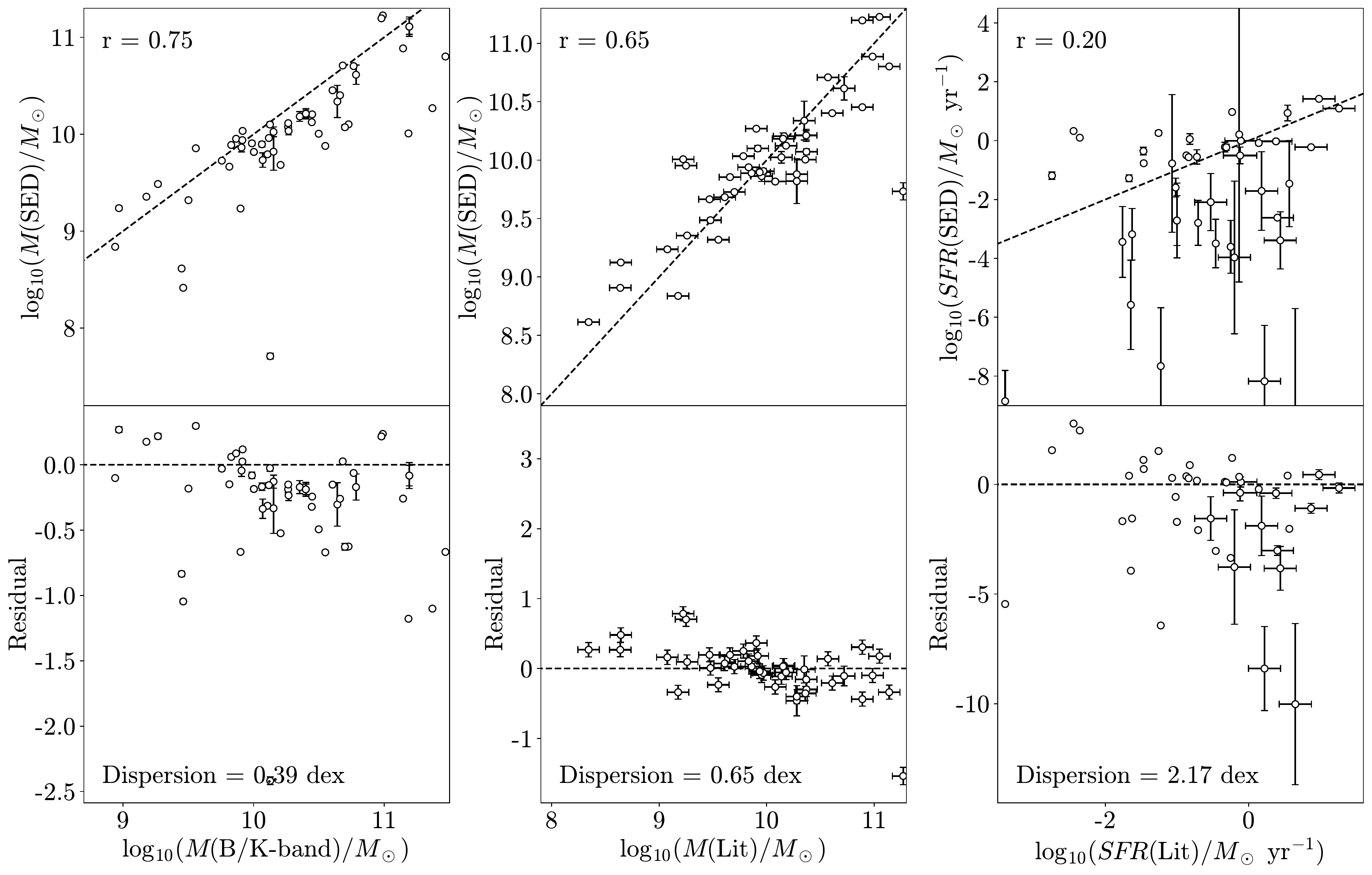}
\caption{\textbf{Upper panels:} Host galaxy stellar masses and SFRs for LOSS from our SED fits to SDSS $ugriz$ photometry compared with stellar masses derived from the $K$-band mass-to-luminosity ratio presented in \citetalias{li11} and with stellar masses and SFRs presented in \citet{Leroy19} and \citet{Karachentseva20}. Correlation coefficients are shown, and the dashed line shows a perfect agreement. \textbf{Lower panels:} Residuals from the perfect agreement in the upper panels.}
\label{LOSS_host_residual}
\end{figure*}

Fig.~\ref{LOSS_host_residual} shows comparisons with our host properties from SED fitting for the $B$/$K$-band stellar masses and host stellar masses and SFRs from \citet{Leroy19} and \citet{Karachentseva20}. These were performed with the full LOSS SN sample, without the selection of events used for the SNe in the luminosity functions -- this gives 56 host galaxies with properties derived from SED fits to SDSS photometry, 96 host galaxies with $B$/$K$-band masses and 71 with previously published literature values for stellar mass and SFR.

Our stellar masses from SED-fitting are consistent  with the $B$/$K$-band masses. The Pearson correlation coefficient ($r$) for these two sets of masses is 0.75 indicating a strong correlation, with a dispersion of 0.39\,dex. Comparing our stellar masses from others in the literature gives $r=0.65$ and a dispersion of 0.65\,dex. Overall, our masses seem broadly consistent with those derived from other methods. However, the correlation between our SFR values derived from SED fits and those from literature is only $r=0.30$ with a dispersion of 2.17\,dex, demonstrating the uncertainties in estimating SFRs from SED fitting which have a stronger dependence on star-formation history. We do not use SFR in our analysis, and instead use rest-frame $U-R$, which is well-constrained by the observed data.

\section{ZTF Magnitude Limit for Malmquist Bias Correction}
\label{appendix:ztf_lim}

\begin{figure}
\centering
\includegraphics[width = \columnwidth]{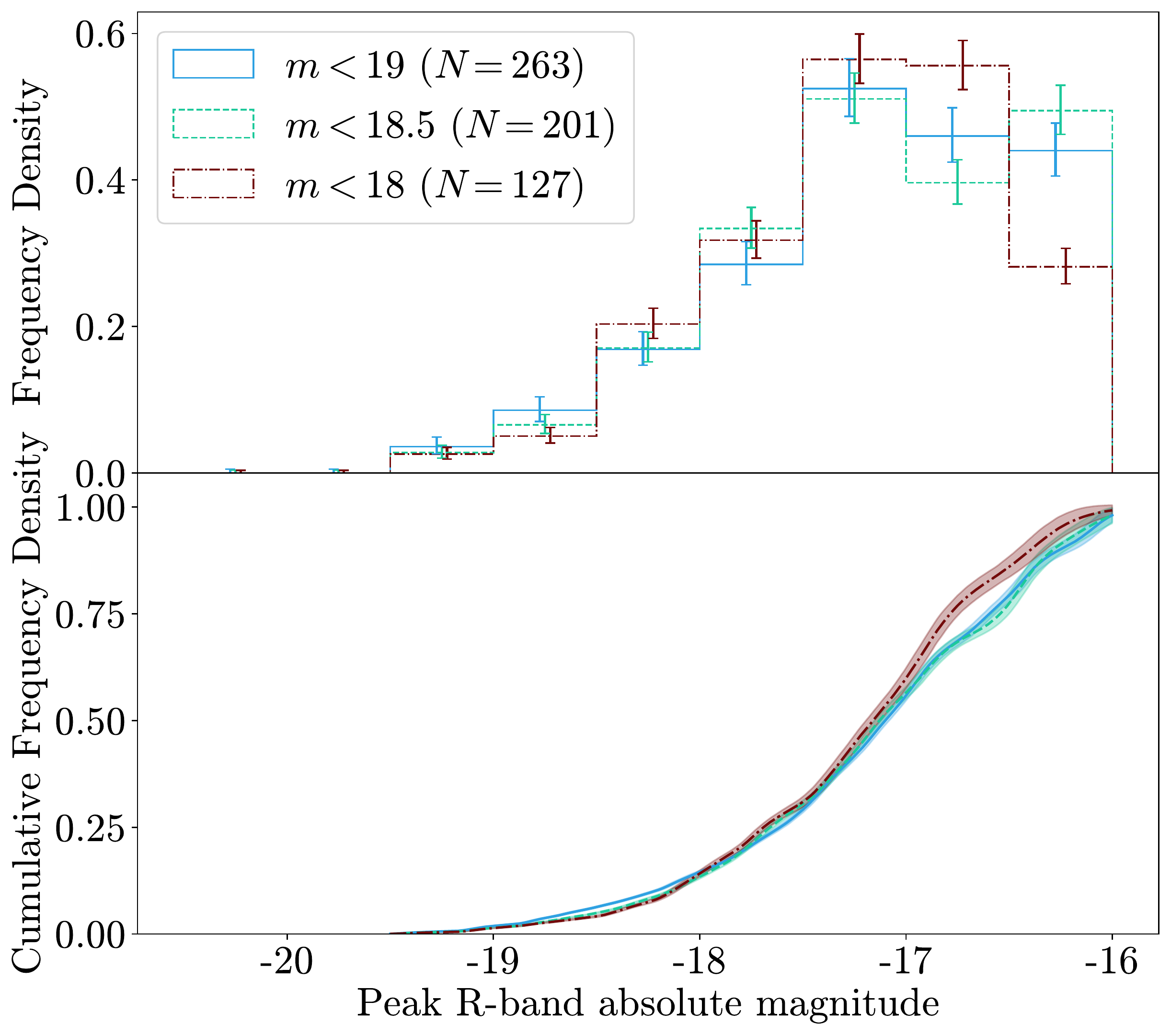}
\caption{As Fig. \ref{lfs}, but for samples only from ZTF with $V_{max}$ corrections calculated for magnitude limits of 18, 18.5 and 19 mag.}
\label{all_ztf_lfs}
\end{figure}

As mentioned in Section \ref{Vmax}, we consider the 97, 93 and 75 per cent spectroscopic completeness limits of ZTF, $18$, $18.5$ and $19$ mag respectively. The luminosity functions for ZTF with these three limits are shown in figure \ref{all_ztf_lfs}. Above -17 mag, these appear consistent with each other, however these luminosity functions diverge between -16 and -17. A limit of 18 mag omits some SNe in this region and appears to bias the sample in favour of brighter objects. The luminosity functions for a limit of 18.5 and 19 appear consistent with each other - as a result, we settle on a limit of 19 mag to maximise the sample size. 

\section{Data Tables}
\label{appendix:data_tables}

Table \ref{long_table} present the peak $R$-band absolute magnitudes and host galaxy properties used for all analysis in this work, for the three different samples. Table \ref{DES_phot_table} presents DECam photometry of the DES sample presented in this paper.

\clearpage
\onecolumn

\begin{ThreePartTable}
\begin{longtable}{llllllllllllll}
\caption{Supernova peak $R$-band absolute magnitudes (M$_\text{max}$) and host galaxy properties for the DES, LOSS and ZTF samples presented in this work, along with their associated errors. Please note, objects denoted with $^*$ correspond to two SNe II in ZTF with M$_\text{max}$ less than -16 which have used only for the fits presented in Section 3.3, to help constrain the peak of the luminosity function, and not in the rest of the analysis.}
\label{long_table}
\\
\hline
SN & Survey & Class & Source$^\text{a}$ & z & M$_\text{max}^\text{b}$ & err & weight$^\text{c}$ & $M^\text{d}$ & err & $U-R^\text{e}$ & err & $B-V^\text{f}$ & err \\ 
\hline
\endfirsthead
\multicolumn{14}{c}{{\bfseries \tablename\ \thetable{} -- continued from previous page}} \\
\hline
SN & Survey & Class & Source$^\text{a}$ & z & M$_\text{max}^\text{b}$ & err & weight$^\text{c}$ & $M^\text{d}$ & err & $U-R^\text{e}$ & err & $B-V^\text{f}$ & err \\  
\hline
\endhead
\hline \multicolumn{14}{c}{{\textbf{Continued on next page}}} \\ 
\endfoot
\hline
\endlastfoot
DES13C2jtx & DES & II & spec & 0.223 & -18.61 & 0.03 & 1.00 & 10.63 & 0.02 & 1.32 & 0.01 & 0.72 & 0.01 \\
DES13X3fca & DES & II & spec & 0.096 & -16.92 & 0.02 & 1.00 & 10.02 & 0.01 & 1.48 & 0.01 & 0.69 & 0.01 \\
DES14C3aol & DES & II & spec & 0.076 & -16.21 & 0.02 & 1.00 & 10.83 & 0.02 & 1.71 & 0.02 & 0.89 & 0.02 \\
DES14X2cy & DES & II & spec & 0.232 & -18.99 & 0.01 & 1.00 & 9.22 & 0.02 & 0.43 & 0.02 & 0.38 & 0.01 \\
DES14X2nzt & DES & II & spec & 0.195 & -18.31 & 0.02 & 1.00 & -- & -- & -- & -- & -- & -- \\
DES15C2eaz & DES & II & spec & 0.062 & -17.49 & 0.03 & 1.00 & 8.10 & 0.01 & 0.78 & 0.03 & 0.39 & 0.02 \\
DES15C2lna & DES & II & spec & 0.069 & -16.72 & 0.02 & 1.39 & 10.02 & 0.03 & 1.07 & 0.04 & 0.62 & 0.02 \\
DES15C2lpp & DES & II & spec & 0.181 & -17.12 & 0.04 & 1.00 & 10.47 & 0.01 & 1.21 & 0.02 & 0.65 & 0.01 \\
DES15C2npz & DES & II & spec & 0.123 & -17.18 & 0.02 & 1.00 & 9.13 & 0.01 & 0.73 & 0.02 & 0.46 & 0.01 \\
DES15E1iuh & DES & II & spec & 0.105 & -17.14 & 0.02 & 1.00 & 7.44 & 0.08 & 0.68 & 0.14 & 0.43 & 0.06 \\
DES15S1by & DES & II & spec & 0.129 & -17.13 & 0.03 & 1.00 & 9.33 & 0.04 & 0.60 & 0.06 & 0.40 & 0.01 \\
DES15S1cj & DES & II & spec & 0.167 & -18.16 & 0.02 & 1.00 & 10.17 & 0.01 & 1.33 & 0.02 & 0.71 & 0.01 \\
DES15S1lrp & DES & II & spec & 0.223 & -18.44 & 0.02 & 1.00 & 8.36 & 0.05 & 0.71 & 0.10 & 0.41 & 0.05 \\
DES15S2eaq & DES & II & spec & 0.068 & -16.37 & 0.03 & 2.09 & 9.22 & 0.01 & 0.69 & 0.02 & 0.44 & 0.01 \\
DES15X2mku & DES & II & spec & 0.09 & -16.67 & 0.02 & 1.47 & 8.73 & 0.04 & -- & -- & 0.36 & 0.02 \\
DES15X3mpq & DES & II & spec & 0.188 & -17.55 & 0.02 & 1.00 & 10.29 & 0.01 & 1.00 & 0.02 & 0.56 & 0.01 \\
DES15X3nad & DES & II & spec & 0.10 & -18.15 & 0.03 & 1.00 & -- & -- & -- & -- & -- & -- \\
DES16C1cbg & DES & II & spec & 0.111 & -16.99 & 0.02 & 1.02 & 9.07 & 0.01 & 0.85 & 0.03 & 0.51 & 0.01 \\
DES16C2cbv & DES & II & spec & 0.109 & -17.37 & 0.03 & 1.00 & 8.08 & 0.02 & 0.10 & 0.02 & 0.25 & 0.01 \\
DES16X2bkr & DES & II & spec & 0.159 & -17.42 & 0.02 & 1.00 & 8.20 & 0.02 & -- & -- & 0.31 & 0.02 \\
DES16X3cpl & DES & II & spec & 0.205 & -17.29 & 0.01 & 1.00 & 9.07 & 0.03 & -- & -- & 0.59 & 0.03 \\
DES17C2pf & DES & II & spec & 0.135 & -17.22 & 0.02 & 1.00 & 7.15 & 0.14 & 0.39 & 0.23 & 0.34 & 0.09 \\
DES17C3aye & DES & II & spec & 0.157 & -18.07 & 0.02 & 1.00 & 9.83 & 0.01 & 0.89 & 0.02 & 0.51 & 0.01 \\
DES17C3bei & DES & II & spec & 0.103 & -17.15 & 0.03 & 1.00 & 10.46 & 0.01 & 1.12 & 0.02 & 0.61 & 0.01 \\
DES17E2bhj & DES & II & spec & 0.186 & -17.09 & 0.03 & 1.00 & 10.44 & 0.01 & 1.03 & 0.03 & 0.57 & 0.01 \\
DES17S2oo & DES & II & spec & 0.23 & -17.82 & 0.03 & 1.00 & 7.95 & 0.16 & 0.53 & 0.22 & 0.39 & 0.11 \\
DES17X1aow & DES & II & spec & 0.139 & -17.65 & 0.04 & 1.00 & 10.56 & 0.06 & -- & -- & 0.91 & 0.01 \\
DES17X1axb & DES & II & spec & 0.139 & -16.77 & 0.02 & 1.30 & 9.44 & 0.02 & 0.39 & 0.03 & 0.35 & 0.01 \\
DES17X1gd & DES & II & spec & 0.189 & -17.59 & 0.03 & 1.00 & 9.02 & 0.03 & 0.31 & 0.03 & 0.31 & 0.01 \\
DES17X2dql & DES & II & spec & 0.143 & -16.66 & 0.03 & 1.48 & -- & -- & -- & -- & -- & -- \\
DES17X3dub & DES & II & spec & 0.123 & -16.11 & 0.02 & 1.00 & 8.53 & 0.06 & 0.45 & 0.09 & 0.36 & 0.02 \\
DES13C1ffj & DES & II & phot & 0.218 & -17.36 & 0.02 & 1.00 & 9.32 & 0.04 & -- & -- & 0.36 & 0.02 \\
DES13C1woj & DES & II & phot & 0.232 & -17.06 & 0.04 & 1.00 & 9.04 & 0.02 & 0.47 & 0.03 & 0.38 & 0.02 \\
DES13C3abtm & DES & II & phot & 0.195 & -16.75 & 0.08 & 1.00 & 9.43 & 0.02 & 0.23 & 0.02 & 0.27 & 0.01 \\
DES13C3avns & DES & II & phot & 0.242 & -17.28 & 0.02 & 1.00 & 9.85 & 0.02 & 0.48 & 0.03 & 0.36 & 0.01 \\
DES13E1ackb & DES & II & phot & 0.216 & -17.55 & 0.03 & 1.00 & 9.04 & 0.03 & -- & -- & 0.37 & 0.02 \\
DES13E1pat & DES & II & phot & 0.223 & -18.54 & 0.03 & 1.00 & 10.80 & 0.06 & 1.86 & 0.02 & 0.95 & 0.01 \\
DES13X1hxq & DES & II & phot & 0.24 & -17.22 & 0.05 & 1.00 & 9.31 & 0.01 & 0.56 & 0.02 & 0.40 & 0.01 \\
DES14C1lnl & DES & II & phot & 0.196 & -17.67 & 0.01 & 1.00 & 8.46 & 0.03 & 0.38 & 0.04 & 0.29 & 0.02 \\
DES14C2rqo & DES & II & phot & 0.216 & -18.17 & 0.02 & 1.00 & 7.88 & 0.06 & 0.41 & 0.09 & 0.34 & 0.05 \\
DES14C2rso & DES & II & phot & 0.223 & -17.41 & 0.02 & 1.00 & 10.04 & 0.02 & 0.79 & 0.02 & 0.47 & 0.01 \\
DES14C3asy & DES & II & phot & 0.174 & -16.20 & 0.02 & 1.00 & 9.05 & 0.01 & 0.57 & 0.02 & 0.40 & 0.01 \\
DES14C3kzd & DES & II & phot & 0.078 & -17.33 & 0.03 & 1.00 & 10.59 & 0.01 & 1.18 & 0.02 & 0.64 & 0.01 \\
DES14E1bdh & DES & II & phot & 0.238 & -16.89 & 0.02 & 1.13 & 9.11 & 0.02 & 0.49 & 0.02 & 0.38 & 0.01 \\
DES14E2fmi & DES & II & phot & 0.21 & -17.56 & 0.03 & 1.00 & 9.50 & 0.01 & 0.73 & 0.02 & 0.44 & 0.01 \\
DES14E2gvo & DES & II & phot & 0.126 & -16.44 & 0.03 & 1.91 & 10.06 & 0.01 & -- & -- & 0.79 & 0.01 \\
DES14S1mkj & DES & II & phot & 0.189 & -17.23 & 0.03 & 1.00 & 9.93 & 0.01 & -- & -- & 0.71 & 0.01 \\
DES14S2dbe & DES & II & phot & 0.201 & -17.81 & 0.03 & 1.00 & 10.67 & 0.01 & 0.91 & 0.02 & 0.52 & 0.01 \\
DES14S2obu & DES & II & phot & 0.18 & -17.49 & 0.03 & 1.00 & 9.59 & 0.01 & -- & -- & 0.45 & 0.01 \\
DES14X2mqu & DES & II & phot & 0.092 & -16.93 & 0.03 & 1.09 & 9.85 & 0.01 & 0.91 & 0.02 & 0.53 & 0.01 \\
DES14X3ddy & DES & II & phot & 0.205 & -17.21 & 0.02 & 1.00 & 9.86 & 0.02 & 0.56 & 0.02 & 0.41 & 0.01 \\
DES14X3ili & DES & II & phot & 0.142 & -17.24 & 0.16 & 1.00 & 9.05 & 0.02 & 0.56 & 0.03 & 0.41 & 0.01 \\
DES15C1ats & DES & II & phot & 0.244 & -17.16 & 0.05 & 1.00 & 9.32 & 0.02 & 0.54 & 0.02 & 0.39 & 0.01 \\
DES15C1lwn & DES & II & phot & 0.206 & -16.84 & 0.02 & 1.20 & 9.03 & 0.02 & 0.28 & 0.02 & 0.31 & 0.01 \\
DES15C2lpm & DES & II & phot & 0.19 & -17.00 & 0.04 & 1.00 & 10.10 & 0.01 & 0.87 & 0.02 & 0.50 & 0.01 \\
DES15X2aso & DES & II & phot & 0.207 & -17.44 & 0.04 & 1.00 & 9.59 & 0.02 & 0.65 & 0.02 & 0.43 & 0.01 \\
DES16C1ftn & DES & II & phot & 0.205 & -17.27 & 0.03 & 1.00 & 9.62 & 0.01 & 1.02 & 0.02 & 0.57 & 0.01 \\
DES16C3ecv & DES & II & phot & 0.244 & -17.69 & 0.02 & 1.00 & -- & -- & -- & -- & -- & -- \\
DES16C3fuk & DES & II & phot & 0.25 & -17.05 & 0.04 & 1.00 & -- & -- & -- & -- & -- & -- \\
DES16E1eab & DES & II & phot & 0.179 & -16.88 & 0.02 & 1.15 & 8.64 & 0.03 & 0.34 & 0.03 & 0.34 & 0.01 \\
DES16E1eke & DES & II & phot & 0.081 & -17.65 & 0.09 & 1.00 & 8.24 & 0.01 & 0.68 & 0.03 & 0.43 & 0.01 \\
DES16S2eam & DES & II & phot & 0.181 & -16.39 & 0.02 & 2.03 & 10.36 & 0.01 & 1.00 & 0.04 & 0.55 & 0.01 \\
DES16X3qx & DES & II & phot & 0.187 & -16.46 & 0.02 & 1.00 & 10.81 & 0.02 & 1.28 & 0.02 & 0.70 & 0.01 \\
DES17C2eoa & DES & II & phot & 0.213 & -16.64 & 0.02 & 1.51 & 10.06 & 0.01 & 0.69 & 0.02 & 0.43 & 0.01 \\
DES17E1dtv & DES & II & phot & 0.199 & -17.25 & 0.02 & 1.00 & 10.11 & 0.01 & 1.21 & 0.02 & 0.65 & 0.01 \\
DES17E2brh & DES & II & phot & 0.222 & -16.97 & 0.03 & 1.04 & 10.00 & 0.01 & -- & -- & 0.48 & 0.01 \\
DES17X1gco & DES & II & phot & 0.203 & -17.59 & 0.02 & 1.00 & 8.84 & 0.01 & 0.38 & 0.02 & 0.35 & 0.01 \\
DES17X2ayo & DES & II & phot & 0.122 & -17.02 & 0.04 & 1.00 & 8.83 & 0.02 & -- & -- & 0.39 & 0.01 \\
DES17X2bxs & DES & II & phot & 0.177 & -16.57 & 0.06 & 1.65 & 10.40 & 0.01 & -- & -- & 0.64 & 0.01 \\
DES13C1feu & DES & Ibc & spec & 0.059 & -17.68 & 0.03 & 1.00 & 10.42 & 0.04 & 1.29 & 0.04 & 0.70 & 0.01 \\
DES14X2fna & DES & Ibc & spec & 0.045 & -19.37 & 0.04 & 1.00 & 8.17 & 0.01 & 0.83 & 0.02 & 0.43 & 0.02 \\
DES15C1mat & DES & Ibc & spec & 0.12 & -17.48 & 0.03 & 1.00 & 10.54 & 0.01 & 0.87 & 0.02 & 0.50 & 0.01 \\
DES15E2mhi & DES & Ibc & spec & 0.208 & -17.51 & 0.04 & 1.00 & 8.93 & 0.01 & 0.64 & 0.02 & 0.43 & 0.01 \\
DES16C1zb & DES & Ibc & spec & 0.13 & -17.18 & 0.02 & 1.00 & 10.70 & 0.01 & 1.18 & 0.02 & 0.64 & 0.01 \\
DES16S1kt & DES & Ibc & spec & 0.068 & -17.31 & 0.04 & 1.00 & 9.90 & 0.04 & 1.02 & 0.05 & 0.58 & 0.02 \\
DES16X2bvf & DES & Ibc & spec & 0.135 & -17.97 & 0.02 & 1.00 & 8.57 & 0.02 & 0.32 & 0.03 & 0.32 & 0.02 \\
DES16X2dqz & DES & Ibc & spec & 0.204 & -17.55 & 0.03 & 1.00 & 8.75 & 0.02 & 0.94 & 0.06 & 0.52 & 0.03 \\
DES16X3bdb & DES & Ibc & spec & 0.229 & -18.08 & 0.02 & 1.00 & 9.68 & 0.01 & -- & -- & 0.50 & 0.01 \\
DES17C1bzd & DES & Ibc & spec & 0.12 & -17.91 & 0.02 & 1.00 & 11.67 & 0.02 & 1.28 & 0.02 & 0.70 & 0.01 \\
DES17C1ffz & DES & Ibc & spec & 0.09 & -17.85 & 0.02 & 1.00 & 9.50 & 0.01 & -- & -- & 0.54 & 0.01 \\
DES13C1anve & DES & Ibc & phot & 0.214 & -17.34 & 0.03 & 1.00 & 9.44 & 0.02 & 0.57 & 0.03 & 0.40 & 0.01 \\
DES13C1hxh & DES & Ibc & phot & 0.161 & -17.18 & 0.03 & 1.00 & 9.98 & 0.01 & 0.87 & 0.04 & 0.50 & 0.01 \\
DES13C2rgr & DES & Ibc & phot & 0.244 & -18.20 & 0.03 & 1.00 & 9.44 & 0.02 & -- & -- & 0.24 & 0.01 \\
DES13C3absw & DES & Ibc & phot & 0.069 & -18.55 & 0.03 & 1.00 & 9.08 & 0.04 & -- & -- & 0.66 & 0.02 \\
DES13C3aeiv & DES & Ibc & phot & 0.174 & -17.76 & 0.02 & 1.00 & 8.36 & 0.03 & -- & -- & 0.28 & 0.01 \\
DES13C3lex & DES & Ibc & phot & 0.109 & -16.88 & 0.02 & 1.00 & 8.48 & 0.01 & -- & -- & 0.20 & 0.01 \\
DES13C3tqe & DES & Ibc & phot & 0.217 & -17.15 & 0.03 & 1.00 & 9.11 & 0.02 & 0.31 & 0.02 & 0.33 & 0.01 \\
DES13S1rww & DES & Ibc & phot & 0.201 & -17.10 & 0.03 & 1.00 & 9.06 & 0.04 & 0.42 & 0.04 & 0.37 & 0.02 \\
DES13S2hxn & DES & Ibc & phot & 0.143 & -16.82 & 0.04 & 1.23 & 9.53 & 0.02 & -- & -- & 0.39 & 0.01 \\
DES13X1atuq & DES & Ibc & phot & 0.084 & -16.28 & 0.06 & 2.31 & 9.98 & 0.03 & 1.16 & 0.04 & 0.65 & 0.02 \\
DES14C2bcu & DES & Ibc & phot & 0.213 & -17.47 & 0.02 & 1.00 & 10.86 & 0.01 & -- & -- & 0.94 & 0.01 \\
DES14C3dzo & DES & Ibc & phot & 0.247 & -17.62 & 0.01 & 1.00 & 9.54 & 0.01 & -- & -- & 0.44 & 0.01 \\
DES14C3guv & DES & Ibc & phot & 0.215 & -17.25 & 0.02 & 1.00 & 9.86 & 0.01 & -- & -- & 0.44 & 0.01 \\
DES14C3ouq & DES & Ibc & phot & 0.107 & -16.74 & 0.03 & 1.00 & 10.03 & 0.01 & 1.10 & 0.02 & 0.60 & 0.01 \\
DES14C3qby & DES & Ibc & phot & 0.219 & -17.65 & 0.07 & 1.00 & 10.50 & 0.01 & 1.28 & 0.01 & 0.69 & 0.01 \\
DES14C3sle & DES & Ibc & phot & 0.144 & -17.25 & 0.02 & 1.00 & 9.48 & 0.01 & -- & -- & 0.58 & 0.01 \\
DES14E1nyv & DES & Ibc & phot & 0.118 & -16.65 & 0.03 & 1.49 & 10.39 & 0.04 & 1.61 & 0.03 & 0.84 & 0.01 \\
DES14X1mhi & DES & Ibc & phot & 0.129 & -17.28 & 0.05 & 1.00 & 9.78 & 0.01 & 1.50 & 0.02 & 0.79 & 0.01 \\
DES14X1qwn & DES & Ibc & phot & 0.151 & -18.28 & 0.02 & 1.00 & 10.74 & 0.02 & -- & -- & 0.91 & 0.01 \\
DES14X1tae & DES & Ibc & phot & 0.046 & -16.70 & 0.07 & 1.42 & 10.77 & 0.01 & 2.02 & 0.02 & 1.02 & 0.02 \\
DES14X3fok & DES & Ibc & phot & 0.094 & -16.66 & 0.03 & 1.00 & 10.18 & 0.01 & 1.08 & 0.02 & 0.59 & 0.02 \\
DES15C3hbo & DES & Ibc & phot & 0.16 & -16.87 & 0.02 & 1.00 & 9.72 & 0.01 & 0.79 & 0.03 & 0.47 & 0.01 \\
DES15C3lrw & DES & Ibc & phot & 0.129 & -17.46 & 0.01 & 1.00 & 9.40 & 0.01 & -- & -- & 0.61 & 0.01 \\
DES15C3mbe & DES & Ibc & phot & 0.181 & -16.41 & 0.02 & 1.00 & 8.94 & 0.02 & 0.52 & 0.03 & 0.37 & 0.01 \\
DES15C3mud & DES & Ibc & phot & 0.182 & -17.85 & 0.06 & 1.00 & 10.02 & 0.01 & 1.10 & 0.01 & 0.60 & 0.01 \\
DES15C3nqt & DES & Ibc & phot & 0.139 & -17.44 & 0.12 & 1.00 & 10.78 & 0.01 & 1.25 & 0.02 & 0.68 & 0.01 \\
DES15X2kzu & DES & Ibc & phot & 0.136 & -17.14 & 0.04 & 1.00 & 9.67 & 0.01 & 0.87 & 0.02 & 0.51 & 0.02 \\
DES15X3nso & DES & Ibc & phot & 0.238 & -17.15 & 0.02 & 1.00 & 9.32 & 0.01 & 0.72 & 0.02 & 0.44 & 0.01 \\
DES16C3aky & DES & Ibc & phot & 0.239 & -16.10 & 0.04 & 1.00 & 8.77 & 0.02 & -- & -- & 0.38 & 0.02 \\
DES16C3byu & DES & Ibc & phot & 0.143 & -18.09 & 0.02 & 1.00 & 10.96 & 0.01 & -- & -- & 0.94 & 0.01 \\
DES16E1bkh & DES & Ibc & phot & 0.115 & -17.19 & 0.02 & 1.00 & 10.13 & 0.01 & -- & -- & 0.75 & 0.01 \\
DES16S1bnj & DES & Ibc & phot & 0.183 & -17.46 & 0.02 & 1.00 & 7.83 & 0.09 & 0.23 & 0.13 & 0.28 & 0.05 \\
DES16S1ku & DES & Ibc & phot & 0.064 & -16.19 & 0.03 & 2.59 & 10.20 & 0.04 & 1.33 & 0.04 & 0.72 & 0.02 \\
DES16X1bax & DES & Ibc & phot & 0.213 & -17.76 & 0.02 & 1.00 & 10.14 & 0.01 & 0.94 & 0.02 & 0.53 & 0.01 \\
DES16X1fmd & DES & Ibc & phot & 0.155 & -16.09 & 0.03 & 2.92 & 9.22 & 0.03 & 0.54 & 0.05 & 0.39 & 0.01 \\
DES16X2eaw & DES & Ibc & phot & 0.151 & -17.52 & 0.05 & 1.00 & 10.74 & 0.02 & 1.44 & 0.03 & 0.77 & 0.01 \\
DES16X3cpc & DES & Ibc & phot & 0.147 & -17.51 & 0.01 & 1.00 & 9.58 & 0.01 & -- & -- & 0.66 & 0.01 \\
DES17C3pt & DES & Ibc & phot & 0.187 & -16.38 & 0.04 & 1.00 & 9.51 & 0.02 & 0.63 & 0.03 & 0.41 & 0.01 \\
DES17X3mr & DES & Ibc & phot & 0.166 & -16.99 & 0.02 & 1.00 & 9.30 & 0.02 & -- & -- & 0.56 & 0.01 \\
SN 1999D & LOSS & II & spec & 0.01 & -16.47 & 0.16 & 1.00 & 10.04 & 0.01 & 0.89 & 0.03 & 0.51 & 0.01 \\
SN 1999an & LOSS & II & spec & 0.005 & -16.09 & 0.32 & 1.00 & 8.83 & 0.01 & 0.90 & 0.02 & 0.47 & 0.02 \\
SN 1999el & LOSS & II & spec & 0.005 & -18.00 & 0.26 & 1.00 & -- & -- & -- & -- & -- & -- \\
SN 1999em & LOSS & II & spec & 0.002 & -16.02 & 0.62 & 1.00 & -- & -- & -- & -- & -- & -- \\
SN 1999go & LOSS & II & spec & 0.013 & -18.32 & 0.18 & 1.00 & -- & -- & -- & -- & -- & -- \\
SN 2000cb & LOSS & II & spec & 0.006 & -16.07 & 0.23 & 1.00 & 9.74 & 0.07 & 1.56 & 0.01 & 0.83 & 0.01 \\
SN 2000dc & LOSS & II & spec & 0.01 & -16.99 & 0.15 & 1.00 & -- & -- & -- & -- & -- & -- \\
SN 2000eo & LOSS & II & spec & 0.01 & -18.16 & 0.24 & 1.00 & -- & -- & -- & -- & -- & -- \\
SN 2001K & LOSS & II & spec & 0.011 & -16.43 & 0.20 & 1.00 & 9.89 & 0.03 & 1.52 & 0.01 & 0.76 & 0.01 \\
SN 2001bq & LOSS & II & spec & 0.008 & -17.11 & 0.22 & 1.00 & -- & -- & -- & -- & -- & -- \\
SN 2001bq & LOSS & II & spec & 0.008 & -17.11 & 0.22 & 1.00 & -- & -- & -- & -- & -- & -- \\
SN 2001cm & LOSS & II & spec & 0.012 & -17.10 & 0.19 & 1.00 & 11.23 & 0.01 & -- & 0.01 & 1.17 & 0.01 \\
SN 2001do & LOSS & II & spec & 0.011 & -17.46 & 0.14 & 1.00 & -- & -- & -- & -- & -- & -- \\
SN 2001hf & LOSS & II & spec & 0.014 & -16.96 & 0.32 & 1.00 & -- & -- & -- & -- & -- & -- \\
SN 2002an & LOSS & II & spec & 0.012 & -17.65 & 0.23 & 1.00 & -- & -- & -- & -- & -- & -- \\
SN 2002ds & LOSS & II & spec & 0.007 & -16.73 & 0.24 & 1.00 & -- & -- & -- & -- & -- & -- \\
SN 2002gw & LOSS & II & spec & 0.009 & -16.25 & 0.21 & 1.00 & -- & -- & -- & -- & -- & -- \\
SN 2003G & LOSS & II & spec & 0.011 & -18.42 & 0.23 & 1.00 & 10.08 & 0.03 & 1.60 & 0.01 & 0.85 & 0.01 \\
SN 2003dv & LOSS & II & spec & 0.008 & -16.45 & 0.19 & 1.00 & 8.91 & 0.01 & 0.87 & 0.02 & 0.47 & 0.02 \\
SN 2003ef & LOSS & II & spec & 0.013 & -16.55 & 0.32 & 1.00 & -- & -- & -- & -- & -- & -- \\
SN 2003hg & LOSS & II & spec & 0.014 & -17.06 & 0.14 & 1.00 & 10.83 & 0.01 & 1.58 & 0.01 & 0.82 & 0.01 \\
SN 2003hl & LOSS & II & spec & 0.008 & -16.42 & 0.18 & 1.00 & 11.11 & 0.00 & 2.14 & 0.01 & 1.06 & 0.01 \\
SN 2003iq & LOSS & II & spec & 0.008 & -17.02 & 0.18 & 1.00 & 11.11 & 0.00 & 2.14 & 0.01 & 1.06 & 0.01 \\
SN 2003ld & LOSS & II & spec & 0.013 & -16.42 & 0.41 & 1.00 & 10.11 & 0.01 & 1.20 & 0.01 & 0.67 & 0.01 \\
SN 2004al & LOSS & II & spec & 0.013 & -16.55 & 0.22 & 1.00 & -- & -- & -- & -- & -- & -- \\
SN 2004ci & LOSS & II & spec & 0.013 & -16.23 & 0.18 & 1.00 & 10.61 & 0.13 & 1.86 & 0.03 & 0.94 & 0.02 \\
SN 2004dd & LOSS & II & spec & 0.013 & -16.16 & 0.18 & 1.00 & 9.96 & 0.03 & 1.29 & 0.02 & 0.71 & 0.02 \\
SN 2004er & LOSS & II & spec & 0.014 & -16.82 & 0.18 & 1.00 & 10.14 & 0.01 & 1.03 & 0.01 & 0.60 & 0.01 \\
SN 2004et & LOSS & II & spec & 0.001 & -16.39 & 1.06 & 1.00 & -- & -- & -- & -- & -- & -- \\
SN 2005H & LOSS & II & spec & 0.012 & -17.31 & 0.23 & 1.00 & 10.12 & 0.02 & 1.02 & 0.03 & 0.59 & 0.02 \\
SN 2005J & LOSS & II & spec & 0.013 & -16.94 & 0.18 & 1.00 & 10.20 & 0.14 & 1.82 & 0.03 & 0.89 & 0.03 \\
SN 2005an & LOSS & II & spec & 0.01 & -16.62 & 0.24 & 1.00 & -- & -- & -- & -- & -- & -- \\
SN 2005aq & LOSS & II & spec & 0.012 & -16.53 & 0.51 & 1.00 & -- & -- & -- & -- & -- & -- \\
SN 2005mg & LOSS & II & spec & 0.013 & -17.07 & 0.32 & 1.00 & 10.70 & 0.03 & 2.16 & 0.02 & 1.07 & 0.01 \\
SN 2006be & LOSS & II & spec & 0.007 & -16.40 & 0.27 & 1.00 & 9.96 & 0.01 & 1.98 & 0.02 & 1.00 & 0.01 \\
SN 2006bp & LOSS & II & spec & 0.004 & -16.10 & 0.33 & 1.00 & 10.69 & 0.01 & 2.32 & 0.01 & 1.13 & 0.01 \\
SN 2006ca & LOSS & II & spec & 0.009 & -17.25 & 0.18 & 1.00 & -- & -- & -- & -- & -- & -- \\
SN 1998dt & LOSS & Ibc & spec & 0.014 & -16.84 & 0.51 & 1.00 & -- & -- & -- & -- & -- & -- \\
SN 1999bu & LOSS & Ibc & spec & 0.009 & -16.22 & 0.52 & 1.00 & 10.68 & 0.03 & 2.02 & 0.01 & 1.01 & 0.01 \\
SN 1999cd & LOSS & Ibc & spec & 0.014 & -16.13 & 0.18 & 1.00 & -- & -- & -- & -- & -- & -- \\
SN 1999dn & LOSS & Ibc & spec & 0.009 & -16.94 & 0.16 & 1.00 & -- & -- & -- & -- & -- & -- \\
SN 2000C & LOSS & Ibc & spec & 0.012 & -17.64 & 0.19 & 1.00 & 10.03 & 0.01 & 0.61 & 0.01 & 0.42 & 0.01 \\
SN 2000H & LOSS & Ibc & spec & 0.012 & -17.18 & 0.23 & 1.00 & -- & -- & -- & -- & -- & -- \\
SN 2000N & LOSS & Ibc & spec & 0.013 & -16.63 & 0.23 & 1.00 & -- & -- & -- & -- & -- & -- \\
SN 2001is & LOSS & Ibc & spec & 0.013 & -16.07 & 0.32 & 1.00 & -- & -- & -- & -- & -- & -- \\
SN 2002J & LOSS & Ibc & spec & 0.012 & -16.31 & 0.15 & 1.00 & 10.80 & 0.02 & 2.02 & 0.01 & 1.01 & 0.01 \\
SN 2002ap & LOSS & Ibc & spec & 0.002 & -17.43 & 0.56 & 1.00 & -- & -- & -- & -- & -- & -- \\
SN 2002jj & LOSS & Ibc & spec & 0.013 & -17.38 & 0.23 & 1.00 & 10.04 & 0.05 & 1.51 & 0.02 & 0.80 & 0.02 \\
SN 2002jz & LOSS & Ibc & spec & 0.005 & -16.20 & 0.33 & 1.00 & -- & -- & -- & -- & -- & -- \\
SN 2003aa & LOSS & Ibc & spec & 0.01 & -16.91 & 0.17 & 1.00 & -- & -- & -- & -- & -- & -- \\
SN 2004al & LOSS & Ibc & spec & 0.013 & -16.85 & 0.22 & 1.00 & -- & -- & -- & -- & -- & -- \\
SN 2004be & LOSS & Ibc & spec & 0.007 & -16.97 & 0.36 & 1.00 & -- & -- & -- & -- & -- & -- \\
SN 2004dk & LOSS & Ibc & spec & 0.005 & -17.23 & 0.25 & 1.00 & 10.23 & 0.00 & 2.25 & 0.00 & 1.10 & 0.01 \\
SN 2004gq & LOSS & Ibc & spec & 0.006 & -16.80 & 0.24 & 1.00 & -- & -- & -- & -- & -- & -- \\
SN 2005U & LOSS & Ibc & spec & 0.011 & -17.76 & 0.24 & 1.00 & 10.28 & 0.01 & 1.29 & 0.01 & 0.71 & 0.01 \\
SN 2005az & LOSS & Ibc & spec & 0.009 & -16.87 & 0.18 & 1.00 & 9.65 & 0.02 & 0.90 & 0.01 & 0.54 & 0.01 \\
SN 2006F & LOSS & Ibc & spec & 0.013 & -16.44 & 0.41 & 1.00 & -- & -- & -- & -- & -- & -- \\
SN 2006T & LOSS & Ibc & spec & 0.007 & -17.34 & 0.24 & 1.00 & -- & -- & -- & -- & -- & -- \\
ZTF18aaaibml & ZTF & II & spec & 0.035 & -17.76 & 0.04 & 1.66 & 9.25 & 0.01 & 0.41 & 0.01 & 0.36 & 0.01 \\
ZTF18aavqdyq & ZTF & II & spec & 0.026 & -16.61 & 0.06 & 7.66 & 8.92 & 0.02 & 0.81 & 0.04 & 0.42 & 0.02 \\
ZTF18aawyjjq & ZTF & II & spec & 0.04 & -17.55 & 0.04 & 2.20 & 10.65 & 0.02 & 1.84 & 0.02 & 0.90 & 0.01 \\
ZTF18abceakp & ZTF & II & spec & 0.03 & -17.50 & 0.11 & 2.34 & 7.52 & 0.07 & 0.63 & 0.15 & 0.36 & 0.08 \\
ZTF18abcezmh & ZTF & II & spec & 0.057 & -18.13 & 0.05 & 1.02 & 9.62 & 0.02 & 0.57 & 0.02 & 0.41 & 0.01 \\
ZTF18abcpmwh & ZTF & II & spec & 0.015 & -17.51 & 0.10 & 2.31 & 10.81 & 0.0 & 1.74 & 0.0 & 0.85 & 0.01 \\
ZTF18abcptmt & ZTF & II & spec & 0.05 & -18.10 & 0.04 & 1.06 & 9.15 & 0.02 & 0.73 & 0.03 & 0.45 & 0.01 \\
ZTF18abltfho & ZTF & II & spec & 0.055 & -18.81 & 0.03 & 1.0 & 10.42 & 0.01 & 0.72 & 0.01 & 0.45 & 0.01 \\
ZTF18abokyfk & ZTF & II & spec & 0.017 & -17.11 & 0.09 & 3.95 & -- & -- & -- & -- & -- & -- \\
ZTF18abqyvzy & ZTF & II & spec & 0.015 & -17.90 & 0.10 & 1.38 & 8.96 & 0.02 & 0.41 & 0.02 & 0.34 & 0.02 \\
ZTF18abrlljc & ZTF & II & spec & 0.05 & -18.57 & 0.03 & 1.0 & -- & -- & -- & -- & -- & -- \\
ZTF18abvmlow & ZTF & II & spec & 0.007 & -16.05 & 0.21 & 16.22 & 8.98 & 0.02 & 1.17 & 0.02 & 0.65 & 0.01 \\
ZTF18abvvmdf & ZTF & II & spec & 0.03 & -17.08 & 0.07 & 4.11 & 10.34 & 0.04 & 1.30 & 0.02 & 0.71 & 0.01 \\
ZTF18abzrgim & ZTF & II & spec & 0.021 & -17.37 & 0.08 & 2.77 & 10.61 & 0.05 & 1.79 & 0.02 & 0.91 & 0.01 \\
ZTF18acebssa & ZTF & II & spec & 0.03 & -18.35 & 0.05 & 1.0 & 10.20 & 0.02 & 1.68 & 0.02 & 0.87 & 0.01 \\
ZTF18acefuhk & ZTF & II & spec & 0.057 & -18.68 & 0.03 & 1.0 & 9.10 & 0.02 & 0.25 & 0.03 & 0.30 & 0.01 \\
ZTF18achtnvk & ZTF & II & spec & 0.04 & -17.73 & 0.04 & 1.72 & 9.97 & 0.02 & 1.58 & 0.04 & 0.82 & 0.03 \\
ZTF18acqwdla & ZTF & II & spec & 0.028 & -18.72 & 0.06 & 1.0 & 8.76 & 0.03 & 0.64 & 0.04 & 0.42 & 0.02 \\
ZTF18acrtvmm & ZTF & II & spec & 0.023 & -17.57 & 0.07 & 2.12 & 10.90 & 0.05 & 2.0 & 0.02 & 0.99 & 0.01 \\
ZTF18acszaiy & ZTF & II & spec & 0.03 & -17.84 & 0.05 & 1.49 & 10.02 & 0.06 & 1.90 & 0.04 & 0.96 & 0.02 \\
ZTF18acuqskr & ZTF & II & spec & 0.045 & -18.35 & 0.05 & 1.0 & 9.86 & 0.02 & 1.42 & 0.02 & 0.75 & 0.01 \\
ZTF18acvwdkk & ZTF & II & spec & 0.023 & -16.11 & 0.07 & 15.01 & -- & -- & -- & -- & -- & -- \\
ZTF18adazblo & ZTF & II & spec & 0.027 & -16.60 & 0.06 & 7.76 & 9.88 & 0.02 & 0.96 & 0.01 & 0.57 & 0.01 \\
ZTF18adbacau & ZTF & II & spec & 0.037 & -17.36 & 0.04 & 2.81 & 10.57 & 0.04 & 1.42 & 0.02 & 0.77 & 0.01 \\
ZTF18adbmrug & ZTF & II & spec & 0.024 & -17.97 & 0.08 & 1.25 & -- & -- & -- & -- & -- & -- \\
ZTF19aadnxog & ZTF & II & spec & 0.02 & -18.13 & 0.08 & 1.02 & 10.03 & 0.02 & 1.23 & 0.02 & 0.66 & 0.02 \\
ZTF19aailepg & ZTF & II & spec & 0.03 & -17.43 & 0.05 & 2.55 & 8.98 & 0.01 & 0.48 & 0.01 & 0.39 & 0.01 \\
ZTF19aakjcxs & ZTF & II & spec & 0.038 & -19.11 & 0.05 & 1.0 & 9.63 & 0.01 & 0.91 & 0.02 & 0.52 & 0.01 \\
ZTF19aaloqmd & ZTF & II & spec & 0.034 & -17.58 & 0.05 & 2.09 & 9.12 & 0.02 & 1.0 & 0.02 & 0.58 & 0.01 \\
ZTF19aamhgwm & ZTF & II & spec & 0.034 & -17.84 & 0.05 & 1.50 & 10.13 & 0.02 & 1.10 & 0.02 & 0.62 & 0.01 \\
ZTF19aamhmsx & ZTF & II & spec & 0.047 & -18.18 & 0.04 & 1.0 & 9.10 & 0.03 & 0.99 & 0.05 & 0.47 & 0.03 \\
ZTF19aamkmxv & ZTF & II & spec & 0.014 & -17.54 & 0.11 & 2.21 & 9.29 & 0.01 & 1.14 & 0.02 & 0.64 & 0.01 \\
ZTF19aamvape & ZTF & II & spec & 0.03 & -17.96 & 0.05 & 1.28 & 7.16 & 0.19 & 1.31 & 0.30 & 0.68 & 0.21 \\
ZTF19aanfnvl & ZTF & II & spec & 0.032 & -17.72 & 0.05 & 1.74 & 11.24 & 0.0 & 2.36 & 0.0 & 1.14 & 0.01 \\
ZTF19aanfqug & ZTF & II & spec & 0.046 & -17.79 & 0.04 & 1.59 & 6.77 & 0.19 & -0.19 & 0.25 & 0.13 & 0.15 \\
ZTF19aanhhal & ZTF & II & spec & 0.026 & -17.15 & 0.06 & 3.72 & 10.96 & 0.04 & 1.87 & 0.01 & 0.95 & 0.01 \\
ZTF19aaniore & ZTF & II & spec & 0.03 & -17.23 & 0.05 & 3.35 & 10.85 & 0.01 & 2.03 & 0.01 & 1.01 & 0.01 \\
ZTF19aanpcep & ZTF & II & spec & 0.031 & -17.74 & 0.05 & 1.71 & 8.99 & 0.02 & 0.94 & 0.05 & 0.46 & 0.02 \\
ZTF19aanqzhm & ZTF & II & spec & 0.049 & -18.33 & 0.05 & 1.0 & -- & -- & -- & -- & -- & -- \\
ZTF19aanrrqu & ZTF & II & spec & 0.024 & -18.01 & 0.06 & 1.19 & 8.80 & 0.02 & 0.11 & 0.02 & 0.25 & 0.03 \\
ZTF19aapafit & ZTF & II & spec & 0.019 & -16.92 & 0.08 & 5.05 & 9.23 & 0.02 & 0.90 & 0.05 & 0.50 & 0.02 \\
ZTF19aapafqd & ZTF & II & spec & 0.032 & -17.56 & 0.05 & 2.17 & 10.73 & 0.02 & 1.90 & 0.01 & 0.93 & 0.01 \\
ZTF19aapbfot & ZTF & II & spec & 0.03 & -16.97 & 0.05 & 4.75 & 8.59 & 0.03 & 0.74 & 0.07 & 0.40 & 0.02 \\
ZTF19aapzbjr & ZTF & II & spec & 0.03 & -17.77 & 0.05 & 1.64 & 9.78 & 0.02 & 0.98 & 0.02 & 0.57 & 0.02 \\
ZTF19aaqdkrm & ZTF & II & spec & 0.034 & -17.83 & 0.05 & 1.52 & 10.13 & 0.02 & 1.10 & 0.02 & 0.62 & 0.01 \\
ZTF19aaqxosb & ZTF & II & spec & 0.019 & -16.56 & 0.08 & 8.23 & 10.48 & 0.05 & 1.47 & 0.01 & 0.79 & 0.01 \\
ZTF19aathllr & ZTF & II & spec & 0.055 & -18.77 & 0.04 & 1.0 & -- & -- & -- & -- & -- & -- \\
ZTF19aatqzim & ZTF & II & spec & 0.05 & -18.22 & 0.04 & 1.0 & 10.65 & 0.01 & 1.72 & 0.02 & 0.89 & 0.01 \\
ZTF19aaugaam & ZTF & II & spec & 0.05 & -18.84 & 0.03 & 1.0 & 10.13 & 0.01 & 1.06 & 0.01 & 0.61 & 0.01 \\
ZTF19aauisdr & ZTF & II & spec & 0.043 & -17.82 & 0.05 & 1.53 & 9.38 & 0.02 & 0.66 & 0.03 & 0.43 & 0.01 \\
ZTF19aauishy & ZTF & II & spec & 0.023 & -16.40 & 0.07 & 10.21 & 9.17 & 0.02 & 0.73 & 0.04 & 0.46 & 0.02 \\
ZTF19aauxxgk & ZTF & II & spec & 0.026 & -18.01 & 0.06 & 1.19 & -- & -- & -- & -- & -- & -- \\
ZTF19aavbjfp & ZTF & II & spec & 0.028 & -17.06 & 0.05 & 4.20 & 8.60 & 0.03 & 0.69 & 0.05 & 0.45 & 0.02 \\
ZTF19aavbkly & ZTF & II & spec & 0.041 & -17.40 & 0.04 & 2.67 & 10.37 & 0.01 & 1.24 & 0.02 & 0.65 & 0.02 \\
ZTF19aavhblr & ZTF & II & spec & 0.05 & -18.37 & 0.06 & 1.0 & 9.0 & 0.02 & 0.93 & 0.04 & 0.53 & 0.02 \\
ZTF19aavjukt & ZTF & II & spec & 0.036 & -18.43 & 0.05 & 1.0 & 9.10 & 0.01 & 0.29 & 0.01 & 0.34 & 0.01 \\
ZTF19aavkptg & ZTF & II & spec & 0.038 & -17.26 & 0.04 & 3.20 & 9.73 & 0.05 & 1.74 & 0.06 & 0.87 & 0.01 \\
ZTF19aawgxdn & ZTF & II & spec & 0.031 & -17.29 & 0.05 & 3.07 & 10.82 & 0.02 & 1.65 & 0.02 & 0.82 & 0.02 \\
ZTF19aazfvhh & ZTF & II & spec & 0.034 & -17.20 & 0.05 & 3.50 & -- & -- & -- & -- & -- & -- \\
ZTF19aazudta & ZTF & II & spec & 0.024 & -16.88 & 0.07 & 5.35 & -- & -- & -- & -- & -- & -- \\
ZTF19aazyvub & ZTF & II & spec & 0.023 & -17.09 & 0.07 & 4.04 & 9.37 & 0.02 & 1.25 & 0.02 & 0.62 & 0.02 \\
ZTF19abacxod & ZTF & II & spec & 0.018 & -16.85 & 0.09 & 5.57 & 9.81 & 0.01 & 0.74 & 0.02 & 0.46 & 0.01 \\
ZTF19abajxet & ZTF & II & spec & 0.015 & -17.68 & 0.10 & 1.84 & 10.25 & 0.02 & 1.05 & 0.01 & 0.61 & 0.01 \\
ZTF19abbnamr & ZTF & II & spec & 0.014 & -17.30 & 0.10 & 3.06 & 10.88 & 0.02 & 1.63 & 0.01 & 0.85 & 0.01 \\
ZTF19abbwfgp & ZTF & II & spec & 0.026 & -18.10 & 0.06 & 1.06 & 10.71 & 0.03 & 1.82 & 0.01 & 0.90 & 0.01 \\
ZTF19abbxykm & ZTF & II & spec & 0.047 & -18.96 & 0.03 & 1.0 & 9.30 & 0.02 & 1.15 & 0.03 & 0.64 & 0.02 \\
ZTF19abcneik & ZTF & II & spec & 0.035 & -17.35 & 0.06 & 2.85 & 9.25 & 0.03 & 1.10 & 0.05 & 0.60 & 0.02 \\
ZTF19abctxhf & ZTF & II & spec & 0.05 & -18.16 & 0.04 & 1.0 & 9.76 & 0.02 & 1.03 & 0.02 & 0.59 & 0.01 \\
ZTF19abddsvk & ZTF & II & spec & 0.058 & -18.54 & 0.05 & 1.0 & 9.93 & 0.02 & 0.86 & 0.04 & 0.51 & 0.02 \\
ZTF19abdviwl & ZTF & II & spec & 0.03 & -17.50 & 0.05 & 2.35 & 7.30 & 0.14 & 0.44 & 0.20 & 0.35 & 0.07 \\
ZTF19abecaca & ZTF & II & spec & 0.032 & -19.45 & 0.05 & 1.0 & 9.68 & 0.01 & 0.78 & 0.02 & 0.49 & 0.01 \\
ZTF19abegizf & ZTF & II & spec & 0.037 & -17.48 & 0.05 & 2.39 & -- & -- & -- & -- & -- & -- \\
ZTF19abfloxk & ZTF & II & spec & 0.016 & -16.84 & 0.11 & 5.61 & -- & -- & -- & -- & -- & -- \\
ZTF19abgndlf & ZTF & II & spec & 0.03 & -17.65 & 0.05 & 1.91 & -- & -- & -- & -- & -- & -- \\
ZTF19abgrmfu & ZTF & II & spec & 0.035 & -18.46 & 0.05 & 1.0 & 8.73 & 0.02 & 0.94 & 0.03 & 0.56 & 0.02 \\
ZTF19abiqfxi & ZTF & II & spec & 0.054 & -19.44 & 0.03 & 1.0 & 9.48 & 0.02 & 0.68 & 0.04 & 0.43 & 0.01 \\
ZTF19abiszoe & ZTF & II & spec & 0.043 & -18.22 & 0.10 & 1.0 & 10.72 & 0.02 & 1.90 & 0.01 & 0.93 & 0.01 \\
ZTF19abjpntj & ZTF & II & spec & 0.055 & -18.14 & 0.03 & 1.01 & 10.67 & 0.02 & 1.23 & 0.02 & 0.67 & 0.02 \\
ZTF19abjsmmv & ZTF & II & spec & 0.02 & -17.19 & 0.08 & 3.53 & 10.47 & 0.08 & 1.75 & 0.02 & 0.90 & 0.02 \\
ZTF19abkfqqp & ZTF & II & spec & 0.03 & -19.32 & 0.05 & 1.0 & 11.25 & 0.02 & 1.42 & 0.02 & 0.75 & 0.01 \\
ZTF19ablfdwt & ZTF & II & spec & 0.026 & -17.31 & 0.06 & 3.03 & 9.15 & 0.03 & 1.04 & 0.04 & 0.58 & 0.02 \\
ZTF19ablojrw & ZTF & II & spec & 0.049 & -18.53 & 0.05 & 1.0 & -- & -- & -- & -- & -- & -- \\
ZTF19abovstj & ZTF & II & spec & 0.045 & -18.22 & 0.06 & 1.0 & 9.49 & 0.01 & 0.90 & 0.02 & 0.53 & 0.02 \\
ZTF19abpidqn & ZTF & II & spec & 0.015 & -16.29 & 0.10 & 11.72 & -- & -- & -- & -- & -- & -- \\
ZTF19abpxiff & ZTF & II & spec & 0.055 & -18.19 & 0.05 & 1.0 & -- & -- & -- & -- & -- & -- \\
ZTF19abpyqog & ZTF & II & spec & 0.031 & -17.12 & 0.05 & 3.85 & -- & -- & -- & -- & -- & -- \\
ZTF19abqgtqo & ZTF & II & spec & 0.036 & -17.44 & 0.04 & 2.54 & 9.08 & 0.03 & 0.68 & 0.05 & 0.44 & 0.02 \\
ZTF19abqhobb & ZTF & II & spec & 0.018 & -17.49 & 0.08 & 2.38 & 9.07 & 0.02 & 0.71 & 0.04 & 0.45 & 0.01 \\
ZTF19abqrhvt & ZTF & II & spec & 0.021 & -18.09 & 0.07 & 1.08 & 9.29 & 0.01 & 0.87 & 0.03 & 0.42 & 0.02 \\
ZTF19abqrhvy & ZTF & II & spec & 0.032 & -18.13 & 0.05 & 1.02 & 11.19 & 0.01 & 1.58 & 0.01 & 0.82 & 0.01 \\
ZTF19abrbmvt & ZTF & II & spec & 0.039 & -17.81 & 0.04 & 1.56 & 11.09 & 0.08 & 2.23 & 0.03 & 1.09 & 0.01 \\
ZTF19abueupg & ZTF & II & spec & 0.02 & -16.05 & 0.07 & 16.29 & 9.19 & 0.06 & 1.25 & 0.04 & 0.70 & 0.02 \\
ZTF19abukakm & ZTF & II & spec & 0.057 & -18.65 & 0.03 & 1.0 & -- & -- & -- & -- & -- & -- \\
ZTF19abuzinv & ZTF & II & spec & 0.02 & -16.45 & 0.07 & 9.46 & -- & -- & -- & -- & -- & -- \\
ZTF19abwsagv & ZTF & II & spec & 0.038 & -18.18 & 0.04 & 1.0 & -- & -- & -- & -- & -- & -- \\
ZTF19abyuzch & ZTF & II & spec & 0.024 & -18.20 & 0.07 & 1.0 & -- & -- & -- & -- & -- & -- \\
ZTF19abzqwpr & ZTF & II & spec & 0.037 & -18.24 & 0.08 & 1.0 & 9.36 & 0.02 & 0.85 & 0.03 & 0.50 & 0.01 \\
ZTF19acanzwg & ZTF & II & spec & 0.047 & -18.29 & 0.04 & 1.0 & -- & -- & -- & -- & -- & -- \\
ZTF19acbhvgi & ZTF & II & spec & 0.031 & -17.36 & 0.05 & 2.81 & 7.79 & 0.07 & 1.06 & 0.13 & 0.60 & 0.08 \\
ZTF19acbmxky & ZTF & II & spec & 0.048 & -18.34 & 0.04 & 1.0 & 9.78 & 0.02 & 0.87 & 0.03 & 0.52 & 0.01 \\
ZTF19acbrzzr & ZTF & II & spec & 0.03 & -17.17 & 0.05 & 3.61 & 7.40 & 0.06 & 0.16 & 0.10 & 0.20 & 0.04 \\
ZTF19acbwejj & ZTF & II & spec & 0.014 & -16.80 & 0.11 & 5.95 & 8.63 & 0.02 & 0.39 & 0.02 & 0.37 & 0.01 \\
ZTF19acbwouf & ZTF & II & spec & 0.03 & -17.03 & 0.05 & 4.39 & 7.14 & 0.22 & -0.04 & 0.31 & 0.16 & 0.15 \\
ZTF19accbeju & ZTF & II & spec & 0.055 & -18.61 & 0.05 & 1.0 & 10.58 & 0.01 & 1.61 & 0.02 & 0.84 & 0.01 \\
ZTF19acchaza & ZTF & II & spec & 0.035 & -17.79 & 0.05 & 1.60 & 8.16 & 0.04 & 0.78 & 0.06 & 0.48 & 0.03 \\
ZTF19acctwpz & ZTF & II & spec & 0.017 & -16.39 & 0.09 & 10.29 & 7.75 & 0.10 & 0.96 & 0.18 & 0.56 & 0.08 \\
ZTF19aceshib & ZTF & II & spec & 0.048 & -18.67 & 0.04 & 1.0 & 9.75 & 0.01 & 0.96 & 0.02 & 0.56 & 0.02 \\
ZTF19acfejbj & ZTF & II & spec & 0.011 & -17.21 & 0.14 & 3.44 & 9.72 & 0.01 & 0.92 & 0.01 & 0.55 & 0.01 \\
ZTF19acftude & ZTF & II & spec & 0.04 & -17.98 & 0.04 & 1.24 & 9.89 & 0.01 & 1.12 & 0.02 & 0.62 & 0.01 \\
ZTF19acgbkzr & ZTF & II & spec & 0.026 & -16.64 & 0.06 & 7.34 & 10.16 & 0.02 & 1.20 & 0.02 & 0.68 & 0.01 \\
ZTF19achjqbk & ZTF & II & spec & 0.05 & -17.93 & 0.04 & 1.32 & -- & -- & -- & -- & -- & -- \\
ZTF19acignlo & ZTF & II & spec & 0.048 & -18.87 & 0.03 & 1.0 & 9.85 & 0.01 & 0.98 & 0.03 & 0.55 & 0.01 \\
ZTF19acjwdnu & ZTF & II & spec & 0.053 & -18.76 & 0.03 & 1.0 & 7.75 & 0.17 & 0.85 & 0.30 & 0.47 & 0.15 \\
ZTF19aclobbu & ZTF & II & spec & 0.012 & -16.68 & 0.12 & 7.01 & 8.29 & 0.02 & 0.98 & 0.02 & 0.57 & 0.01 \\
ZTF19acrcxri & ZTF & II & spec & 0.027 & -18.55 & 0.06 & 1.0 & 8.76 & 0.01 & 0.24 & 0.01 & 0.32 & 0.01 \\
ZTF19acryurj & ZTF & II & spec & 0.022 & -18.08 & 0.07 & 1.08 & 10.53 & 0.04 & 1.71 & 0.02 & 0.87 & 0.02 \\
ZTF19acwrrvg & ZTF & II & spec & 0.027 & -18.59 & 0.06 & 1.0 & -- & -- & -- & -- & -- & -- \\
ZTF19acxowrr & ZTF & II & spec & 0.051 & -18.54 & 0.03 & 1.0 & 10.94 & 0.02 & 1.65 & 0.02 & 0.85 & 0.01 \\
ZTF19acyjviz & ZTF & II & spec & 0.022 & -17.34 & 0.07 & 2.88 & 9.40 & 0.02 & 1.01 & 0.01 & 0.59 & 0.01 \\
ZTF19aczlldp & ZTF & II & spec & 0.028 & -18.0 & 0.06 & 1.21 & 7.96 & 0.05 & 0.15 & 0.05 & 0.25 & 0.02 \\
ZTF19adannbl & ZTF & II & spec & 0.048 & -18.93 & 0.05 & 1.0 & -- & -- & -- & -- & -- & -- \\
ZTF19adavzew & ZTF & II & spec & 0.04 & -17.85 & 0.04 & 1.47 & 9.16 & 0.01 & 1.03 & 0.02 & 0.57 & 0.01 \\
ZTF19adccrca & ZTF & II & spec & 0.045 & -17.82 & 0.03 & 1.53 & 9.20 & 0.01 & 0.99 & 0.02 & 0.56 & 0.01 \\
ZTF20aabconi & ZTF & II & spec & 0.043 & -18.07 & 0.04 & 1.11 & -- & -- & -- & -- & -- & -- \\
ZTF20aabqiav & ZTF & II & spec & 0.05 & -18.82 & 0.04 & 1.0 & 8.92 & 0.03 & 0.88 & 0.04 & 0.52 & 0.02 \\
ZTF20aacbyec & ZTF & II & spec & 0.036 & -19.13 & 0.04 & 1.0 & 10.93 & 0.01 & 1.31 & 0.02 & 0.70 & 0.01 \\
ZTF20aadchdd & ZTF & II & spec & 0.04 & -17.92 & 0.04 & 1.34 & 6.50 & 0.09 & -0.63 & 0.08 & 0.03 & 0.09 \\
ZTF20aaeoqqd & ZTF & II & spec & 0.05 & -18.62 & 0.05 & 1.0 & 10.73 & 0.06 & 1.88 & 0.03 & 0.94 & 0.02 \\
ZTF20aaetrle & ZTF & II & spec & 0.02 & -17.34 & 0.07 & 2.91 & -- & -- & -- & -- & -- & -- \\
ZTF20aafckit & ZTF & II & spec & 0.031 & -17.88 & 0.05 & 1.41 & 9.01 & 0.02 & 0.52 & 0.02 & 0.39 & 0.01 \\
ZTF20aahbamv & ZTF & II & spec & 0.045 & -19.15 & 0.03 & 1.0 & 9.62 & 0.02 & 1.21 & 0.02 & 0.67 & 0.01 \\
ZTF20aahqbsr & ZTF & II & spec & 0.022 & -16.44 & 0.07 & 9.59 & 10.77 & 0.03 & 1.54 & 0.02 & 0.81 & 0.02 \\
ZTF20aaieyup & ZTF & II & spec & 0.012 & -16.91 & 0.12 & 5.13 & -- & -- & -- & -- & -- & -- \\
ZTF20aamamnp & ZTF & II & spec & 0.05 & -18.70 & 0.04 & 1.0 & 8.04 & 0.05 & 0.08 & 0.05 & 0.28 & 0.03 \\
ZTF20aamazzl & ZTF & II & spec & 0.056 & -18.05 & 0.03 & 1.14 & 9.73 & 0.01 & 0.88 & 0.02 & 0.51 & 0.01 \\
ZTF20aamoaim & ZTF & II & spec & 0.048 & -18.06 & 0.09 & 1.12 & 8.31 & 0.06 & 0.54 & 0.13 & 0.33 & 0.05 \\
ZTF20aamxuwl & ZTF & II & spec & 0.037 & -17.99 & 0.06 & 1.23 & 10.43 & 0.05 & 1.74 & 0.03 & 0.89 & 0.02 \\
ZTF20aaoldej & ZTF & II & spec & 0.026 & -16.91 & 0.06 & 5.12 & 9.57 & 0.02 & 0.73 & 0.03 & 0.46 & 0.01 \\
ZTF20aaophpu & ZTF & II & spec & 0.017 & -16.01 & 0.16 & 17.11 & -- & -- & -- & -- & -- & -- \\
ZTF20aatqesi & ZTF & II & spec & 0.041 & -19.03 & 0.04 & 1.0 & -- & -- & -- & -- & -- & -- \\
ZTF20aauqhka & ZTF & II & spec & 0.039 & -18.29 & 0.04 & 1.0 & 9.64 & 0.02 & 0.87 & 0.04 & 0.52 & 0.02 \\
ZTF20aaurfhs & ZTF & II & spec & 0.035 & -17.96 & 0.05 & 1.27 & 10.28 & 0.02 & 2.05 & 0.02 & 1.02 & 0.01 \\
ZTF20aaurjbj & ZTF & II & spec & 0.043 & -18.71 & 0.04 & 1.0 & 8.88 & 0.01 & -0.03 & 0.02 & 0.21 & 0.02 \\
ZTF20aavhixe & ZTF & II & spec & 0.051 & -18.98 & 0.04 & 1.0 & 10.07 & 0.01 & 0.96 & 0.02 & 0.54 & 0.02 \\
ZTF20aavptjf & ZTF & II & spec & 0.03 & -17.58 & 0.05 & 2.10 & 8.54 & 0.02 & -0.04 & 0.02 & 0.20 & 0.03 \\
ZTF20aavvaup & ZTF & II & spec & 0.033 & -18.08 & 0.05 & 1.10 & 8.88 & 0.01 & 0.52 & 0.02 & 0.36 & 0.02 \\
ZTF20aawgrcu & ZTF & II & spec & 0.042 & -18.36 & 0.04 & 1.0 & 9.41 & 0.02 & 1.48 & 0.06 & 0.73 & 0.02 \\
ZTF20aawijco & ZTF & II & spec & 0.025 & -17.26 & 0.06 & 3.20 & -- & -- & -- & -- & -- & -- \\
ZTF20aawjbsf & ZTF & II & spec & 0.022 & -16.97 & 0.08 & 4.73 & 8.70 & 0.02 & 0.90 & 0.02 & 0.54 & 0.01 \\
ZTF20aaxunbm & ZTF & II & spec & 0.055 & -18.28 & 0.03 & 1.0 & 10.04 & 0.01 & 1.45 & 0.04 & 0.73 & 0.01 \\
ZTF20aaynrrh & ZTF & II & spec & 0.005 & -17.32 & 0.29 & 2.97 & -- & -- & -- & -- & -- & -- \\
ZTF20aazrxef & ZTF & II & spec & 0.033 & -17.91 & 0.05 & 1.36 & 11.47 & 0.01 & 1.63 & 0.01 & 0.85 & 0.01 \\
ZTF20aazswwk & ZTF & II & spec & 0.032 & -16.89 & 0.05 & 5.24 & -- & -- & -- & -- & -- & -- \\
ZTF20aazycgy & ZTF & II & spec & 0.03 & -18.10 & 0.06 & 1.06 & 10.31 & 0.02 & 1.39 & 0.01 & 0.75 & 0.01 \\
ZTF20abbpkpa & ZTF & II & spec & 0.033 & -16.98 & 0.05 & 4.66 & 9.20 & 0.01 & 1.08 & 0.02 & 0.60 & 0.02 \\
ZTF20abccixp & ZTF & II & spec & 0.044 & -18.78 & 0.04 & 1.0 & 8.94 & 0.04 & 0.66 & 0.05 & 0.43 & 0.02 \\
ZTF20abcgkom & ZTF & II & spec & 0.055 & -18.49 & 0.04 & 1.0 & 9.72 & 0.01 & 0.45 & 0.01 & 0.38 & 0.01 \\
ZTF20abekbzp & ZTF & II & spec & 0.04 & -17.43 & 0.04 & 2.56 & 9.24 & 0.02 & 0.75 & 0.02 & 0.46 & 0.01 \\
ZTF20abekcdt & ZTF & II & spec & 0.049 & -17.88 & 0.04 & 1.42 & 10.03 & 0.01 & 1.16 & 0.02 & 0.64 & 0.01 \\
ZTF20abjaapj & ZTF & II & spec & 0.03 & -17.80 & 0.05 & 1.57 & 7.92 & 0.12 & 1.71 & 0.16 & 0.88 & 0.11 \\
ZTF20abjatqy & ZTF & II & spec & 0.026 & -17.75 & 0.06 & 1.68 & -- & -- & -- & -- & -- & -- \\
ZTF20abjonjs & ZTF & II & spec & 0.016 & -16.82 & 0.10 & 5.78 & -- & -- & -- & -- & -- & -- \\
ZTF20abjuxoy & ZTF & II & spec & 0.027 & -17.19 & 0.06 & 3.53 & -- & -- & -- & -- & -- & -- \\
ZTF20abliiex & ZTF & II & spec & 0.039 & -18.62 & 0.07 & 1.0 & -- & -- & -- & -- & -- & -- \\
ZTF20ablklei & ZTF & II & spec & 0.025 & -17.74 & 0.06 & 1.71 & 10.12 & 0.02 & 1.45 & 0.02 & 0.77 & 0.01 \\
ZTF20ablygyy & ZTF & II & spec & 0.017 & -18.11 & 0.09 & 1.05 & 10.07 & 0.03 & 1.60 & 0.02 & 0.79 & 0.02 \\
ZTF20abonvte & ZTF & II & spec & 0.053 & -18.80 & 0.04 & 1.0 & 9.21 & 0.03 & 0.62 & 0.04 & 0.42 & 0.01 \\
ZTF20abwdaeo & ZTF & II & spec & 0.021 & -17.15 & 0.07 & 3.71 & 9.62 & 0.01 & 1.37 & 0.02 & 0.75 & 0.01 \\
ZTF20abxmwwd & ZTF & II & spec & 0.03 & -16.95 & 0.06 & 4.88 & -- & -- & -- & -- & -- & -- \\
ZTF20abyylgi & ZTF & II & spec & 0.029 & -18.44 & 0.05 & 1.0 & -- & -- & -- & -- & -- & -- \\
ZTF20abyzomt & ZTF & II & spec & 0.022 & -17.69 & 0.07 & 1.82 & 9.48 & 0.01 & 0.41 & 0.01 & 0.36 & 0.01 \\
ZTF20abyzprl & ZTF & II & spec & 0.053 & -18.50 & 0.03 & 1.0 & 10.70 & 0.03 & 1.66 & 0.02 & 0.86 & 0.01 \\
ZTF20accrldu & ZTF & II & spec & 0.038 & -17.56 & 0.04 & 2.16 & -- & -- & -- & -- & -- & -- \\
ZTF20acedqis & ZTF & II & spec & 0.05 & -18.84 & 0.04 & 1.0 & 9.82 & 0.01 & 1.41 & 0.02 & 0.75 & 0.01 \\
ZTF19abudlps* & ZTF & II & spec & 0.013 & -15.87 & 0.11 & 20.76 & -- & -- & -- & -- & -- & -- \\
ZTF20aapycrh* & ZTF & II & spec & 0.015 & -15.60 & 0.10 & 29.76 & -- & -- & -- & -- & -- & -- \\
ZTF18aakkrjm & ZTF & Ibc & spec & 0.021 & -17.01 & 0.07 & 4.48 & 9.44 & 0.01 & 1.30 & 0.01 & 0.71 & 0.01 \\
ZTF18aaxiuyp & ZTF & Ibc & spec & 0.03 & -17.17 & 0.05 & 3.61 & 10.08 & 0.01 & 1.17 & 0.01 & 0.65 & 0.01 \\
ZTF18abdkkwa & ZTF & Ibc & spec & 0.025 & -17.24 & 0.07 & 3.29 & 10.09 & 0.03 & 1.28 & 0.02 & 0.71 & 0.01 \\
ZTF18abfzfcv & ZTF & Ibc & spec & 0.038 & -17.51 & 0.05 & 2.31 & 10.16 & 0.01 & 1.02 & 0.02 & 0.57 & 0.02 \\
ZTF18abfzhct & ZTF & Ibc & spec & 0.04 & -18.43 & 0.04 & 1.0 & 9.93 & 0.03 & 1.11 & 0.03 & 0.63 & 0.01 \\
ZTF18abojpnr & ZTF & Ibc & spec & 0.038 & -17.58 & 0.06 & 2.11 & -- & -- & -- & -- & -- & -- \\
ZTF18acbzvpg & ZTF & Ibc & spec & 0.026 & -16.78 & 0.07 & 6.08 & 10.54 & 0.04 & 1.42 & 0.02 & 0.76 & 0.01 \\
ZTF18achcpwu & ZTF & Ibc & spec & 0.055 & -18.45 & 0.04 & 1.0 & -- & -- & -- & -- & -- & -- \\
ZTF18acnncve & ZTF & Ibc & spec & 0.044 & -18.02 & 0.08 & 1.18 & 8.74 & 0.03 & 0.86 & 0.05 & 0.52 & 0.03 \\
ZTF18aczqzrj & ZTF & Ibc & spec & 0.043 & -18.68 & 0.04 & 1.0 & 9.51 & 0.01 & 0.97 & 0.02 & 0.56 & 0.02 \\
ZTF18adasisj & ZTF & Ibc & spec & 0.034 & -18.49 & 0.05 & 1.0 & -- & -- & -- & -- & -- & -- \\
ZTF19aaejtof & ZTF & Ibc & spec & 0.038 & -17.44 & 0.08 & 2.54 & 9.44 & 0.02 & 1.36 & 0.05 & 0.70 & 0.02 \\
ZTF19aafmyow & ZTF & Ibc & spec & 0.026 & -18.37 & 0.06 & 1.0 & 8.06 & 0.04 & 0.54 & 0.06 & 0.36 & 0.02 \\
ZTF19aailcgs & ZTF & Ibc & spec & 0.02 & -16.46 & 0.08 & 9.35 & 10.94 & 0.17 & 1.96 & 0.04 & 0.98 & 0.02 \\
ZTF19aailsge & ZTF & Ibc & spec & 0.036 & -17.34 & 0.06 & 2.91 & 10.63 & 0.02 & 1.26 & 0.02 & 0.70 & 0.01 \\
ZTF19aakirwj & ZTF & Ibc & spec & 0.032 & -17.14 & 0.05 & 3.77 & 10.34 & 0.03 & 1.35 & 0.02 & 0.73 & 0.02 \\
ZTF19aaknate & ZTF & Ibc & spec & 0.012 & -16.53 & 0.13 & 8.49 & -- & -- & -- & -- & -- & -- \\
ZTF19aakpcuw & ZTF & Ibc & spec & 0.032 & -17.77 & 0.06 & 1.64 & 9.79 & 0.01 & 0.55 & 0.01 & 0.40 & 0.01 \\
ZTF19aalouag & ZTF & Ibc & spec & 0.055 & -18.83 & 0.03 & 1.0 & 10.28 & 0.01 & 1.22 & 0.02 & 0.66 & 0.01 \\
ZTF19aamgghn & ZTF & Ibc & spec & 0.03 & -17.91 & 0.06 & 1.35 & 9.09 & 0.05 & 1.27 & 0.05 & 0.71 & 0.02 \\
ZTF19aamsetj & ZTF & Ibc & spec & 0.028 & -18.12 & 0.06 & 1.04 & 10.39 & 0.02 & 1.51 & 0.02 & 0.74 & 0.02 \\
ZTF19aanfukh & ZTF & Ibc & spec & 0.028 & -17.34 & 0.07 & 2.90 & 10.85 & 0.01 & 1.54 & 0.0 & 0.75 & 0.01 \\
ZTF19aanijpu & ZTF & Ibc & spec & 0.05 & -19.33 & 0.04 & 1.0 & 8.86 & 0.04 & 0.57 & 0.08 & 0.38 & 0.02 \\
ZTF19aaoxvfe & ZTF & Ibc & spec & 0.036 & -17.24 & 0.05 & 3.30 & 9.18 & 0.02 & 1.04 & 0.04 & 0.54 & 0.02 \\
ZTF19aapadxs & ZTF & Ibc & spec & 0.035 & -17.69 & 0.05 & 1.82 & -- & -- & -- & -- & -- & -- \\
ZTF19aaugupw & ZTF & Ibc & spec & 0.041 & -17.52 & 0.07 & 2.26 & 9.17 & 0.04 & 0.64 & 0.06 & 0.43 & 0.02 \\
ZTF19aavoweu & ZTF & Ibc & spec & 0.036 & -17.66 & 0.05 & 1.88 & 10.65 & 0.01 & 1.26 & 0.01 & 0.69 & 0.01 \\
ZTF19aawqcgy & ZTF & Ibc & spec & 0.021 & -16.99 & 0.07 & 4.59 & -- & -- & -- & -- & -- & -- \\
ZTF19aaxfcpq & ZTF & Ibc & spec & 0.038 & -18.41 & 0.05 & 1.0 & 9.84 & 0.01 & 1.21 & 0.02 & 0.67 & 0.01 \\
ZTF19aaxzdtw & ZTF & Ibc & spec & 0.041 & -17.43 & 0.05 & 2.55 & 8.78 & 0.02 & -0.04 & 0.02 & 0.20 & 0.02 \\
ZTF19abafmwj & ZTF & Ibc & spec & 0.034 & -16.95 & 0.05 & 4.89 & 10.60 & 0.01 & 1.76 & 0.03 & 0.86 & 0.02 \\
ZTF19abamqxo & ZTF & Ibc & spec & 0.057 & -19.10 & 0.05 & 1.0 & 9.18 & 0.02 & 0.74 & 0.03 & 0.46 & 0.01 \\
ZTF19abdoior & ZTF & Ibc & spec & 0.047 & -18.27 & 0.05 & 1.0 & 10.24 & 0.01 & 1.47 & 0.02 & 0.77 & 0.01 \\
ZTF19abfsxpw & ZTF & Ibc & spec & 0.029 & -18.08 & 0.07 & 1.08 & 9.48 & 0.02 & 0.93 & 0.02 & 0.55 & 0.01 \\
ZTF19abgfuhh & ZTF & Ibc & spec & 0.035 & -17.60 & 0.05 & 2.06 & 8.17 & 0.03 & 0.78 & 0.10 & 0.44 & 0.04 \\
ZTF19ablesob & ZTF & Ibc & spec & 0.056 & -19.09 & 0.04 & 1.0 & -- & -- & -- & -- & -- & -- \\
ZTF19abqmsnk & ZTF & Ibc & spec & 0.036 & -17.65 & 0.07 & 1.91 & 9.34 & 0.02 & 0.89 & 0.03 & 0.50 & 0.02 \\
ZTF19abqqrgy & ZTF & Ibc & spec & 0.03 & -17.59 & 0.06 & 2.07 & 9.43 & 0.02 & 1.12 & 0.03 & 0.61 & 0.02 \\
ZTF19abqshry & ZTF & Ibc & spec & 0.031 & -17.12 & 0.06 & 3.89 & 10.04 & 0.01 & 1.07 & 0.01 & 0.61 & 0.01 \\
ZTF19abqwtfu & ZTF & Ibc & spec & 0.014 & -18.14 & 0.10 & 1.01 & 9.41 & 0.04 & 1.86 & 0.02 & 0.94 & 0.02 \\
ZTF19abtsnyy & ZTF & Ibc & spec & 0.04 & -18.28 & 0.04 & 1.0 & -- & -- & -- & -- & -- & -- \\
ZTF19abupned & ZTF & Ibc & spec & 0.05 & -19.08 & 0.03 & 1.0 & 9.07 & 0.02 & 0.74 & 0.04 & 0.46 & 0.01 \\
ZTF19abvdgqo & ZTF & Ibc & spec & 0.037 & -18.37 & 0.04 & 1.0 & -- & -- & -- & -- & -- & -- \\
ZTF19abxjrge & ZTF & Ibc & spec & 0.022 & -17.39 & 0.08 & 2.72 & -- & -- & -- & -- & -- & -- \\
ZTF19abxtcio & ZTF & Ibc & spec & 0.016 & -16.14 & 0.10 & 14.46 & 8.57 & 0.03 & 1.13 & 0.03 & 0.65 & 0.02 \\
ZTF19abztknu & ZTF & Ibc & spec & 0.054 & -19.05 & 0.03 & 1.0 & -- & -- & -- & -- & -- & -- \\
ZTF19acbmojx & ZTF & Ibc & spec & 0.027 & -16.69 & 0.07 & 6.93 & 9.50 & 0.02 & 1.17 & 0.02 & 0.66 & 0.01 \\
ZTF19ackjene & ZTF & Ibc & spec & 0.044 & -19.12 & 0.05 & 1.0 & -- & -- & -- & -- & -- & -- \\
ZTF19ackjjwf & ZTF & Ibc & spec & 0.016 & -16.86 & 0.09 & 5.50 & 10.26 & 0.04 & 1.27 & 0.02 & 0.71 & 0.01 \\
ZTF19acmbekd & ZTF & Ibc & spec & 0.05 & -18.39 & 0.08 & 1.0 & 8.81 & 0.06 & 1.18 & 0.10 & 0.64 & 0.04 \\
ZTF19acmelor & ZTF & Ibc & spec & 0.027 & -17.40 & 0.06 & 2.68 & 10.32 & 0.02 & 1.25 & 0.01 & 0.69 & 0.01 \\
ZTF19acxxwvi & ZTF & Ibc & spec & 0.011 & -16.59 & 0.13 & 7.90 & 9.75 & 0.02 & 1.01 & 0.02 & 0.58 & 0.01 \\
ZTF19acyogrm & ZTF & Ibc & spec & 0.02 & -17.40 & 0.08 & 2.68 & 9.82 & 0.16 & 1.92 & 0.03 & 0.96 & 0.03 \\
ZTF19adcfsad & ZTF & Ibc & spec & 0.025 & -17.15 & 0.07 & 3.71 & 10.35 & 0.02 & 1.39 & 0.02 & 0.76 & 0.01 \\
ZTF20aaekkuv & ZTF & Ibc & spec & 0.02 & -16.43 & 0.10 & 9.82 & 8.29 & 0.03 & 0.86 & 0.04 & 0.52 & 0.02 \\
ZTF20aaelulu & ZTF & Ibc & spec & 0.005 & -17.96 & 0.28 & 1.27 & -- & -- & -- & -- & -- & -- \\
ZTF20aaertpj & ZTF & Ibc & spec & 0.029 & -17.02 & 0.06 & 4.45 & 9.58 & 0.01 & 0.90 & 0.01 & 0.54 & 0.01 \\
ZTF20aahgejq & ZTF & Ibc & spec & 0.046 & -17.66 & 0.03 & 1.90 & 9.40 & 0.02 & 0.79 & 0.03 & 0.48 & 0.02 \\
ZTF20aaiftgi & ZTF & Ibc & spec & 0.034 & -17.85 & 0.06 & 1.47 & -- & -- & -- & -- & -- & -- \\
ZTF20aaiqiti & ZTF & Ibc & spec & 0.025 & -16.70 & 0.07 & 6.81 & 10.30 & 0.02 & 1.49 & 0.02 & 0.78 & 0.02 \\
ZTF20aajcdad & ZTF & Ibc & spec & 0.018 & -17.44 & 0.08 & 2.54 & 10.23 & 0.03 & 1.52 & 0.02 & 0.79 & 0.02 \\
ZTF20aalcyih & ZTF & Ibc & spec & 0.027 & -17.69 & 0.06 & 1.82 & 7.44 & 0.0 & -0.68 & 0.0 & 0.02 & 0.01 \\
ZTF20aalxlis & ZTF & Ibc & spec & 0.025 & -18.83 & 0.06 & 1.0 & 10.12 & 0.01 & 1.34 & 0.01 & 0.71 & 0.01 \\
ZTF20aammtwx & ZTF & Ibc & spec & 0.027 & -17.71 & 0.06 & 1.78 & -- & -- & -- & -- & -- & -- \\
ZTF20aamqmhj & ZTF & Ibc & spec & 0.048 & -18.04 & 0.04 & 1.15 & 10.56 & 0.03 & 1.95 & 0.02 & 0.97 & 0.02 \\
ZTF20aatzhhl & ZTF & Ibc & spec & 0.008 & -17.17 & 0.20 & 3.63 & 10.88 & 0.02 & 1.99 & 0.02 & 1.0 & 0.01 \\
ZTF20aavcvrm & ZTF & Ibc & spec & 0.055 & -18.15 & 0.03 & 1.0 & 6.56 & 0.14 & -0.60 & 0.14 & 0.04 & 0.21 \\
ZTF20aavgcnu & ZTF & Ibc & spec & 0.018 & -16.42 & 0.10 & 9.90 & -- & -- & -- & -- & -- & -- \\
ZTF20aavhyel & ZTF & Ibc & spec & 0.025 & -17.45 & 0.06 & 2.49 & 8.72 & 0.02 & 0.74 & 0.03 & 0.46 & 0.01 \\
ZTF20aavzffg & ZTF & Ibc & spec & 0.005 & -17.05 & 0.28 & 4.24 & 10.23 & 0.0 & 2.25 & 0.0 & 1.10 & 0.01 \\
ZTF20aaxhzhc & ZTF & Ibc & spec & 0.037 & -17.47 & 0.05 & 2.43 & 8.94 & 0.03 & 0.69 & 0.06 & 0.42 & 0.02 \\
ZTF20aaxvzja & ZTF & Ibc & spec & 0.033 & -16.95 & 0.06 & 4.86 & 10.64 & 0.02 & 1.46 & 0.02 & 0.77 & 0.01 \\
ZTF20aazkjfv & ZTF & Ibc & spec & 0.037 & -18.15 & 0.05 & 1.0 & 9.08 & 0.02 & 0.35 & 0.02 & 0.33 & 0.01 \\
ZTF20abbpkng & ZTF & Ibc & spec & 0.037 & -17.44 & 0.05 & 2.53 & 10.59 & 0.01 & 1.13 & 0.02 & 0.63 & 0.01 \\
ZTF20abbplei & ZTF & Ibc & spec & 0.031 & -18.62 & 0.05 & 1.0 & 10.82 & 0.02 & 1.88 & 0.02 & 0.92 & 0.01 \\
ZTF20abfcrzj & ZTF & Ibc & spec & 0.023 & -17.13 & 0.07 & 3.80 & -- & -- & -- & -- & -- & -- \\
ZTF20abhlncz & ZTF & Ibc & spec & 0.031 & -17.06 & 0.05 & 4.19 & 9.93 & 0.02 & 0.98 & 0.02 & 0.56 & 0.01 \\
ZTF20abjpvce & ZTF & Ibc & spec & 0.031 & -17.36 & 0.05 & 2.82 & -- & -- & -- & -- & -- & -- \\
ZTF20abqdkne & ZTF & Ibc & spec & 0.028 & -18.23 & 0.05 & 1.0 & 9.67 & 0.03 & 0.73 & 0.05 & 0.46 & 0.02 \\
ZTF20abswdbg & ZTF & Ibc & spec & 0.03 & -17.42 & 0.05 & 2.58 & -- & -- & -- & -- & -- & -- \\
ZTF20abtkjfw & ZTF & Ibc & spec & 0.017 & -16.96 & 0.08 & 4.80 & -- & -- & -- & -- & -- & -- \\
ZTF20abvquuo & ZTF & Ibc & spec & 0.03 & -17.95 & 0.06 & 1.29 & 8.25 & 0.03 & 0.85 & 0.05 & 0.50 & 0.03 \\
ZTF20abvvnqh & ZTF & Ibc & spec & 0.037 & -17.54 & 0.05 & 2.23 & -- & -- & -- & -- & -- & -- \\
ZTF20abwxywy & ZTF & Ibc & spec & 0.017 & -17.86 & 0.09 & 1.45 & 9.78 & 0.02 & 1.13 & 0.02 & 0.64 & 0.01 \\
ZTF20abwzqzo & ZTF & Ibc & spec & 0.023 & -17.21 & 0.07 & 3.42 & 10.46 & 0.03 & 1.78 & 0.02 & 0.88 & 0.02 \\
ZTF20abxpoxd & ZTF & Ibc & spec & 0.022 & -17.77 & 0.07 & 1.64 & 11.04 & 0.01 & -- & 0.01 & -- & 0.01 \\
ZTF20abywsut & ZTF & Ibc & spec & 0.031 & -17.65 & 0.05 & 1.92 & 10.45 & 0.02 & 1.61 & 0.02 & 0.79 & 0.02 \\
ZTF20abyznqs & ZTF & Ibc & spec & 0.05 & -18.22 & 0.05 & 1.0 & 10.25 & 0.01 & 1.25 & 0.03 & 0.66 & 0.01 \\
ZTF20abzjcdg & ZTF & Ibc & spec & 0.046 & -18.09 & 0.06 & 1.08 & -- & -- & -- & -- & -- & -- \\
\end{longtable}
\begin{tablenotes}
\item Notes: \\ $^\text{a}$Source of classification; spectroscopic or photometric. $^\text{b}$Peak $R$-band absolute magnitude. $^\text{c}$Weight from V$_\text{max}$ correction. $^\text{d}$Host galaxy stellar mass, expressed as $\log_{10}(M/M_\odot)$. $^\text{e}$Host galaxy rest-frame $U-R$ colour. $^\text{f}$Host galaxy rest-frame $B-V$ colour.
\end{tablenotes}
\end{ThreePartTable}

\newpage

\begin{table}
    \label{DES_phot_table}
    \centering
    \caption{DECam photometry of the DES CCSN sample presented in this paper. Values quoted are flux densities in f$_\lambda$. A full, machine-readable version of this table can be found in the electronic version of the article.}
    \begin{tablenotes}
    \item Notes: \\ $^\text{a}$Date of explosion. $^\text{b}$Phase with respect to explosion date. $^\text{c}$Values are $f_\lambda$ quoted in terms of $10^{-19}$ erg s$^{-1}$ cm$^{-2}$ \AA$^{-1}$, can be converted into AB magnitudes with $m = -2.5\log(f_\lambda) - ZP$ where $ZP$ is the zero point 20.802, 21.436, 21.866 and 22.214 for $griz$ bands respectively.
    \end{tablenotes}
    \begin{tabular}{|l|l|l|l|l|l|l|l|l|l|l|l|l|}
        \hline
    	SN & UTC & MJD & t$_\text{exp}^\text{a}$ & Phase$^\text{b}$ & g$^\text{c}$ & err & r$^\text{c}$ & err & i$^\text{c}$ & err & z$^\text{c}$ & err \\ 
    	& & & & & \multicolumn{6}{c}{($10^{-19}$ erg s$^{-1}$ cm$^{-2}$ \AA$^{-1}$)} \\
    	\hline
    	DES13C1anve & 20131111 & 56607.06 & 56614.59 & -6.20 & -3.86 & 6.33 & -1.72 & 3.34 & -0.99 & 2.43 & 3.34 & 2.16 \\
        DES13C1anve & 20131118 & 56614.07 & 56614.59 & -0.43 & 2.95 & 6.92 & -- & -- & -- & -- & -- & -- \\
        DES13C1anve & 20131118 & 56614.08 & 56614.59 & -0.42 & -- & -- & 6.91 & 3.21 & 5.98 & 1.83 & 2.08 & 1.23 \\
        DES13C1anve & 20131119 & 56615.10 & 56614.59 & 0.42 & -4.59 & 8.37 & 8.58 & 3.95 & 9.66 & 1.92 & 0.665 & 1.51 \\
        DES13C1anve & 20131202 & 56628.08 & 56614.59 & 11.11 & 22.8 & 2.47 & 17.6 & 1.67 & 13.7 & 1.34 & 10.7 & 1.29 \\
        DES13C1anve & 20131209 & 56635.13 & 56614.59 & 16.92 & 23.4 & 3.01 & 21.9 & 1.9 & 18.1 & 1.5 & -- & -- \\
        DES13C1anve & 20131209 & 56635.14 & 56614.59 & 16.93 & -- & -- & -- & -- & -- & -- & 11.3 & 1.2 \\
        DES13C1anve & 20131216 & 56642.09 & 56614.59 & 22.65 & -- & -- & -- & -- & 19 & 6.24 & 8.7 & 2.27 \\
        DES13C1anve & 20131219 & 56645.08 & 56614.59 & 25.12 & 17.4 & 5.12 & 15.1 & 2.77 & 16.6 & 1.65 & 11.6 & 1.11 \\
        DES13C1anve & 20131223 & 56649.17 & 56614.59 & 28.49 & 6.82 & 4.36 & 16.6 & 2.46 & 12.3 & 1.73 & 10.3 & 1.4 \\
        \hline \multicolumn{13}{c}{{\textbf{Full machine readable version in electronic version of article}}}
    \end{tabular}
\end{table}


\bsp	
\label{lastpage}
\end{document}